\documentclass[%
 reprint,
 amsmath,amssymb,
aps,
floatfix,
]{revtex4-2}

\usepackage{graphicx}
\usepackage{dcolumn}
\usepackage{bm}

\usepackage[colorlinks,citecolor=red,urlcolor=blue,bookmarks=false,hypertexnames=true]{hyperref} 
\usepackage[inline]{enumitem}

\usepackage[caption=false]{subfig}

\usepackage{siunitx}
\usepackage[version=4,arrows=pgf-filled,
mathfontname=mathsf]{mhchem}

\usepackage{tikz}
\usetikzlibrary{arrows}

\usepackage{xspace}
\usepackage{multirow}

\newcommand{\me}{\ensuremath{m_\mathrm{e}}\xspace}
\newcommand{\we}{\ensuremath{\omega_\mathrm{e}}\xspace}

\newcommand{\wwe}{\ensuremath{\Omega_\mathrm{e}}\xspace}
\newcommand{\wwI}{\ensuremath{\Omega_\mathrm{I}}\xspace}

\begin{document}

\preprint{APS/123-QED}

\title{Roadmap to planar electron-ion point Paul trap}

\author{Niklas Vilhelm Lausti}
\author{Vineet Kumar}%
 \author{Ivan Hud\'ak}%
 \altaffiliation[Also at ]{Institute of Photonics and Electronics CAS, v.v.i., Chaberská 1014/57, Prague 8, Czech Republic.}
\author{Michal Hejduk}%
 \email{michal.hejduk@matfyz.cuni.cz}
\affiliation{%
 Charles University, Faculty of Mathematics and Physics, Dept.\ of Surface and Plasma Science, Prague 8, Czech Republic
}%
\author{Michal Tarana}
\affiliation{J. Heyrovský Institute of Physical Chemistry of the Czech Academy of Sciences, Dolejškova 2155/3, 182 23 Prague 8, Czech Republic}

\begin{abstract}
We present a technical guide to developing a quantum-mechanical system with co-trapped laser-cooled ions and electrons, aiming to utilize this mixed-species system in quantum computing and sensing. We outline a method to control the system's quantum state and provide a blueprint for a forward-compatible design for containing it. The proposed technical solution features a planar configuration with a large trapping volume located at a considerable height above the electrode plane. We detail a manufacturing method using copper-coated, laser-machined glass substrates suitable for a high-power microwave drive signal. We discuss electron state decoherence in this trap and suggest that using superconductive films could enhance trapping abilities, though initial experiments are feasible with the current design.
\end{abstract}

\maketitle


\section{\label{sec:intro}Introduction}

Ion traps are an essential tool for experiments in atomic and quantum physics: they are used in the research of ultra-cold chemistry \cite{petralia_strong_2020}, quantum sensing\cite{gilmore_quantum-enhanced_2021}, metrology\cite{pyka_high-precision_2014} and computing\cite{bruzewicz_trapped-ion_2019}. Trapped ion qubits boast long coherence times that exceed one hour\cite{wang2021single}. Additionally, ion qubits can have high fidelities: some 99.9\% for a two-qubit gate and over 99.99\% for a single-qubit gate have been reported \cite{ballance2016high}. However, the gate operation time of a trapped-ion entanglement gate, a parameter important for some quantum algorithms\cite{fowler_surface_2012,gidney_how_2021}, is on the order of \SI{1}{\micro \second} at least, and is significantly longer than in superconducting quantum gates\cite{kandala_demonstration_2021} or those made of cold Rydberg atoms\cite{chew_ultrafast_2022}. This is because, although there have been some advancements\cite{wang_fast_2022,saner_breaking_2023}, the entanglement process is predominantly facilitated by the vibrational modes of the ion chain. The narrow energy gap between these vibrational modes, resulting from the charge-to-mass ratio and, hence, the slow response time to changes in local electric fields, indicates a low quantum speed limit\cite{deffner_quantum_2017}. The charge-to-mass ratio  also remains a constraining factor in quantum computing chips, where ions need to be moved from one location to another (so called QCCD chips \cite{lekitsch_blueprint_2017,pino_demonstration_2021}).

To overcome these limitations, entangling the ions via electrical dipoles induced by Rydberg excitation has been proposed and demonstrated.\cite{zhang_submicrosecond_2020} Indeed, Rydberg ions are excellent candidates for qubits as they offer states with long lifetimes that scale as \(n^3l^2\)\cite{flannery_quantal_2003} (\(n\) and \(l\) being principal and angular momentum quantum numbers, respectively) and large electrical and optionally magnetic\cite{cohen_quantum_2021} dipole moments. The dipole-entangled gate mentioned above reaches the operation time \SI{700}{\nano \second}. The question is whether there is any other entangling channel that would offer shorter gate operation times.

If one exceeds the ionisation threshold in the process of electronic excitation of the trapped ions, a free unbound electron is produced. Trapped electrons react to changes in the electric field more rapidly thanks to their mass, which is 16,000 times smaller than the lightest ions laser-cooled so far\cite{wineland_laser-fluorescence_1983}. That can be exploited in Paul traps to obtain single particle quantum harmonic oscillator energy gaps on the order of hundreds of \si{MHz}, if realistic trap geometries and parameters such as driving frequency of \(2 \pi \times \SI{2}{GHz}\), amplitude of the driving signal on the order of \SI{100}{V}, and electrode separation of \SI{1}{mm} are used.\cite{matthiesen_trapping_2021} Imitating existing ion trap quantum computers, a single qubit can be encoded using the electron spin, with entanglement facilitated by the collective motion.\cite{yu_feasibility_2022} This should result in gate operation times that are several orders of magnitude faster than those in trapped-ion quantum computers.

However, the whole field is still in its infancy. So far, state preparation and readout protocols have been formulated. It has been proposed to couple the collective state of electrons with the high-quality superconducting resonator\cite{kotler_hybrid_2017} or transmon qubits\cite{osada_feasibility_2022} to open an electronic interface for cooling and state readout. To satisfy rather stringent requirements for the coupling strength, the accuracy of phonon number measurement\cite{kotler_hybrid_2017} and the signal-to-noise ratio, temperature well below \SI{1}{K} must be maintained, which conflicts with the scalability of quantum devices.

In the prospective electron trap quantum computer, the spin-state degeneration of the electrons is removed in the presence of the magnetic field. Using miniature electromagnets to tune its magnitude locally, the resonance frequency of the spin-flip transition\cite{hammond_spin_2012} can be set to the frequency of the control radio frequency wave\cite{siegele-brown_fabrication_2022}. The gradient of the magnetic field introduces a spin-state-dependent force, which couples to the motion of an electron -- different spins cause different phase responses of a current in a pick-up electrode induced by the electron located in a \SI{10}{\micro\meter} scale distance.\cite{peng_spin_2017} Here again, temperature of the trap as low as \SI{4}{K} is required to increase the signal-to-noise ratio, which can be a challenge in the presence of \SI{1}{A} current that needs to be supplied to the electromagnet.\cite{yu_feasibility_2022}

At this moment, no technical solution for combining the manipulation of individual spins with the control of collective states in a setting with multiple electrons is available. (Recently, a concept of QCCD-like architecture that relies on cryogenic cooling methods was presented.\cite{huangNumericalInvestigationsElectron2025}) One of the biggest achievements in the field, laser cooling\cite{itano_cooling_1995} that is able to bring ions to submillikelvin temperatures is not naturally applicable in pure ``electron-only'' devices. This is a significant setback, since recent advancements indicate that laser systems are more scalable compared to cryogenic devices. This is evident from the wafer-level integration of traps with light guides\cite{mehta_integrated_2020}, the development of fibre-based optical cavities\cite{gulati_fiber_2017}, and micro-optics\cite{doherty_multi-resonant_2023}. Commercial rack-based trapped-ion computers are smaller in total than superconducting qubit ones.\cite{pogorelov2021compact}

One of the ways in which the laser-cooling method can be employed in the service of free electron qubits is to couple the electron motion with the quantum system of a laser-cooled ion. For example, a ``tin can telephone'' connecting an ion trap and an electron trap by a wire can be constructed to couple the two corresponding quantum harmonic oscillators.\cite{yu_strong_2024} The surface electric field noise has to be reduced by cryogenic cooling here as well to reach the quantum-coherent coupling.\cite{an_coupling_2022} By removing the wire connection and partially overlapping electron and ion Paul traps, a coupling between the two quantum systems can be mediated by the Coulomb interaction.\cite{osada_feasibility_2022} However, the technical solution as presented by \citet{osada_feasibility_2022} exposes the electrons to a strong electric field coming from the ion trap electrodes because electric field minima of both traps do not overlap by definition.

In this paper, we show a feasible way of engineering a quantum-mechanical system, in which the particle trap, the ions and the electrons are coupled to provide laser-accessible transitions for manipulating motional and spin states of ``quasi-free'' electrons (see Section \ref{sec:exp}). The core of the idea lies in the ability to store electrons and ions simultaneously in the same space, which is not a trivial problem. The trapping of electrons without ions has been demonstrated in Penning\cite{fan_one-electron_2022} and Paul traps\cite{matthiesen_trapping_2021}, and above the liquid helium\cite{kawakami_blueprint_2023} and solid neon\cite{zhou_single_2022} surfaces. The simultaneous trapping of protons, \ce{H_2^+} ions and electrons with undefined temperatures has been achieved in a combined Paul/Penning trap\cite{walz_combined_1995}. Our approach exploits the idea of the two-frequency Paul trap,\cite{foot_two-frequency_2018} which has been proposed to be used in generating antihydrogen atoms by recombination\cite{leefer_investigation_2016} or in quantum sensing using nanoparticles.\cite{bykov_nanoparticle_2024,gonzalez-ballestero_levitodynamics_2021} From our perspective, the method allows us to avoid using the magnetic field that causes Zeeman splitting of energy levels (and hence more complicated laser cooling and excitation schemes) and centrifugal separation introducing additional unknown variables to the electron-ion collision system.\cite{oneil_centrifugal_1981}

Our technical concept of the trap is based on a planar design, as opposed to the ``three-dimensional'' one. Although we are successfully developing a three-dimensional 3D-printed trap\cite{kumar_3d-printed_2025} based on the design of \citet{jefferts_coaxial-resonator-driven_1995} and have contributed to testing of the trap \cite{mikhailovskii_trapping_2025} based on the design of \citet{matthiesen_trapping_2021}, we believe that planar models offer benefits in the long term. Large (in the sense of few \si{cm}) three-dimensional structures are difficult to manufacture in optimal shape with precision (which is limited by the methods of machining and assembly, typically significantly worse than \SI{1}{\micro\meter}). Due to their enclosed shape, integration with optical and electronic elements is difficult. On the other hand, planar traps are easier to manufacture on a large scale with sub-micrometre features.\cite{ragg_segmented_2019} Moreover, integration with photonic layers\cite{mehta_integrated_2020} and superconducting or conventional electronic layers is an active field of development.\cite{tolpygo_progress_2023} Note that miniaturisation is important to increase the coupling between trapped electrons and electronic components, as well as the scalability of the design.

In the design described here, we have addressed the challenge of maintaining a symmetric trapping field that can confine \ce{Ca+} ions and electrons at a relatively large distance ($\approx \SI{1}{mm}$) from the trap electrodes. These two requirements often conflict with one another because ensuring the uniformity of electric field above all trap electrodes is difficult if their sizes are comparable to the wavelength of the driving signal. This issue is described in Section \ref{sec:trap}. Section \ref{sec:heating} examines how effectively we can reduce electron heating by placing them at such a considerable distance from the electrodes. This challenge is associated with the manufacturing method, which is discussed in Section \ref{sec:manufacturing}. However, before all that, we would like to elaborate on a concrete physical system that we would like to examine in our proof-of-concept experiments in Section \ref{sec:exp}.

\section{Physical concept}
\label{sec:exp}

Currently, we are preparing a series of theoretical studies and experiments, in which we aim to understand and investigate a quantum-mechanical system of one \ce{Ca^{2+}} ion and one electron trapped in the potential of a two-frequency Paul trap where one radio-frequency potential effectively traps the ion and the other traps the electron in the highly excited state.

\begin{figure}
    \centering
    \includegraphics[width=\linewidth]{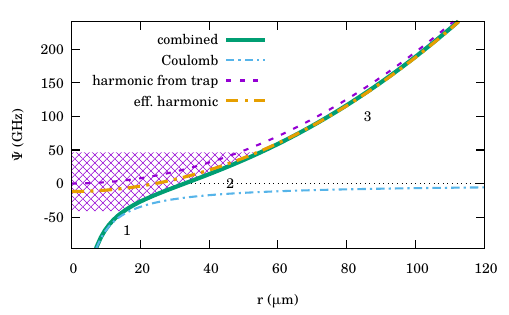}    
    \caption{Radial part of the potential experienced by the electron, and its regions, marked as enumerated in Section~\ref{sec:categories}. The boundaries of the transition region 2 are selected as energies, at which the combined potential starts deviating from the Coulomb or effective harmonic profile by more than 10\%. The figure neglects the influence of slowly varying ion trapping field, for that can be dealt as a perturbation under certain conditions mentioned in Section \ref{sec:trappingpotentials}.}
    \label{fig:potentials}
\end{figure}

\subsection{Categories of electronic states}
\label{sec:categories}

In order to understand a qualitative nature of different electronic states of the trapped \ce{Ca+}, an approximative and computationally tractable model is presented in this subsection. In further discussion, it is assumed that the precursor ion \ce{Ca^+} is initially laser-cooled to the Doppler limit and it is located in the centre of the Paul trap at all times. The potential $\Psi$ experienced by the electron is a superposition of the fields generated by the ion and the trap: specifically, the Coulomb potential $V_C(r) = -{k_\mathrm{e} Z Q_\mathrm{e}^2}/{r}$, where $k_\mathrm{e}$ is the Coulomb constant, $Q_\mathrm{e}$ is the elementary charge, and $Z = 2$ is the charge number of the ion; and the time-independent pseudopotential $V_P(r) = {k r^2}/{2} = {\me \we^2 r^2}/{2}$, where $\me$ is the electron mass and $\we$ is the secular frequency of the electron in the Paul trap. Since the goal of this model is only to distinguish qualitatively different categories of the states, we employ a spherical approximation of the trap now, as is illustrated in Figure \ref{fig:potentials}. The more realistic non-spherical trapping potential is discussed in Section~\ref{sec:trappingpotentials}.

\begin{figure}
    \centering
    \includegraphics[width=\linewidth]{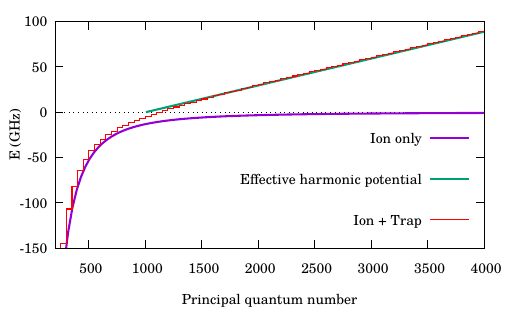}
    \caption{Energy levels of an electron-\ce{Ca^{2+}} system confined within a spherically symmetric trapping potential. For clarity, only every 50th energy level of states with zero angular momentum is plotted. As the Coulomb well deforms the harmonic potential, the extrapolation of the linear dependence does not intersect the origin.}
    \label{fig:energylevels}
\end{figure}

The quantum mechanical Hamiltonian of the electron in the potential discussed above ($\omega_e=2\pi\times26.87\thinspace$MHz) has been expressed in a B-spline basis and diagonalized. The obtained energy levels, considering only the $s$ states, are plotted in Figure~\ref{fig:energylevels}.

We notice that three qualitatively different categories of states of such a system are recognized:
\begin{enumerate*}
    \item The most straightforward case is where the electron is in its ground state or in a low excited state of the atomic ion (\ce{Ca+}) at an energy level well below the ionization threshold. Then the effect of the trap can be either entirely neglected, or treated as a weak perturbation.
    
    \item\label{it:medium} As the energy of the excited atomic ion increases, the classical turning point of the electron moves further from the \ce{Ca^{2+}} core, causing the electronic wave function to become more delocalized and exhibit Rydberg characteristics. The gaps between energy levels narrow as the excitation energy nears the ionization threshold of a free \ce{Ca+} ion, but never get as small as in the case of a Rydberg ion because of the deformation of the potential landscape by the trap. At a certain level, not necessarily at the ionization threshold, the gaps start widening to align with the spectrum in the subsequent regime.
    
    \item\label{it:far} Further above the ionization potential of \ce{Ca^{+}}, the electron is bound by the potential of the trap rather than by the Coulomb interaction with the ion core. The amplitudes and driving frequencies of the trap determine the spatial extent of the electron wave function. Therefore, the character of the states of the electron above the ionization threshold resembles the motional states of the trapped ions \cite{leibfried_quantum_2003}. Nevertheless, the electron dynamics remains influenced by the ion core, as well. Fitting the long-range region of the potential $\Psi$ for $r\gtrsim\qty{100}{\micro\meter}$ to an ``effective'' harmonic potential $V_{\mathrm{eff}}(r)=V_S + k_{\mathrm{eff}}r^2/2$ yields a negative shift $V_S=\qty{-6.6}{GHz}$ and an effective spring constant $k_{\mathrm{eff}}$ approximately 1\% higher than the value in the harmonic potential term $V_P(r)$ of the potential $\Psi$ (shown as eff. harmonic in Figure \ref{fig:potentials}).
\end{enumerate*}

 Figure~\ref{fig:energylevels} reveals a gradual transition from hydrogen-like energy level spacing to equidistant spacing characteristic of a quantum harmonic oscillator. The transitional region, with principal quantum numbers ($n$) ranging from \numrange{700}{2000} (corresponding to region 2 in Figure \ref{fig:potentials}), exhibits an abundance of $\Delta n = 1$ transitions in the \qty{10}{MHz} range, with frequency differences of \qtyrange{1}{10}{kHz}. As discussed in Section \ref{sec:coupling}, these transitions (tuneable by parameters of the trapping field) could potentially be exploited for quantum computing applications. The harmonic oscillator-like region, conversely, may facilitate high-resolution photon counting.

Although trapped Rydberg ions have been extensively studied previously (see reference \cite{andrijauskas_rydberg_2021} and the references therein), the understanding of states in cases \ref{it:medium} and \ref{it:far} (so-called ``trap-induced'' states) remains limited. The primary objective of the experimental approach discussed in this paper, once fully developed, is to excite Rydberg states and trap-induced states, as well as investigate their mutual coupling.

\begin{figure}
\centerline{
  \vspace{5mm}
  \resizebox{7cm}{!}{
    \begin{tikzpicture}[
      scale=1,
      level/.style={thick},
      virtual/.style={thick,densely dashed},
      hostate/.style={thick, dash dot},
      trans/.style={thick,<->,shorten >=2pt,shorten <=2pt,>=stealth},
      upwards/.style={thick,->,shorten >=0pt,shorten <=0pt,>=stealth},
      upwardsd/.style={thin,double,->,shorten >=0pt,shorten <=0pt,>=stealth},
      classical/.style={thin,double,<->,shorten >=4pt,shorten <=4pt,>=stealth}
    ]
    \draw[level] (1cm,-2.4em) -- (0cm,-2.4em) node[left] {3$^2$D$_{3/2}$};
    \draw[level] (2cm,0em) -- (1cm,0em) node[left] {4$^2$P$_{1/2}$};
    \draw[level] (2cm,-5em) -- (3cm,-5em) node[right] {4$^2$S$_{1/2}$};
    \draw[level] (0cm,9em) -- (3cm,9em) node[right] {6$^2$S$_{1/2}$};
    \draw[hostate] (-1.2cm,16em) -- (4.2cm,16em) node[right] {};
    \draw[hostate] (-1.4cm,16.8em) -- (4.4cm,16.8em) node[right] {};
    \draw[hostate] (-1cm,15.2em) -- (4cm,15.2em) node[right] {TI};
    
    \draw[trans] (2.5cm,-5em) -- (1.5cm,0em) node[midway,right] {{1}};
    \draw[trans] (0.5cm,-2.4em) -- (1.5cm,0em) node[midway,right] {{2}};
    \draw[trans] (1.5cm,0em) -- (1.5cm,9em) node[midway,left] {{3}};
    \draw[upwards] (2.5cm,9em) -- (2.5cm,15.2em) node[midway,left] {{4}};
    \draw[upwards] (0.5cm,16.8em) -- (0.5cm,9em) node[midway,left] {{6}};
    \draw[upwardsd] (1.5cm,15.4em) -- (1.5cm,16.8em) node[above] {{5}};
    \end{tikzpicture}
  }
}
\caption{Energy levels and transitions used in the experiments with the \ce{Ca^{2+}}-electron system. The wavelength values and nomenclature of the beams are given in Table \ref{tab:beams}. 
\textit{TI}, also dash dot lines -- trap-induced levels. 
Not to scale.}
\label{fig:levels}
\end{figure}
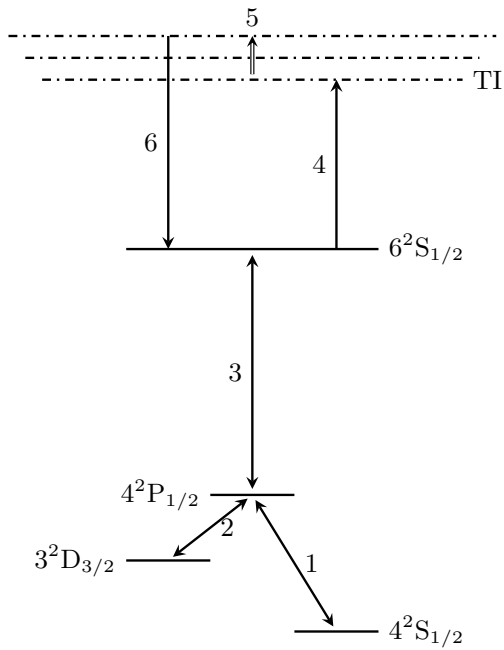

Experimentally, the discussed particle system can be generated by photoionization of a Doppler-cooled \ce{Ca+} ion in a cascade of transitions illustrated in Figure \ref{fig:levels}. In this way, we avoid electron trapping instability caused by anharmonic potentials near the electrodes\cite{matthiesen_trapping_2021}.  Beams 1 and 2 secure the Doppler-cooling of \ce{Ca+} ions using a standard three-level scheme \cite{hayasaka_laser_2000}. Then, a \SI{220}{nm} laser excites the singly charged ion from the 4$^2$P$_{1/2}$ state to 6$^2$S$_{1/2}$. The necessary wavelength can be achieved by frequency-doubling a blue laser light with a wavelength of 440 nm \cite{tangtrongbenchasil_219-nm_2006}. From this state, beam 4 promotes the electron to a state above the ionisation threshold of \ce{Ca^{+}}.
Single-photon transitions between different trap-induced states with energy separations on the rf-scale is in Figure \ref{fig:levels} denoted by the coupling 5.

Beam 6 causes the stimulated emission from a trap-induced state to a bound state (6$^2$S$_{1/2}$ in this case), resembling the laser-induced recombination \cite{wolz_stimulated_2020}. The ion can be brought to the laser-cooling states by stimulated emission induced by beam 3.

\begin{table}
    \centering
    \begin{tabular}{clr}
         $n$ & name & $\lambda_n$ (nm) \\
         \hline
         1 & cooling & 397\\
         2 & repumping & 866\\
         3 & excitation & 220\\
         4 & ionisation & $\approx 399$\\
         5 & control    & \textit{see the text} \\
         6 & recombination & \textit{optical or near-UV range}
    \end{tabular}
    \caption{Electromagnetic waves used in the experiment. The wavelength values ($\lambda_n$ where $n$ is the beam number from Figure \ref{fig:levels}) are rounded to nanometres.}
    \label{tab:beams}
\end{table}

\subsection{Trapping potentials}
\label{sec:trappingpotentials}

To accommodate the ion alongside the electron, the Paul trap is driven by an rf-voltage signal
\begin{equation}
\label{eq:waveform}
V(t) = V_{\mathrm{e}0}\cos(\Omega_\mathrm{e} t) + V_{\mathrm{I}0} \cos(\Omega_\mathrm{I}t + \phi)
\end{equation}
where subscripts indicate to which particles the terms primarily attribute -- either electrons (the subscript $\mathrm{e}$) or much heavier ions ($\mathrm{I}$). The voltage amplitudes and angular frequencies are denoted by $V_{\mathrm{e,I}0}$ and $\Omega_{\mathrm{e,I}}$, respectively. The phase $\phi$ is typically not controlled. The spatial component of the time-dependent potential inside the trap is considered quadrupole near the centre
\begin{equation}
    U(x,y,z)=\frac{1}{2r_0^2}(x^2+y^2-2z^2).
\end{equation}
 Here, $r_0$ stands for characteristic trap dimension, which can be assumed as trap height for a harmonic planar trap.

To facilitate the formulation of the physical concept of the electron trapping, the motion of the ion in the trap will be neglected in the text below, as well as its interaction with the electron in the trap-induced motional states of the electron. While the range of the Coulomb potential of the ion core is fundamentally infinite, the spatial scale on which the electron is confined by the potential of the trap is also large. This decreases the relative influence of the ion on the wave function of the electron with increasing energy of its motional state. Classical simulations of the electron-ion system in the trap \cite{mahmoudi_tarana_hejduk} show that  the Coulomb interaction of the electron with the ion can be safely neglected  at initial electron energies on the order of \qty{1}{meV} and higher.

With these approximations, the quantum mechanical Hamiltonian of the electron in the trap is
\begin{equation}
    \hat{H}(t)=-\frac{\hbar^2}{2m_{\mathrm{e}}}\nabla^2+eV(t)U(x,y,z),
\end{equation}
where $e$ is the charge of the electron.

In the rest of the text, the discussion is restricted to one dimension for the simplicity of the equations. Generalization to three dimensions is straightforward. The stability of a single frequency trap is generally evaluated in terms of the stability parameter $q$ 
\cite{leibfried_quantum_2003,foot_atomic_2005}. Defining it for the electron and ion signal individually, as
\begin{equation}
\label{eq:qs}
  q_{\mathrm{e}} = \frac{2eV_{\mathrm{e}0}}{m_{\mathrm{e}}r_0^2\Omega_{\mathrm{e}}^2},\;\;\; q_{\mathrm{I}} = \frac{2eV_{\mathrm{I}0}}{m_{\mathrm{e}}r_0^2\Omega_{\mathrm{I}}^2},
\end{equation}
respectively, allows for the characterization of the regime of stability for each trap in the absence of the other.

For the trapping of the electrons discussed in this paper to work, we require $V_{\mathrm{e}0} \gg V_{\mathrm{I}0}$ and $\Omega_\mathrm{e} \gg \Omega_\mathrm{I}$ \cite{foot_two-frequency_2018}. Under such conditions, the electron motion can be approximated by three periodic components: One is the secular motion with the angular frequency
\begin{equation}
\label{eq:secular}
\omega_{\mathrm{e}}\approx\frac{q_\mathrm{e}\Omega_e}{2\sqrt{2}},
\end{equation}
the remaining two high-frequency components are induced by the electron and ion trap with the angular frequencies $\gamma_{\mathrm{e}}(n_\mathrm{e})=\omega_\mathrm{e}+n_\mathrm{e}\Omega_\mathrm{e}$ and $\gamma_{\mathrm{I}}(n_\mathrm{I})=\omega_\mathrm{e}+n_\mathrm{I}\Omega_\mathrm{I}$, respectively, where $n_\mathrm{e}$ and $n_\mathrm{I}$ are integers.

In an analogy with the motional states of a single cold ion in a single-frequency Paul trap, in the situations when   $q_\mathrm{e},q_\mathrm{I}\lesssim1$, only the amplitude of the secular component of the motion is dominant, and the interaction of the electron with the trap can be approximated by a harmonic time-independent pseudopotential $\hat{V}_P=m_\mathrm{e}\omega_\mathrm{e}^2\hat{z}^2/2$, assuming an averaging over the small amplitudes of the micromotions due to the ion and electron trap. However, since the electrons are much lighter than the ions, it is realistic to anticipate that the higher-frequency components with $\gamma_{\mathrm{e}}(n_\mathrm{e}>1)$ and $\gamma_{\mathrm{I}}(n_\mathrm{I}>1)$ will not be negligible.

Another view of the applicability of the time-independent pseudopotential approximation is provided by \cite{Cook_1985} in terms of the parametric resonances. When the driving angular frequencies $\Omega_\mathrm{e}$ and $\Omega_\mathrm{I}$ are chosen so that they are not close to the parametric resonances
\begin{equation}
\label{eq:parametric}
\Omega^{\text{res}}_\mathrm{e,I}(n)=\sqrt{\frac{\sqrt{2}eV_\mathrm{e0,I0}}{r_0^2 m_\mathrm{e}n}},
\end{equation}
where $n$ is an integer, then the transitions between the harmonic eigenstates of $\hat{V}_P$ due to the non-secular components of the motion are minimal. Consequently, the pseudopotential approximation is valid and the energy levels form the ladder structure as depicted in Figure \ref{fig:levels} provides a good picture.

We aim for the saddle of the field generated by $V_\mathrm{I}$ to overlap with the minimum of the trapping field originating from $V_\mathrm{e}$. This is not guaranteed as long as $V_\mathrm{e}$ and $V_\mathrm{I}$ are brought to separate electrodes, for example, in a geometry where the electron trapping electrodes are surrounded by the ion trapping electrodes \cite{osada_feasibility_2022}.

\subsection{Optical couplings}
\label{sec:coupling}

During the state-manipulation process, it is crucial to maintain the ion at its initial position, which means that cooling must continue. This can be achieved through the fluorescence caused by beams 1 to 4 as shown in Figure \ref{fig:levels}.

\begin{figure}
    \centering
    \includegraphics[width=\linewidth]{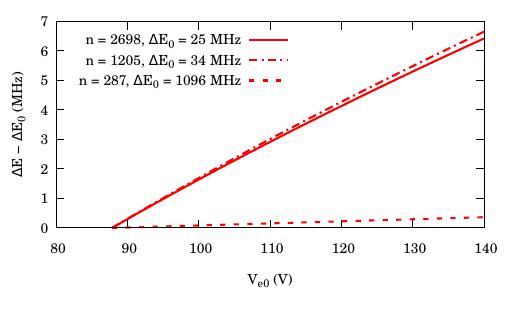}
    \caption{Tuning transition 5 (as depicted in Figure \ref{fig:levels}) adjusting electron trapping voltage amplitude $V_\mathrm{e0}$. The variation of energy required to alter the quantum number of the model system from Figure \ref{fig:energylevels} by one is demonstrated for states located in the harmonic-potential region ($n=2698$), transition region ($n = 1205$) and the Coulomb potential well ($n=287$) in relation to the trapping parameter. $\Delta E_0$ refers to the transition energy at the initial value of the amplitude $V_\mathrm{e0} = \qty{88}{V}$.}
    \label{fig:ediftuning}
\end{figure}

An experiment designed with this coupling approach involves detecting photons at frequencies suitable to provide resonant coupling 5. This technique achieves this by observing variations in electromagnetically induced transparency in transition 2 \cite{qiElectromagneticallyInducedTransparency2009,finkelstein_practical_2023} caused by the presence or absence of coupling 5. Figure \ref{fig:ediftuning} demonstrates, utilizing the spherically symmetric simplified model from Section \ref{sec:categories}, how the highest transition can be adjusted by altering the amplitude of the electron trapping voltage signal $V_\mathrm{e0}$. Even though the plotted energy differences are only between the $s$-states, and, therefore, do not correspond to single-photon transitions, they provide an insight into scaling of the transition energies with the trapping parameter, as well as the position of the initial quantum state (prepared through excitation 4): transitions within \qtyrange{10}{1000}{MHz} can be targeted by a proper choice of the potential region 1, 2 or 3 from Figure \ref{fig:potentials}. More realistic model in the future will elucidate which transitions (and by how many quanta) are permitted. This is an instance of applying the proposed experimental method without needing electron spin state manipulation.

In other applications, where manipulation of the electron spin state is desired, a magnetic field has to be superimposed on the trapping field. At the flux density of $\gtrsim\SI{1}{mT}$, the spin state splitting is comparable to $\we$ of our trap described below. The same field also provides clearly distinguishable optical transitions from spin-up and spin-down states of the magnetic-split state 6$^2$S$_{1/2}$ in the optical or near-UV wavelength range (around $\lambda_4$). This could be exploited to encode quantum information without the need to use microwaves.

Further interesting modifications of the experiment include the generation of a three-particle system of \ce{Ca^{2+}}-\ce{Ca^{2+}}-electron, in which the electron appears in a Schr\"odinger-cat state, a superposition of states of electron ``being above ion 1'' and ``being above ion 2'' \cite{lesanovsky_trap-assisted_2009}. This category of experiments promises coupling of both ions via sharing one electron, specifically by a laser coupling of the electronic states above the ionization threshold (where the electron reaches both cores) with the lower states (where the electron is localized only to one of the cores). This can create a coherent superposition of the states $\mathrm{Ca}^{2+}+\mathrm{Ca}^+$ and $\mathrm{Ca}^++\mathrm{Ca}^{2+}$, which may be interesting in quantum information technologies.
So far, the closest approach to our goal has been made in experiments in Rydberg ion excitation \cite{andrijauskas_rydberg_2021}.

Here, we note that the trap-induced states of the electron above the ionization threshold are defined by the trap geometry and the driving signal. These parameters are under the experimentalist's control and can be tailored to the specific needs of the quantum technological application. One such is a converter of a radio-frequency to an optical signal that follows the example of microwave counterparts frequently implemented using the Rydberg atoms \cite{vogt_efficient_2019,borowka_continuous_2024}. In principle, the design of the converter based on trapped electrons can be adapted to signals with a wide range of frequencies ranging from high to ultrahigh (in terms of the definitions of the International Telecommunication Union \cite{ITU2015}), simply by changing the trap dimensions and driving signal frequencies.

The common denominator of all the applications mentioned above is the simultaneous trapping of the laser-cooled ions and electrons detached from them. This is the main goal of this article. As no experimental approach to the laser cooling of the \ce{Ca^{2+}}-electron system has been proposed so far, we choose such parameters of the trap that provide the confinement volume for tens of ions to be able to achieve the sympathetic cooling of \ce{Ca^{2+}} by \ce{Ca+} ions. That comes at the expense of narrow energy gaps between the trap-induced states of the electrons that complicate the preparation of specific states. Furthermore, the stability of the electrons within the ion component of the two-frequency trap restricts $V_{\mathrm{I}0}$ to low values. Consequently, the trapping field for the ions becomes shallow, and the ion-ion separations become large. That can be suboptimal in the experiments involving multiple ions described above. On the other hand, the large size of the trap means that the trapping volume is distant from the the circuitry -- we describe the benefits of this fact in Section \ref{sec:heating}.

\section{Planar trap design}
\label{sec:trap}

\begin{figure}
    \centering
 \vspace{5mm}\includegraphics[width=0.6\linewidth]{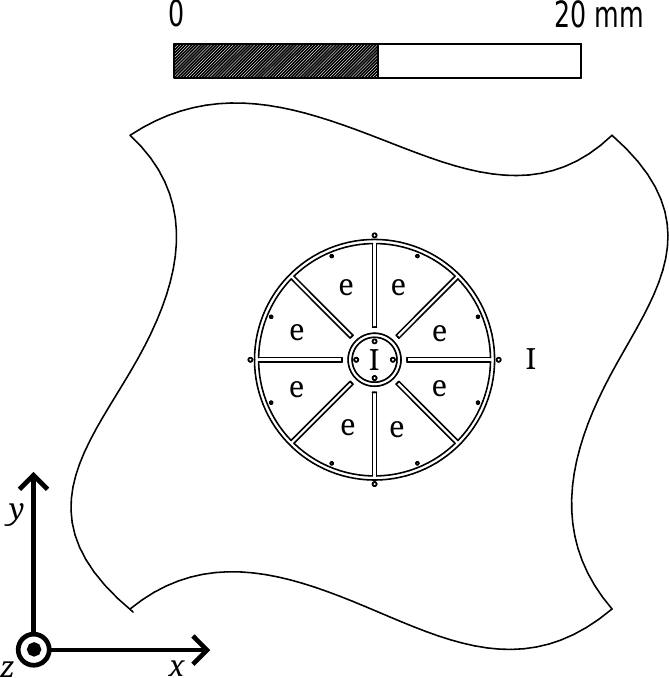}
    \caption{Electrode configuration of ring trap, a.k.a. point Paul trap used in this study. View from top. The driving signal with frequency \wwe\ is delivered to electrode sectors marked ``e''. The signal with frequency \wwI\ is delivered to electrodes marked ``I''. The outermost electrode extends to the edges of the wafer.}
    \label{fig:electrodes}
\end{figure}

We implemented a ring trap design\cite{kim_surface-electrode_2010,clark_ideal_2013,wang_surface_2015} illustrated in Figure \ref{fig:electrodes}, which negates the necessity of static electric fields that could impede the confinement of oppositely charged particles. The component of the driving rf-signal  with the angular frequency \wwe in Equation (\ref{eq:waveform}) is transmitted to the segments of the ring electrode. In theory, the entire waveform from Equation (\ref{eq:waveform}) can be applied to these segments; however, as we discuss below, this is challenging to achieve. Consequently, the driving signal with the angular frequency \wwI is introduced to the regions both inside and outside the ring electrodes, commonly referred to as ground electrodes.

\begin{figure}
\label{fig1}
    \centering
    \includegraphics[width=60mm]{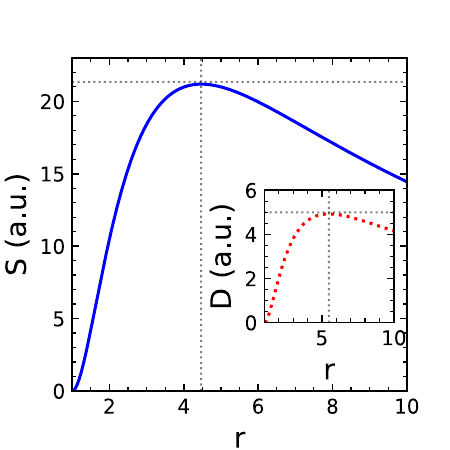}
\caption{\label{fig:1} Dependence of trap steepness $S$ on ratio of RF ring outer and inner radius ($r$). The inset shows the corresponding dependence for the trap depth $D$. $S$ and $D$ are in arbitrary units, as the absolute values depend on other parameters, such as applied voltage and frequency.}

\end{figure}

This geometry has previously been optimised to maximise the depth of the ion trap for a fixed distance from the centre of the ion trapping potential from the surface \cite{kim_surface-electrode_2010}. For our experiments, we require clear separation of the energies of the electronic states induced by the trap. Therefore, our objective is to maximise the steepness of the confining pseudopotential ($V_P$) 
\[S = \frac{D}{z_\mathrm{max} - h}\]
measured along the rotational symmetry axis of the trap $z$.  Here, $z_\mathrm{max}$ and $h$ denote the turning point and the location of the centre of the trapping field $V_P$, respectively.  The trap depth $D$, defined as $V_P(z_\mathrm{max}) - V_P(h)$, has previously been expressed as a function of the inner radius of the electrode $a$, the outer radius $b$ and $z_\mathrm{max}$. Similarly, expressions for $z_\mathrm{max}$ and $h$ have also been derived \cite{kim_surface-electrode_2010}. Here, we look for a ratio $b/a = r > 1$ that optimises the steepness $S$ at a given value of $h$. In our particular electrode geometry, the values of $a$ and $h$ are linked as
\begin{equation}
\label{eqn:Eq.1}
a = \frac{\sqrt{1 + r^{2/3}}}{r^{2/3}}h\,.
\end{equation}
All terms ($a$, $b$, $z_\mathrm{max}$, $D$) can now be expressed in terms of $r$, with $h$ appearing in scaling factors not relevant to optimisation. The steepness $S$ is then expressed as
\begin{widetext}
\begin{equation}
\label{eqn:Eq.2}
S \sim\frac{r^{8/5}\left( r^{4/5} - 1 \right)^{11/2}}{\left( r^{2/3} + 1 \right)\left( r^{2} - 1 \right)^{3}\left( \left( \left( r^{2/3} + 1 \right)\left( r^{6/5} - 1 \right) \right)^{1/2} - r^{4/15}\left( r^{4/5} - 1 \right)^{1/2} \right)}\,
\end{equation}
\end{widetext}
and has its real maximum at $r = 4.47$. Trap depth can also be written in a similar way,
\begin{equation}
\label{eqn:Eq.3}
D \sim\frac{r^{4/3}(r^{4/5}-1)^5}{(r^{2/3}+1)(r^2-1)^3}\,.
\end{equation}
This gives a maximum trap depth for $r = 5.49$. The dependences of $S$ and $D$ on $r$ are plotted in Figure \ref{fig:1}.

If the real trap height is desired to be $h=$ \SI{1.8}{mm}, the radii should be about $a=$ \SI{1.3}{mm} and $b=$ \SI{5.7}{mm}, which gives $r = 4.38$ -- a value not far from the optimum for the steepness. We have verified with the SurfacePattern software package\cite{schmied_optimal_2009,schmied_surfacepattern_nodate} that this design creates a stronger potential than a combination of several rings at a comparable radius of the electrode assembly, which may be explained by a larger electrode surface area. 

The selection of a significant height, in contrast to other planar electron traps \cite{yu_feasibility_2022}, is motivated by the objective of mitigating the adverse effects of surface charge irregularities on trapping stability. Although this choice renders electron cooling and state readout via interaction with cryogenically cooled electric circuits unachievable, our focus lies on employing laser-cooled ions as the cooling and state-readout mechanism. The large trapping volume given by the trap radii allows for confinement of tens of ions alongside the electron, which will aid in the ion-mediated quantum control of trapped electrons.

We fabricate the electrode structure on a dielectric wafer that is \SI{500}{\micro\meter} thick and measures three inches (\SI{76.2}{mm}) in diameter, with a \SI{30}{\micro\meter} conductive layer applied to both faces. To prevent interference from the electromagnetic field of the feeding lines, these lines are located on the side opposite to where the trap electrodes are positioned. The electrodes are linked to the transmission lines via microvias, each with a diameter of \SI{200}{\micro\meter}.

The electron-trapping signal transmission lines are realized through the use of coplanar waveguides. These waveguides allow the transmission lines to be positioned in proximity while minimizing interference from mutual inductance or capacitance. In addition, they provide easier manufacturing compared to microstrip lines, as discussed in Section \ref{sec:manufacturing}.

The simulation of electromagnetic wave propagation within the circuitry was carried out using the finite element method with a commercial software package. During the design process, we came to the conclusion that a straightforward solution to feed the electrodes directly by an electric signal is not possible, as the ground planes and coplanar waveguides easily cause enough inductance and capacitance to decrease the RLC resonance frequency to the MHz range. The impedance is high at the desired GHz-scale frequencies, so the trapping potential remains relatively weak. As a result, the trap must be integrated into a structure that accumulates the electrical power from the ``electron trapping'' driving signal, which, in our scenario, is resonant with the signal with frequency \wwe.

\begin{figure}
    \centering
    \includegraphics[width=0.4\textwidth]{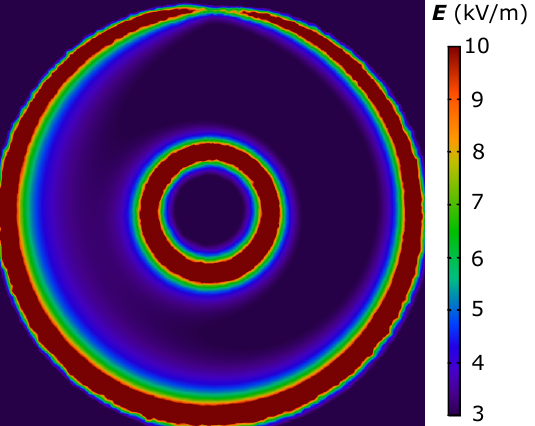}
\caption{\label{fig:onevia}Electric field norm distribution with \SI{10}{W} input power on the top side of the device in the case of only one electrode. Such a field has significant asymmetry, which causes major disruption to harmonic trapping potential and prevents trapping effectively. The RF electrode has inner radius of \SI{1.3}{mm} and outer radius of \SI{5.7}{mm}.}
\end{figure}

\begin{figure}
\label{fig4}
    \centering
    \includegraphics[width=0.5\textwidth]{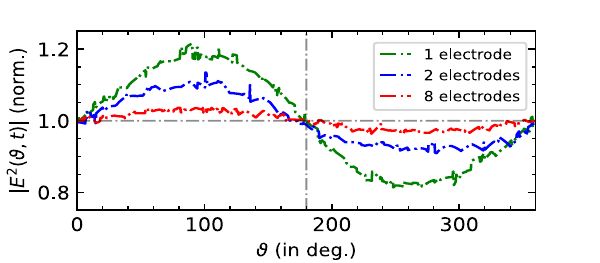}
\caption{\label{fig:angulardist}Square of electric field magnitude dependence on the angle around the electrode system. The series represent corresponding design with the circular electrode divided to one, two or eight sectors. The values have been normalised as 1.0 for average value, and the angle has been optimised to give a sinusoidal dependence without phase shift.}
\label{fig:4}
\end{figure}

Another benefit of incorporating the electrodes to the resonator is that the electric field oscillates with the same phase everywhere above the electrodes. However, the problem of an uneven distribution of the electric field remains to be resolved. This is pronounced especially when the ring electrode is fed only by one resonator through one microvia connection as illustrated in Figure \ref{fig:onevia}. The trapping potential becomes asymmetric: about 40 \% differences in the trapping potential were found in different sides of the trap in the simulation. To increase the symmetry, it is necessary to divide the electrode into sectors and feed them in parallel. The angular distribution of the squared electric field amplitude over the ring electrode is represented in Figure \ref{fig:angulardist}. Dividing the electrode into two parts makes the field more uniform, but an even better result is achieved with eight electrode sectors.

\begin{figure}
    \centering
    \includegraphics[width=0.4\textwidth]{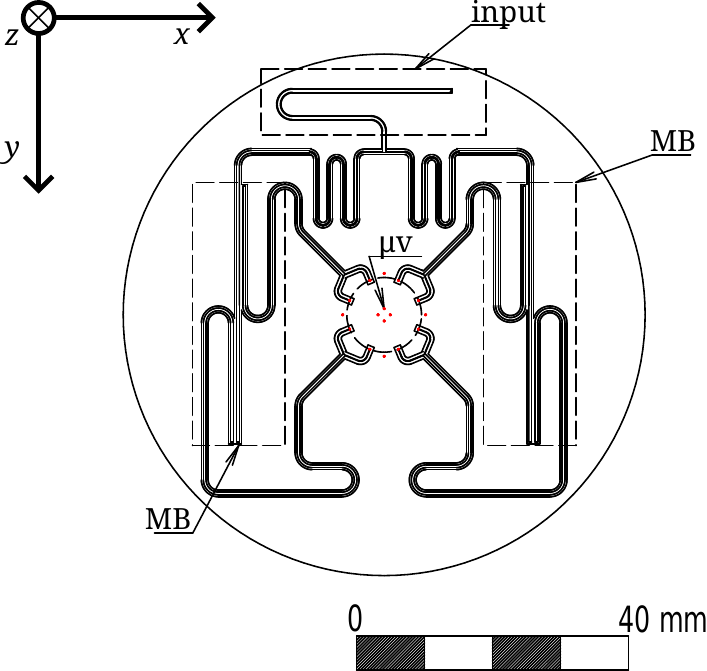}
\caption{\label{fig:feeding} The structure of the transmission line system. $MB$: Marchand balun, {\textmu}v: microvias, drawn as red dots. The microvias on the dashed circle connect the ends of the transmission lines with electrodes marked as ``e'' in Figure \ref{fig:electrodes}.}
\end{figure}

\begin{figure*}
\label{fig6}
    \centering
    \includegraphics[width=0.8\textwidth]{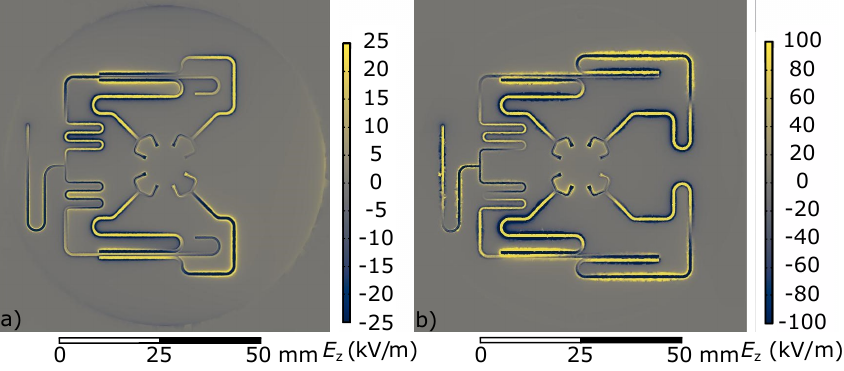}%
\caption{\label{fig:6}The \emph{z}-component of electric field on the bottom side of the device. a) Without Marchand baluns. The resonances exist, but remains weak, and the sign of the electric field is not the same in all lines. b) With Marchand baluns. The electrodes are not connected to each other in this figure. All lines support resonance with the same phase. The input power is \SI{10}{W}.}
\end{figure*}

The challenge of feeding these eight sectors with signals with equal amplitudes and phases has been solved by the structure of the feeding lines depicted in Figure \ref{fig:feeding}. The lines are separated into an input line and a resonating line part and connected by capacitive couplings. The input line is divided into two branches that have \(5\lambda/4\) resonance in the operating frequency \wwe. Each input line is coupled with two resonating lines using a Marchand balun \cite{zhang_new_2005} with the coupling length \(\lambda/4\). Figure \ref{fig:6}a shows the problems in the field distribution when a fork coupling is used instead of the Marchand balun. All resonating lines also have two branches, dividing the supply lines to eight in total. The distances between the start of the resonating lines and its division are \(3\lambda/4\) for the first coupled side and \(5\lambda/4\) for the secondary side in the resonance frequency, and the branches going to the microvias have \(\lambda/4\) resonances. Therefore, there are four resonating parts between the input and the electrodes in total, which amplifies the signal at the fundamental frequency \wwe. Figure \ref{fig:6}b depicts \emph{z}-component of the electric field on the bottom side of the device in a maximum electric field phase of the standing wave.

\begin{figure*}
    \centering
    \includegraphics[width=0.8\textwidth]{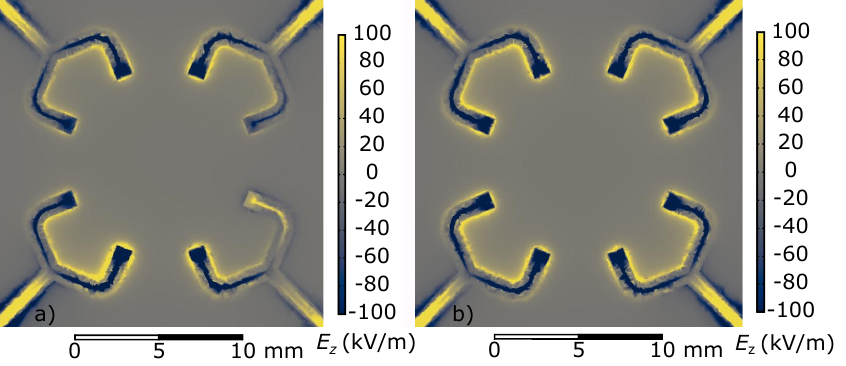}%
\caption{\label{fig:eldistribution} The \emph{z}-component of electric field on the bottom side of the device. The electrodes are separated in a) and connected with each other in b). The connections between the electrodes cause the field to have an increased symmetry, which is required for the formation of a trapping potential. The input power is \SI{10}{W}.}
\end{figure*}

The trap electrodes are surrounded by a ground plane and, in the middle of them, there is a circular ground electrode. The gaps between the ring electrode sectors, and the ring electrode structure and the ground electrodes are \SI{200}{\micro\meter}. The gaps have been designed relatively wide to decrease the Coulomb interactions between ring electrode sectors and ground parts, to suppress the charge concentration to inner and outer edges of the ring electrode segments, and therefore to improve the uniformity of their surface charge density.

\begin{figure*}
    \centering
    \includegraphics[width=0.8\textwidth]{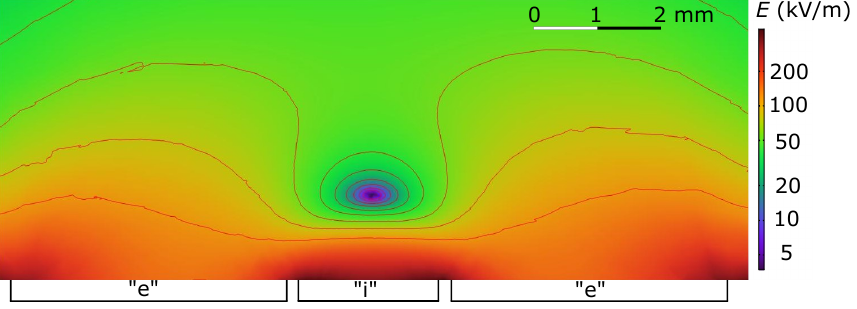}
\caption{Depiction of the electric field amplitude over the trap electrodes. The central (``I'') and surrounding (``e'') electrodes are labelled at the figure's bottom. The area depicted is a vertical cross-section along the $z$-axis, with the trapping region situated centrally above the electrode arrangement. Parts of the equipotential lines that are not smooth are artefacts of the simulation method and appear in regions that are not interesting for the dynamics of trapped particles.}
\label{fig:fieldabove}
\end{figure*}

Even under this condition, electrode segments tend to have different charge densities or even opposite charges for electrostatic reasons, as can be seen in Figure \ref{fig:eldistribution}a. The field configuration formed by such charge distribution does not support trapping. To ensure the equal signs of the charges of the resonators, the electrode segments are connected to each other on their inner boundary with a circular conductor part having \SI{200}{\micro\meter} width. Without this connection, the charge distribution on the ring electrode structure is unstable -- even the slightest imbalance in the radio-frequency signal supply results in one or more RF electrodes having opposite phase than the others, preferred by electrostatics. However, the distribution of the electric field is very symmetric with the connection, as can be seen in Figure \ref{fig:eldistribution}b. The connection between the RF electrodes is the crucial upgrade to the electrode solution previously given \cite{rendek_simulations_2022}.

The initial simulations were performed by assuming that the conductors had infinitely conductive surfaces and no surface roughness. Furthermore, no dielectric losses were assumed. These assumptions have been found to reduce computation time significantly. In this case, the trap depth was calculated to be about \SI{7300}{K} with \SI{10}{W} input power. The value of the input power is rather high, but not uncommon in the electron trapping field\cite{matthiesen_trapping_2021}. The trap center lies at about \SI{1.9}{mm} height from the electrode surface, which is only slightly higher than the analytically derived \SI{1.8}{mm}. We assume that the difference is caused by image currents induced to the ground electrodes in the case of the computer simulations. The distribution of the electric field in a vertical plane with the assumption of infinite conduction is depicted in Figure \ref{fig:fieldabove}.

\section{Manufacturing}
\label{sec:manufacturing}

Section \ref{sec:trap} references the necessity of a \SI{10}{W} input power to achieve sufficient trap depth. Cryogenic cooling is vital for maintaining state population, but necessitates minimal power loss due to microwave heating of the substrate. Thus, the substrate should have a low dielectric loss tangent, and the conduction layers must exceed the microwave signal's skin depth. As the field gradient of around \qty{100}{V} per pairs of conductors separated by less than \qty{100}{\micro\meter} gaps is expected, a material with a high breakdown voltage is needed. A low dielectric constant is essential for uniform electric field distribution above trap electrodes. Strong adhesion between the conductive layer and substrate is crucial because of small feature sizes and thermal stress tolerance\cite{salahouelhadj_reliability_2014}.

Silicone wafers are unsuitable as they become semiconducting at temperatures exceeding \SI{50}{K}, ruling out cost-effective liquid nitrogen cooling. In this temperature regime, its loss tangent is also relatively high on the order of $10^{-2}$ \cite{krupka_measurements_2006}. Glass is a preferable option, with fused silica offering a loss tangent below $2\times 10^{-4}$ and a relatively low dielectric constant of $\sim 4$\cite{rodriguez-cano_broadband_2023}, leading to low amplitude variations at the electrodes. Additionally, its optical transparency could be used to improve laser access in potential future design upgrades.

However, coating tens of microns thick conduction layers with strong adhesion remains challenging. We developed a reliable method using laser microstructuring before magnetron sputtering copper onto glass wafers, tested successfully on fused silica and borosilicate\cite{antony_laser-assisted_2023}. Borosilicate was chosen in the current prototyping stage for cost-effectiveness despite its higher loss tangent ($\sim 6 \times 10^{-3}$).

\begin{figure*}
    \centering
    \subfloat[Side with feeding lines.]{\label{fig:feedings}\includegraphics[width=0.4\textwidth]{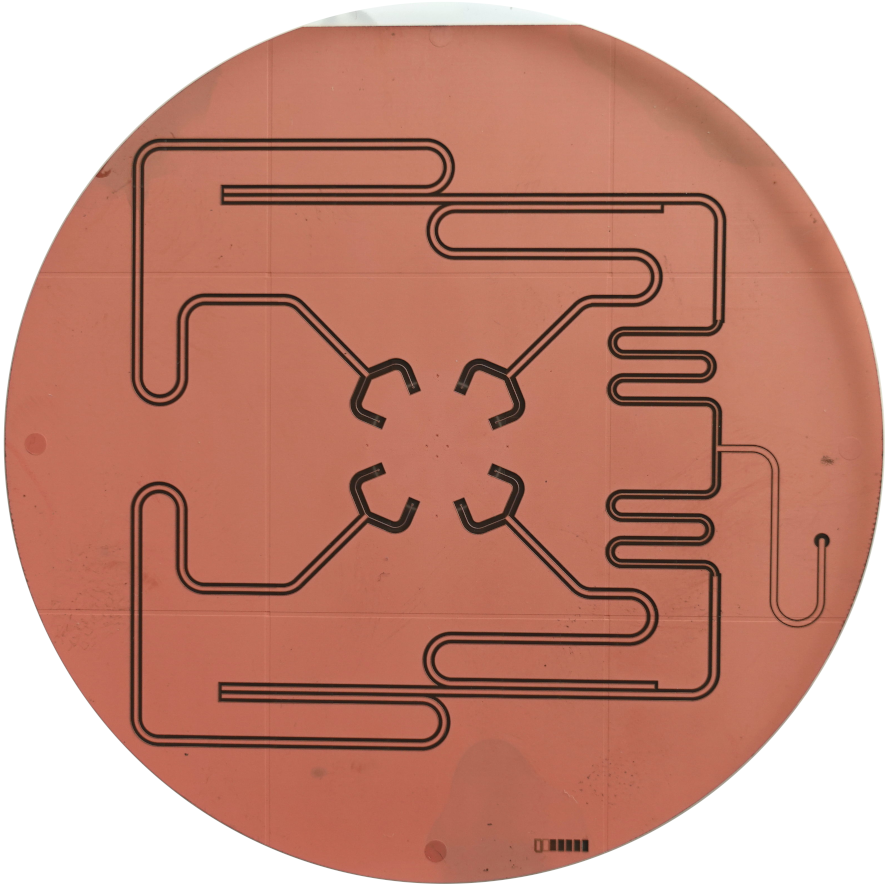}}
    \hspace{10mm}
    \subfloat[Side with trap electrodes.]{\label{fig:trappings}\includegraphics[width=0.4\textwidth]{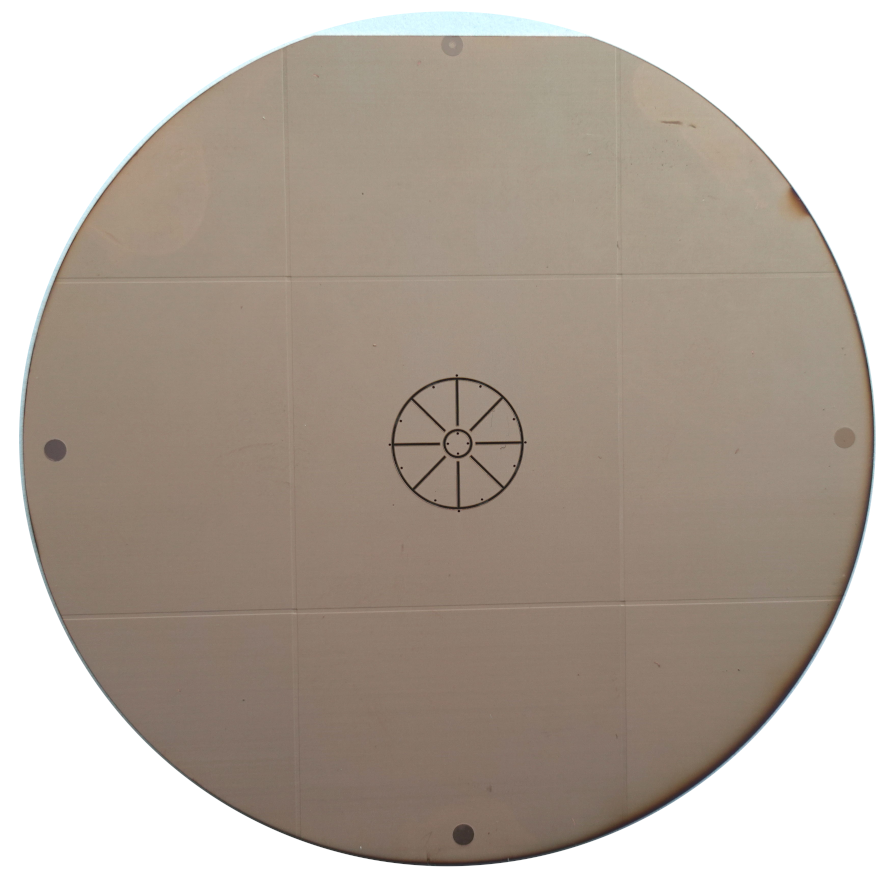}}
    \caption{\label{fig:photos}Copper-coated prototype of the electron-ion trap on a 3-inch laser-treated borosilicate wafer. Side \textit{b} appears darker due to being coated first and resting on a heated platform during the coating of side \textit{a}. This discolouration can be avoided using a magnetron sputtering device where wafers are hung. This version of the prototype has \SI{30}{\micro\meter} deep insulation trenches.}
\end{figure*}

\begin{figure}
    \centering
    \includegraphics[width=0.4\textwidth]{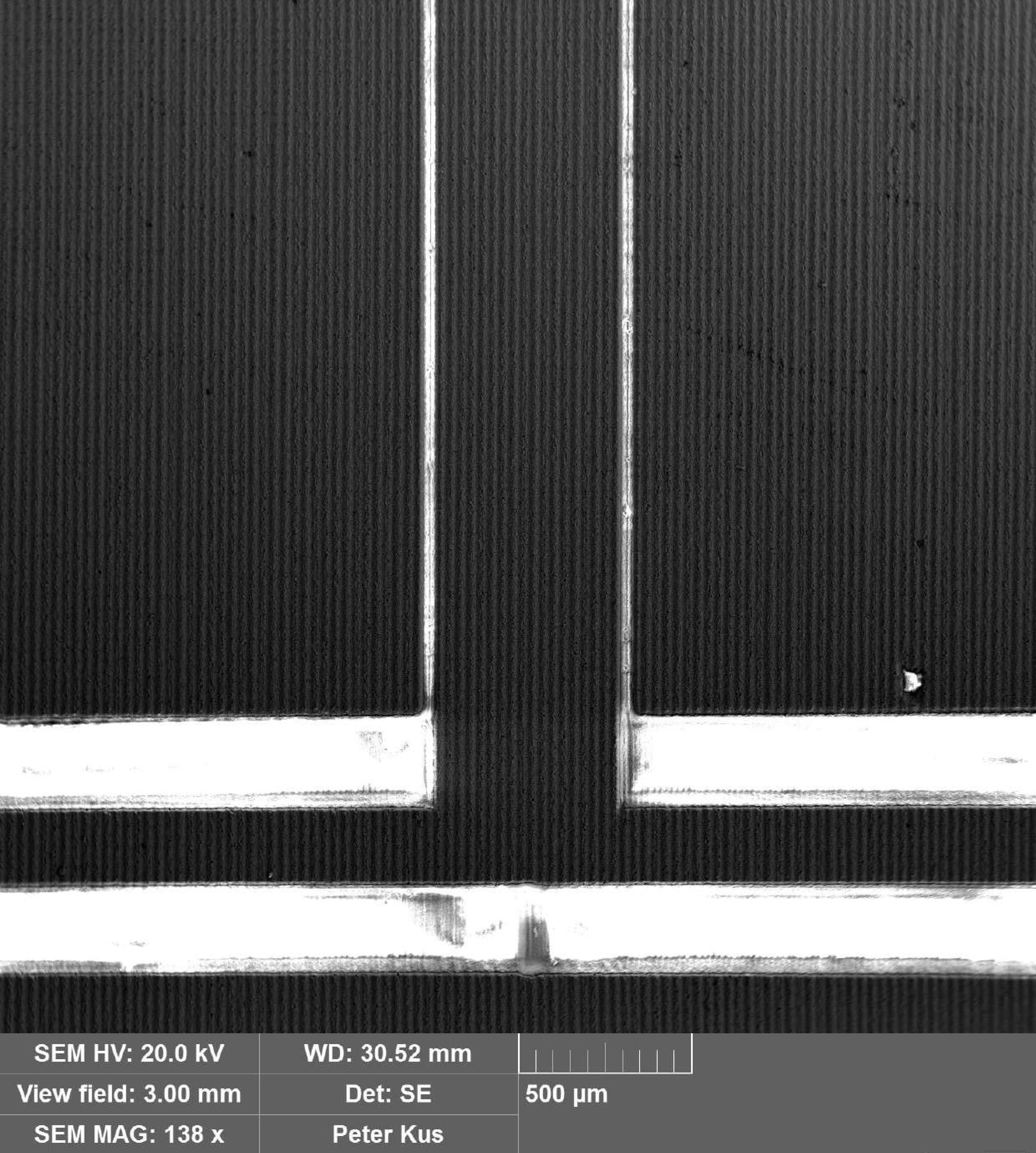}
    \caption{Scanning electron microscope image of manufactured transmission lines, showing gaps with a width of \SI{80}{\micro\meter}.}\label{fig:tee}
\end{figure}

Figure \ref{fig:photos} shows a copper-coated prototype on a 3-inch borosilicate wafer, where microvias were laser-drilled before copper coating all surfaces. Trenches, which are up to \SI{70}{\micro\meter} deep and are created through laser sputtering of copper and substrate material, function to prevent short circuits and enhance the length of the breakdown path\cite{wilson_situ_2022}. The scanning electron microscope of a T-junction in the feeding lines demonstrates precise laser machining capabilities (e.g., \SI{80}{\micro\meter}-wide trenches in Figure \ref{fig:tee}).

\begin{figure}
   \centering
    \includegraphics[width=\linewidth]{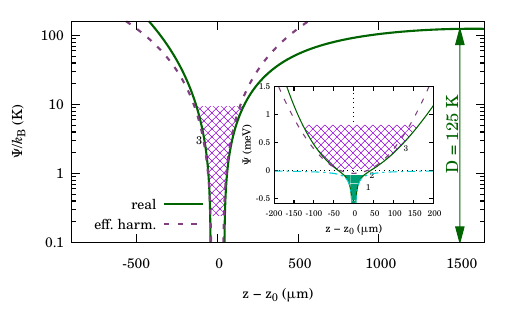}
\caption{\label{fig:9} Simulated trapping pseudopotential (full line) in superconductive trap. The trap depth is about \qty{125}{K}. The approximation by quantum harmonic oscillator (dashed line) is valid at electron temperatures of approximately \qtyrange{0.3}{10}{K} with 25\% accuracy. The inset figure shows other potential regions (1, 2, and 3), including the Coulomb potential from the ion (dash-dotted line), following the description in Section \ref{sec:categories}. The symbol $k_\mathrm{B}$ stands for the Boltzmann  constant.}
\end{figure}

The electrical circuit can also be fabricated using superconductors that offer the benefit of reduced resistive losses, improved resonator's quality\cite{dahan_measurement_2021} factor and trap depth. However, when a roughened glass substrate is used, surface roughness becomes the primary factor that limits the depth of the trap, especially in the GHz range, where the skin depth of copper, for example, is approximately \SI{1.3}{\micro\meter} at about \SI{2.4}{GHz}. Simulations incorporating realistic surface roughness values of \SI{1.5}{\micro\meter} for glass and \SI{1.2}{\micro\meter} for copper \cite{antony_laser-assisted_2023}, as well as those for a superconducting configuration, show that our copper prototype achieves a trap depth of \SI{22}{K} while the superconducting solution increases the depth to \SI{124}{K}. The pseudopotential along the $z$ axis for the superconducting case is depicted in Figure \ref{fig:9}. The asymmetry along the $z$ axis induces deviations from the idealized potential depicted in Figure \ref{fig:potentials}. This permits the approximation by Coulomb or harmonic potentials in the asymptotic regimes (1 and 3 from Section \ref{sec:categories}) only with 25\% accuracy. Furthermore, the regime 3 has also the upper bound, above which the approximation by the harmonic potential is not possible.

\begin{figure}
    \centering
    \includegraphics[width=0.4\textwidth]{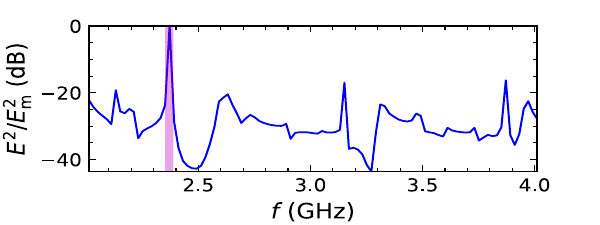}
\caption{\label{fig:10} Simulated resonance curve of the superconductive planar trap between 2 GHz and 4 GHz. The vertical axis is for square of the electric field magnitude, and it has been normalized to give 0 dB for the trapping resonance peak.}
\end{figure}

Figure \ref{fig:10} illustrates the simulated resonance curve for the superconductive trap design. The targeted resonance frequency for electron capture is \SI{2.37}{GHz}, with the resonance peak exceeding any other in the \SI{2}{GHz} to \SI{4}{GHz} range by over \SI{15}{dB}. The quality factor, defined as the resonance frequency divided by its full width at half maximum, is approximately 1,000.

Since our primary objective is to decrease resistive losses, high-temperature superconductors are sufficient for our purposes. YBCO layers can be manufactured on amorphous substrates by spin coating\cite{moriya_growth_2018} or  metal organic chemical vapour deposition\cite{studebaker_liquid_1995}. In terms of technology necessary to reach the critical temperatures, YBCO is the best candidate because a simple liquid nitrogen thermostat suffices. Other materials such as \ce{NbN} or \ce{MoC_x} with the critical temperature well above the liquid helium temperature can also be used\cite{deyu_recent_2025}. Furthermore, the design presented here can be adapted to crystalline substrates such as \ce{MgO} or sapphire that can accommodate YBCO and other high-temperature superconductor films at the expense of higher material cost and smaller trap volume.

\begin{table*}
    \begin{ruledtabular}
    \begin{tabular}{lllllll}
    Trap model & \mbox{BBS} & \mbox{JN} & \mbox{SAd} & \mbox{SR} & \mbox{Total}\\
    & \mbox{(\qty{E-20}{\volt\squared\per\meter\squared\second})}& \mbox{(\qty{E-17}{\volt\squared\per\meter\squared\second})}&
               \mbox{(\qty{E-19}{\volt\squared\per\meter\squared\second})}&
               \mbox{(\qty{E-30}{\volt\squared\per\meter\squared\second})}&
               \mbox{(\qty{E-17}{\volt\squared\per\meter\squared\second})}\\
               \hline
    Copper (\qty{300}{K})& 25 & 610 & 80 & 1.2 & 610 \\
    Copper (\qty{0.4}{K})& 0.024 & 0.31 & 0.07 & 560 & 0.31\\
    YBCO (\qty{93}{K}) & 3 & 0.8 & 5 & 0.8 & 0.9 \\
    \end{tabular}
    \end{ruledtabular}
    \caption{Sources of decoherence in the trap and their spectral densities $S_E$ for three different trap models. \text{BBS} -- black body radiation over the trap surface, \text{JN} -- Johnson-Nyquist, \text{SAd} -- surface adatom, \text{SR} -- surface roughness. Due to the large uncertainties in the material properties of YBCO films, the values for the YBCO trap are rounded to the first significant digit.}
    \label{tab:noise}
\end{table*}
 
\section{Heating rates of electrons}
\label{sec:heating}

The quantum sensing scheme of Section \ref{sec:coupling} is based on changes in the probability density of the trap-induced state occupation. When external signal 5 arrives to excite the trap-induced state, we need to be able to distinguish such an event from a random excitation caused by other factors intrinsic to the trapping device. One of the sources of such decoherence can be the collision with the background gas. Under a background pressure of \SI{2.5E-8}{Pa}, a free electron experiences a momentum transfer from helium atoms roughly once every \SI{30}{s}
at the temperature of \SI{300}{K}.\cite{goldenComparisonLowEnergyTotal1966} This gives a maximum estimate of the decoherence time. 

Within this time frame, the population of the initial trap-induced state may dissipate through dipole interactions with fluctuations in the electric field of the apparatus and the ambient environment. In the bound state region 1 of Figure \ref{fig:9}, we can expect decoherence time on the order of \qty{1}{\micro\second} for Rydberg states with principal quantum number $n' \sim 10$ \cite{higginsCoherentControlSingle2017} and it gets longer as the distance of the electron from the ion core grows following the $n^3 l^2$ scaling rule\cite{flannery_quantal_2003}.

In the region 3, where the system can be described by a quantum harmonic oscillator with the average motional mode occupation $\overline{n}$, the decoherence rate is given by\cite{yu_feasibility_2022, savard_1997, brownnutt_ion-trap_2015}
\begin{equation}
\label{eqn:Eq.4}
\Gamma \approx 2 \overline{n} \frac{e^{2}}{4\me\hslash\we}S_{E}\,,
\end{equation}
where $S_E$ is a spectral density of noise.  The formula accounts for both state excitation and photon emission. It is reasonably accurate for $\overline{n} \sim 1000$ that corresponds to the energies in the region. Eq.~\eqref{eqn:Eq.4} is a generalization of the heating rate in the reference \cite{brownnutt_ion-trap_2015}. The original assumption of the reference \cite{brownnutt_ion-trap_2015} that the initial state is ground is lifted here. It is also assumed that the transition dipole moments between the states coupled by the heating source scale as $\sqrt{n}$.

In Table \ref{tab:noise}, we aggregate the noise contributions from various sources and assess the overall heating rates for three potential models of the trap: the copper-coated trap at temperatures of \SI{300}{K} and \SI{0.4}{K}, and the superconductive trap cooled to \SI{93}{K} (the presumed critical temperature of the YBCO layer). The individual noise sources are described in Appendix. The secular frequencies $\omega_\mathrm{e}$ employed in these calculations are approximately $2\pi\times$\SI{7.3}{MHz}, $2\pi\times$\SI{8.8}{MHz}, and $2\pi\times$\SI{18.8}{MHz} for the respective configurations. The trap heights for these scenarios also exhibit variations, measuring approximately \SI{1.68}{mm}, \SI{1.78}{mm}, and \SI{1.95}{mm}, respectively. We hypothesize that the differences in trap height can be attributed to image currents on adjacent electrodes, which are influenced by variations in electrode conductivity. 

Table \ref{tab:noise} shows that the Johnson-Nyquist noise is the main contributor to the overall decoherence rate in all configurations. Cooling the copper design to \SI{0.4}{K} or using a YBCO-based design in the superconducting regime is equally effective in reducing this noise source. However, we still obtain the decoherence time on the order of \qty{100}{\micro\second} for $\overline{n} \sim 1000$ following the equation (\ref{eqn:Eq.4}).

Based on our preliminary simulations, this temporal period is sufficient for conducting detection experiments aided by EIT\cite{hudak_microcavity_2025}. It also considerably exceeds the coherence time of states in quantum sensing experiments involving ensembles of laser-cooled alkali metal atoms\cite{tu_approaching_2024}. Our trap-based sensor, predictably, experiences substantial quantum projection noise attributed to the ensemble size\cite{degen_quantum_2017}. To mitigate this noise, an array of traps could be created. The total number of traps may not need to be very large, considering the extended coherence time and significant dipole moment of the electron. A cautious estimate is approximately 100 traps, assuming the absence of a mechanism to entangle the electron with multiple ions within a single trap to exceed the standard quantum limit\cite{gilmore_quantum-enhanced_2021}. To achieve this, the design discussed here would require downscaling to fit such a number of traps on one wafer. This modification would naturally decrease the trap height, leading to a reduced decoherence time. Consequently, a detailed design optimization would be required. Nonetheless, preliminary proof-of-concept experiments can be performed with the device described in this work.

\section{Design improvements}
\label{sec:improvement}

Interestingly, the main obstacle that the proposed design encounters is providing the ion trapping signal within the \si{MHz} range. A transmission line directly delivering the frequency \wwI\ signal cannot be connected to the ring electrode segments without significantly disrupting the resonator's resonance. Clearly, supplying the dual-frequency signal (\ref{eq:waveform}) directly to the microwave resonator is fundamentally unfeasible. Consequently, we propose channelling the ion trapping signal to the ``ground'' planes (indicated as ``I'' in Figure \ref{fig:electrodes}) situated both inside and outside the ring electrode assembly. 

The structure of the resonator is inherently floating, since it lacks connections to both the ground and the electrodes labelled ``I''. In such a case, two complications appear. One is the crosstalk between electrodes ``I'' and ``e'', which results in the synchronous oscillation of both the electrodes and hence the decrease in the depth of the ion trapping field. Another is long-term charge accumulation in the dielectric beneath the floating electrode that results in the build-up of a potential offset of the resonator\cite{gao_investigation_2023}.

We have performed measurements of scattering parameters on a planar trap chip without microvia connections to determine the effects of crosstalk in the relevant frequency region (\SIrange{1}{10}{MHz}). As a result, we measure only a small magnitude of the forward voltage gains of the signal propagating from electrode ``I'' to ``e'' (and vice versa) on the order of less than $10^{-3}$ and similarly between a resonator branch in Figure \ref{fig:feeding} and ``I''. In addition, the phase shift falls within the range of approximately $-50$ to $-90$ degrees, which plays a role in sustaining the potential gradient and thus forms the ion trapping field. However, the departure from the perfect $180$ degree phase shift naturally reduces the trap depth, indicating that further optimization of the electrode design is required.

Building on the research conducted by \citet{koszewski_physical_2013}, we predict that charge build-up in the glass substrate will occur significantly slower than observed in their experiment, due to the fields being two orders of magnitude weaker. This delay could be further extended if the circuit is cooled to liquid nitrogen temperatures. Since the actual rate of charge accumulation is also influenced by the geometry of the electrodes and the glass, we can only estimate it to be much lower than \SI{0.1}{\volt\per\hour}. For instance, electrodes labeled ``e'' and ``I'' are separated by trenches tens of microns wide, deeply embedded in the glass, lengthening the charge travel paths within and creating interfaces between vacuum, glass, and conductor that need investigation. Should this issue prove persistent, we may need to adopt measures such as applying specialized waveforms \cite{bahl_charge-drift_2010} to the electrodes, irradiation by UV light\cite{ziemba_removal_2025}, or enhancing the isolation between electrodes ``e'' and ``I''.

In Section \ref{sec:manufacturing}, we have discussed the effect of replacing the metal conductive layer material with a superconductor on the trap depth. Another straightforward method to improve this parameter is to use a substrate material with a reduced dielectric loss compared to borosilicate glass. For instance, fused silica glass exhibits a loss tangent nearly ten times smaller, as highlighted in Section \ref{sec:manufacturing}. A design with superconductors, micrometer-scale surface roughness, and fused silica as a substrate results in a trap depth of more than \SI{250}{K}, which is approximately twice the value computed for borosilicate glass. Consequently, the quality factor of the resonator system also increases from the level of 1,000 to 1,800.

Adjustments in the geometry can also enhance the trap depth. Our simulations have demonstrated that reducing the central ground electrode's height by \SI{250}{\micro\meter}—thus lowering it beneath the ring electrode structure—can elevate the superconductive trap depth to \SI{160}{K} (when the borosilicate is used). Further displacement, such as printing the ring electrode structure\cite{quinn_geometries_2022,xu_3d-printed_2025-nature}, further intensifies the trapping field; however, particular care must be taken to maintain unobstructed optical access to the ions. Moreover, simulating such three-dimensional configurations, especially with support structures, becomes increasingly complex. Completely encasing the trapping field with an endcap electrode, akin to the method in \citet{ragg_segmented_2019}, is inadequate without ensuring optical access: our simulations indicate that suspending a vertical conductive rod of \SI{1.0}{mm} diameter \SI{1.0}{mm} above the trap centre can boost the trap depth to \SI{400}{K}, but a carefully designed support for this rod is crucial to avoid impairing fluorescence imaging of the trapped ions. A drawback of this approach is the resulting more asymmetric trapping field along the vertical $z$ axis.

An alternative geometric approach is to reduce the trap's dimensions. This suggests a potential decrease in the number of electrodes required. However, the trap depth decreases with the square of the trap's size, and a smaller size might necessitate a higher operating frequency, which could partially affect outcomes. Furthermore, proximity to the electrodes increases the decoherence rate of the particles' state (see Section \ref{sec:heating}). Such a device would not be suitable for housing large numbers of ions, thus not aiding in the study of many-particle dynamics, but it could be beneficial for applications in trapped electron quantum technology. The modifications of the feeding lines for such smaller traps with fewer segments are described in the patent by \citet{hejduk_supply_2024}.

\section{Conclusion}
\label{sec:conclusion}

This paper presents a comprehensive technical guide for realizing a hybrid quantum technology device combining levitating electrons and laser-cooled ions in a two-frequency Paul trap. We systematically identify optically addressable transitions within this mixed-particle system that can be precisely tuned via trap parameters, enabling applications in radio-frequency-field sensing and electron qubit manipulation.

The proposed architecture adopts a planar configuration to satisfy scalability demands while maintaining compatibility with external systems. We provide detailed specifications for the microwave feeding circuit and electrode geometry, optimized to minimize surface charge effects, improve field isotropy, and suppress decoherence mechanisms that disrupt electron state preparation.

Additionally, we outline fabrication protocols for an early prototype capable of delivering high-power microwave signals with straightforward implementation requirements. While further refinements (such as implementation of superconducting films) are needed to enhance trapping performance (depth/stability), our guide offers complete technical specifications to facilitate experimental exploration of this promising quantum platform.

\begin{acknowledgments}
This work is supported by the Czech Science Foundation (GAČR: GA24-10992S) and by the Charles University Grant Agency (GAUK 295023, GAUK 131224). We also acknowledge the support from the Czech Ministry of Education, Youth and Sports (project QM4ST, reg. no. EH22\_008/0004572). IH thanks to the
Technology Agency of the Czech Republic (TA\v{C}R: TN02000020) for the support. We also acknowledge previous funding from the University -- the Primus Research Programme (PRIMUS/21/SCI/005). We would like to thank Dr. Peter K\'{u}\v{s} (Charles University) for assistance with scanning electron microscopy, Prof. Ladislav Cvr\v{c}ek (Czech Technical University) and Dr. Petr Hauschwitz (HiLASE Centre) for collaboration in the development of the manufacturing method.
\end{acknowledgments}

\section*{Data Availability}

The data that support the findings of this article are openly available\cite{niklaslaustiSupportingDataRoadmap}.

\section*{Author Declarations}
\subsection*{Conflict of Interest}
MH and NL have Patent CZ310234B6 issued.

\subsection*{Author Contributions}
\textbf{Niklas Vilhelm Lausti}: Conceptualization (supporting), Data curation (lead), Formal Analysis (lead), Investigation (lead), Methodology (supporting), Software (equal), Visualization (equal), Writing – original draft (lead); \textbf{Vineet Kumar}: Conceptualization (supporting), Formal Analysis (supporting), Investigation (supporting), Visualization (equal); \textbf{Ivan Hud\'{a}k}: Conceptualization (supporting); \textbf{Michal Hejduk}: Conceptualization (equal), Funding acquisition (equal), Methodology (lead), Project administration (lead), Resources (equal), Supervision (lead), Visualization (equal), Writing – review \& editing (equal); \textbf{Michal Tarana}: Conceptualization (equal), Funding acquisition (equal), Software (equal), Validation (lead), Writing – review \& editing (equal)

\begin{center} \textbf{ORCID IDs} \end{center}

Niklas Vilhelm Lausti

\url{https://orcid.org/0000-0001-9906-6971}

Vineet Kumar

\url{https://orcid.org/0000-0001-8668-3663}

Ivan Hud\'ak

\url{https://orcid.org/0009-0004-4162-6154}

Michal Hejduk

\url{https://orcid.org/0000-0002-4417-4817}

Michal Tarana

\url{https://orcid.org/0000-0001-5086-7276}

\appendix*
\section{Overview of noise sources}
\label{sec:appendixh}
\subsection{Black-body radiation}
\label{sec:heatingbbr}

 The fundamental source of the heating inside the trap is the black-body radiation from the surroundings. An amplification of the rf-signal by resonators decreases the required input power, maintaining the lower temperatures of the trap assembly that subsequently reduces the radiative heating of the trapped particles. Its further decrease is anticipated by cooling the trap  with liquid nitrogen or a closed-cycle cryostat. The presence of at least liquid nitrogen as a coolant is also necessary for the operation of the superconducting components.

As the temperature of our electrodes \(T_e\) always satisfies the condition $k_\mathrm{B}T_\mathrm{e} \gg \hbar\we$, the spectral density for heating by blackbody radiation from the electrode surface can be expressed as \cite{brownnutt_ion-trap_2015}
\begin{equation}
\label{eqn:Eq.6}
{S_{E}}^{\text{BBS}} = \frac{k_\mathrm{B}T_\mathrm{e}\rho_\mathrm{e}}{2\pi d^{3}}\left( s_{\eta} + \frac{d}{\delta_{s}} \right)\,,
\end{equation}
where $\rho_\mathrm{e}$ is the resistivity of the electrodes, \emph{d} is the minimal height of the trapped particles above the electrode plane, and \(\delta_{s}\) is the electric-skin depth on the electrodes for the secular frequency. The parameter \(s_{\eta}\) represents the perpendicular \((s_{\eta} = 1)\) or parallel \((s_{\eta} = 1/2)\) fluctuations on the electrode surface. For the total spectral density, \(s_{\eta} = 2\). Assuming the resistivity of copper \(\rho_e \approx\ \SI{2.4E-8}{\ohm\meter}\) for our sputtered coating, considering the electrodes to be at room temperature (\(T_\mathrm{e} \approx \SI{300}{\kelvin}\)), the spectral density is \({S_{E}}^{\text{BBS}} \approx \SI{2.5E-19}{\volt\squared\per\meter\squared\second}\), and decoherence rate $\Gamma \approx\ \qty{700}{\per\second}$ for $\overline{n} \sim 1000$. For the electron secular frequency of approximately $2 \pi \times \qty{18.8}{MHz}$ in the superconductive trap with $\rho_\mathrm{e} \approx \qty{E-10}{\ohm\meter}$\cite{xin_comparison_2000}, the black-body radiation from the circuitry cooled to \qty{93}{K} yields $\Gamma\approx \qty{30}{\per\second}$ for the same average state.

\subsection{Electrical noise}
Decoherence by the rf-signal itself is also an issue in the Paul traps \cite{brownnutt_ion-trap_2015}.
It is caused by several factors, such as random fluctuations of the electric field (including Johnson-Nyquist noise) \cite{turchette_2000}, and field disruptions caused by electrode surface roughness \cite{lin_effects_2016}.

Johnson--Nyquist noise has spectral density \cite{brownnutt_ion-trap_2015}
\begin{equation}
\label{eqn:Eq.7}
{S_{E}}^{\text{JN}} = \frac{4k_{B}T_{c}R(\omega,T)}{d^{2}}.
\end{equation}
\emph{R} is resistance (frequency-dependent real component of the impedance) of the electrode and transmission system. $T_c$ is temperature of the circuit, and as the main part of the resistance lies on the trap chip, one can assume $T_c=T_e$. With an approximate calculation of secular frequency $\omega = \we = 2\pi \times \SI{7.3}{MHz}$ in the copper trap of room temperature, \(R \approx \SI{1.04}{\ohm}\), we obtain spectral density \({S_{E}}^{\text{JN}} \approx \SI{6.1E-15}{\volt\squared m^{-2}s}\) and heating rate for electrons from the ground state \(\dot{\overline{n}} \approx \num{2E4}\ \text{quanta/s}\). Although a previous study by \citet{wang_superconducting_2010} did not find significant differences in the  heating rates between superconducting and non-superconducting traps, we estimate that superconductivity would decrease Johnson-Nyquist noise by reducing resistance in our case. Our calculated decoherence rate from $\overline{n} \sim 1000$ would correspond \(\Gamma \approx \qty{2E7}{\per\second}\) in the copper trap potential at room temperature, while operation of YBCO trap at \qty{93}{K} would result in \(\Gamma \approx \qty{1E4}{\per\second}\). Based on our calculations, Johnson-Nyquist noise seems to be the predominant cause of decoherence in our trap.

Spectral density of the electric field noise caused by surface adatoms in the case of ideally smooth surface is approximated to be \SI{8.0E-18}{\volt\squared\per\meter\squared\second} at room temperature and less in lower temperatures\cite{brownnutt_ion-trap_2015}. This is almost three orders of magnitude less than Johnson-Nyquist noise. However, surface roughness has also a contribution to this decoherence rate. We apply the average rate for the case of uniform surface-adatom distribution. An equation for ratio of the field noise spectral density (which is straightly proportional to heating rate) with roughness and without it has been previously derived \cite{lin_effects_2016}. We assume values of \qty{1.2}{\micro\meter} for rms surface roughness, \qty{30}{\micro\meter} for characteristic length of surface roughness, \qty{169}{\pico\meter} for surface-adatom equilibrium position and \qty{31.2}{THz} for surface-adatom vibrational transition frequency. Based on these, the ratio deviates less than $10^{-7}$ from one at the liquid helium-4 temperature (for possible future experiments) and even less at higher temperatures. Therefore, the surface roughness does not have a significant effect on the heating rate in our trap assembly.

In a previous study for heating of a trapped electron \cite{yu_feasibility_2022} (not to be confused with trapping in the two frequency potential or in the potentials from Figures \ref{fig:potentials} and \ref{fig:9}), heating rate from the vibrational ground state caused by Johnson-Nyquist noise was estimated to be 140 quanta/s in a cryogenic temperature (\qty{0.4}{\kelvin} in the article). Calculating copper residual resistivity $\SI{9.3E-9}{\ohm\meter}$ for our transmission line design, $\we = \qty{8.8}{MHz}$ and temperature of \qty{0.4}{K}, we can estimate combined heating rate from Johnson-Nyquist noise, blackbody radiation and surface effects to be 7.5 quanta/s from the ground state, which would significantly improve the previous result. Heating rate of less than 140 quanta/s would be achievable at \qty{10}{\kelvin} for the copper trap and up to \qty{93}{\kelvin} (transition temperature) for the superconducting one.

\bibliography{processed}

\begin{thebibliography}{97}%
\makeatletter
\providecommand \@ifxundefined [1]{%
 \@ifx{#1\undefined}
}%
\providecommand \@ifnum [1]{%
 \ifnum #1\expandafter \@firstoftwo
 \else \expandafter \@secondoftwo
 \fi
}%
\providecommand \@ifx [1]{%
 \ifx #1\expandafter \@firstoftwo
 \else \expandafter \@secondoftwo
 \fi
}%
\providecommand \natexlab [1]{#1}%
\providecommand \enquote  [1]{``#1''}%
\providecommand \bibnamefont  [1]{#1}%
\providecommand \bibfnamefont [1]{#1}%
\providecommand \citenamefont [1]{#1}%
\providecommand \href@noop [0]{\@secondoftwo}%
\providecommand \href [0]{\begingroup \@sanitize@url \@href}%
\providecommand \@href[1]{\@@startlink{#1}\@@href}%
\providecommand \@@href[1]{\endgroup#1\@@endlink}%
\providecommand \@sanitize@url [0]{\catcode `\\12\catcode `\$12\catcode
  `\&12\catcode `\#12\catcode `\^12\catcode `\_12\catcode `\%12\relax}%
\providecommand \@@startlink[1]{}%
\providecommand \@@endlink[0]{}%
\providecommand \url  [0]{\begingroup\@sanitize@url \@url }%
\providecommand \@url [1]{\endgroup\@href {#1}{\urlprefix }}%
\providecommand \urlprefix  [0]{URL }%
\providecommand \Eprint [0]{\href }%
\providecommand \doibase [0]{https://doi.org/}%
\providecommand \selectlanguage [0]{\@gobble}%
\providecommand \bibinfo  [0]{\@secondoftwo}%
\providecommand \bibfield  [0]{\@secondoftwo}%
\providecommand \translation [1]{[#1]}%
\providecommand \BibitemOpen [0]{}%
\providecommand \bibitemStop [0]{}%
\providecommand \bibitemNoStop [0]{.\EOS\space}%
\providecommand \EOS [0]{\spacefactor3000\relax}%
\providecommand \BibitemShut  [1]{\csname bibitem#1\endcsname}%
\let\auto@bib@innerbib\@empty
\bibitem [{\citenamefont {Petralia}\ \emph {et~al.}(2020)\citenamefont
  {Petralia}, \citenamefont {Tsikritea}, \citenamefont {Loreau}, \citenamefont
  {Softley},\ and\ \citenamefont {Heazlewood}}]{petralia_strong_2020}%
  \BibitemOpen
  \bibfield  {author} {\bibinfo {author} {\bibfnamefont {L.~S.}\ \bibnamefont
  {Petralia}}, \bibinfo {author} {\bibfnamefont {A.}~\bibnamefont {Tsikritea}},
  \bibinfo {author} {\bibfnamefont {J.}~\bibnamefont {Loreau}}, \bibinfo
  {author} {\bibfnamefont {T.~P.}\ \bibnamefont {Softley}},\ and\ \bibinfo
  {author} {\bibfnamefont {B.~R.}\ \bibnamefont {Heazlewood}},\ }\bibfield
  {title} {\bibinfo {title} {Strong inverse kinetic isotope effect observed in
  ammonia charge exchange reactions},\ }\href
  {https://doi.org/10.1038/s41467-019-13976-8} {\bibfield  {journal} {\bibinfo
  {journal} {Nature Communications}\ }\textbf {\bibinfo {volume} {11}},\
  \bibinfo {pages} {173} (\bibinfo {year} {2020})}\BibitemShut {NoStop}%
\bibitem [{\citenamefont {Gilmore}\ \emph {et~al.}(2021)\citenamefont
  {Gilmore}, \citenamefont {Affolter}, \citenamefont {Lewis-Swan},
  \citenamefont {Barberena}, \citenamefont {Jordan}, \citenamefont {Rey},\ and\
  \citenamefont {Bollinger}}]{gilmore_quantum-enhanced_2021}%
  \BibitemOpen
  \bibfield  {author} {\bibinfo {author} {\bibfnamefont {K.~A.}\ \bibnamefont
  {Gilmore}}, \bibinfo {author} {\bibfnamefont {M.}~\bibnamefont {Affolter}},
  \bibinfo {author} {\bibfnamefont {R.~J.}\ \bibnamefont {Lewis-Swan}},
  \bibinfo {author} {\bibfnamefont {D.}~\bibnamefont {Barberena}}, \bibinfo
  {author} {\bibfnamefont {E.}~\bibnamefont {Jordan}}, \bibinfo {author}
  {\bibfnamefont {A.~M.}\ \bibnamefont {Rey}},\ and\ \bibinfo {author}
  {\bibfnamefont {J.~J.}\ \bibnamefont {Bollinger}},\ }\bibfield  {title}
  {\bibinfo {title} {Quantum-enhanced sensing of displacements and electric
  fields with two-dimensional trapped-ion crystals},\ }\href
  {https://doi.org/10.1126/science.abi5226} {\bibfield  {journal} {\bibinfo
  {journal} {Science}\ }\textbf {\bibinfo {volume} {373}},\ \bibinfo {pages}
  {673} (\bibinfo {year} {2021})}\BibitemShut {NoStop}%
\bibitem [{\citenamefont {Pyka}\ \emph {et~al.}(2014)\citenamefont {Pyka},
  \citenamefont {Herschbach}, \citenamefont {Keller},\ and\ \citenamefont
  {Mehlstäubler}}]{pyka_high-precision_2014}%
  \BibitemOpen
  \bibfield  {author} {\bibinfo {author} {\bibfnamefont {K.}~\bibnamefont
  {Pyka}}, \bibinfo {author} {\bibfnamefont {N.}~\bibnamefont {Herschbach}},
  \bibinfo {author} {\bibfnamefont {J.}~\bibnamefont {Keller}},\ and\ \bibinfo
  {author} {\bibfnamefont {T.~E.}\ \bibnamefont {Mehlstäubler}},\ }\bibfield
  {title} {\bibinfo {title} {A high-precision segmented {Paul} trap with
  minimized micromotion for an optical multiple-ion clock},\ }\href
  {https://doi.org/10.1007/s00340-013-5580-5} {\bibfield  {journal} {\bibinfo
  {journal} {Applied Physics B}\ }\textbf {\bibinfo {volume} {114}},\ \bibinfo
  {pages} {231} (\bibinfo {year} {2014})}\BibitemShut {NoStop}%
\bibitem [{\citenamefont {Bruzewicz}\ \emph {et~al.}(2019)\citenamefont
  {Bruzewicz}, \citenamefont {Chiaverini}, \citenamefont {McConnell},\ and\
  \citenamefont {Sage}}]{bruzewicz_trapped-ion_2019}%
  \BibitemOpen
  \bibfield  {author} {\bibinfo {author} {\bibfnamefont {C.~D.}\ \bibnamefont
  {Bruzewicz}}, \bibinfo {author} {\bibfnamefont {J.}~\bibnamefont
  {Chiaverini}}, \bibinfo {author} {\bibfnamefont {R.}~\bibnamefont
  {McConnell}},\ and\ \bibinfo {author} {\bibfnamefont {J.~M.}\ \bibnamefont
  {Sage}},\ }\bibfield  {title} {\bibinfo {title} {Trapped-ion quantum
  computing: {Progress} and challenges},\ }\href
  {https://doi.org/10.1063/1.5088164} {\bibfield  {journal} {\bibinfo
  {journal} {Applied Physics Reviews}\ }\textbf {\bibinfo {volume} {6}},\
  \bibinfo {pages} {021314} (\bibinfo {year} {2019})}\BibitemShut {NoStop}%
\bibitem [{\citenamefont {Wang}\ \emph {et~al.}(2021)\citenamefont {Wang},
  \citenamefont {Luan}, \citenamefont {Qiao}, \citenamefont {Um}, \citenamefont
  {Zhang}, \citenamefont {Wang}, \citenamefont {Yuan}, \citenamefont {Gu},
  \citenamefont {Zhang},\ and\ \citenamefont {Kim}}]{wang2021single}%
  \BibitemOpen
  \bibfield  {author} {\bibinfo {author} {\bibfnamefont {P.}~\bibnamefont
  {Wang}}, \bibinfo {author} {\bibfnamefont {C.-Y.}\ \bibnamefont {Luan}},
  \bibinfo {author} {\bibfnamefont {M.}~\bibnamefont {Qiao}}, \bibinfo {author}
  {\bibfnamefont {M.}~\bibnamefont {Um}}, \bibinfo {author} {\bibfnamefont
  {J.}~\bibnamefont {Zhang}}, \bibinfo {author} {\bibfnamefont
  {Y.}~\bibnamefont {Wang}}, \bibinfo {author} {\bibfnamefont {X.}~\bibnamefont
  {Yuan}}, \bibinfo {author} {\bibfnamefont {M.}~\bibnamefont {Gu}}, \bibinfo
  {author} {\bibfnamefont {J.}~\bibnamefont {Zhang}},\ and\ \bibinfo {author}
  {\bibfnamefont {K.}~\bibnamefont {Kim}},\ }\bibfield  {title} {\bibinfo
  {title} {Single ion qubit with estimated coherence time exceeding one hour},\
  }\href {https://doi.org/10.1038/s41467-020-20330-w} {\bibfield  {journal}
  {\bibinfo  {journal} {Nature Communications}\ }\textbf {\bibinfo {volume}
  {12}},\ \bibinfo {pages} {233} (\bibinfo {year} {2021})}\BibitemShut
  {NoStop}%
\bibitem [{\citenamefont {Ballance}\ \emph {et~al.}(2016)\citenamefont
  {Ballance}, \citenamefont {Harty}, \citenamefont {Linke}, \citenamefont
  {Sepiol},\ and\ \citenamefont {Lucas}}]{ballance2016high}%
  \BibitemOpen
  \bibfield  {author} {\bibinfo {author} {\bibfnamefont {C.}~\bibnamefont
  {Ballance}}, \bibinfo {author} {\bibfnamefont {T.}~\bibnamefont {Harty}},
  \bibinfo {author} {\bibfnamefont {N.}~\bibnamefont {Linke}}, \bibinfo
  {author} {\bibfnamefont {M.}~\bibnamefont {Sepiol}},\ and\ \bibinfo {author}
  {\bibfnamefont {D.}~\bibnamefont {Lucas}},\ }\bibfield  {title} {\bibinfo
  {title} {High-fidelity quantum logic gates using trapped-ion hyperfine
  qubits},\ }\href {https://doi.org/10.1103/PhysRevLett.117.060504} {\bibfield
  {journal} {\bibinfo  {journal} {Physical Review Letters}\ }\textbf {\bibinfo
  {volume} {117}},\ \bibinfo {pages} {060504} (\bibinfo {year}
  {2016})}\BibitemShut {NoStop}%
\bibitem [{\citenamefont {Fowler}\ \emph {et~al.}(2012)\citenamefont {Fowler},
  \citenamefont {Mariantoni}, \citenamefont {Martinis},\ and\ \citenamefont
  {Cleland}}]{fowler_surface_2012}%
  \BibitemOpen
  \bibfield  {author} {\bibinfo {author} {\bibfnamefont {A.~G.}\ \bibnamefont
  {Fowler}}, \bibinfo {author} {\bibfnamefont {M.}~\bibnamefont {Mariantoni}},
  \bibinfo {author} {\bibfnamefont {J.~M.}\ \bibnamefont {Martinis}},\ and\
  \bibinfo {author} {\bibfnamefont {A.~N.}\ \bibnamefont {Cleland}},\
  }\bibfield  {title} {\bibinfo {title} {Surface codes: {Towards} practical
  large-scale quantum computation},\ }\href
  {https://doi.org/10.1103/PhysRevA.86.032324} {\bibfield  {journal} {\bibinfo
  {journal} {Physical Review A}\ }\textbf {\bibinfo {volume} {86}},\ \bibinfo
  {pages} {032324} (\bibinfo {year} {2012})}\BibitemShut {NoStop}%
\bibitem [{\citenamefont {Gidney}\ and\ \citenamefont
  {Ekerå}(2021)}]{gidney_how_2021}%
  \BibitemOpen
  \bibfield  {author} {\bibinfo {author} {\bibfnamefont {C.}~\bibnamefont
  {Gidney}}\ and\ \bibinfo {author} {\bibfnamefont {M.}~\bibnamefont
  {Ekerå}},\ }\bibfield  {title} {\bibinfo {title} {How to factor 2048 bit
  {RSA} integers in 8 hours using 20 million noisy qubits},\ }\href
  {https://doi.org/10.22331/q-2021-04-15-433} {\bibfield  {journal} {\bibinfo
  {journal} {Quantum}\ }\textbf {\bibinfo {volume} {5}},\ \bibinfo {pages}
  {433} (\bibinfo {year} {2021})}\BibitemShut {NoStop}%
\bibitem [{\citenamefont {Kandala}\ \emph {et~al.}(2021)\citenamefont
  {Kandala}, \citenamefont {Wei}, \citenamefont {Srinivasan}, \citenamefont
  {Magesan}, \citenamefont {Carnevale}, \citenamefont {Keefe}, \citenamefont
  {Klaus}, \citenamefont {Dial},\ and\ \citenamefont
  {McKay}}]{kandala_demonstration_2021}%
  \BibitemOpen
  \bibfield  {author} {\bibinfo {author} {\bibfnamefont {A.}~\bibnamefont
  {Kandala}}, \bibinfo {author} {\bibfnamefont {K.}~\bibnamefont {Wei}},
  \bibinfo {author} {\bibfnamefont {S.}~\bibnamefont {Srinivasan}}, \bibinfo
  {author} {\bibfnamefont {E.}~\bibnamefont {Magesan}}, \bibinfo {author}
  {\bibfnamefont {S.}~\bibnamefont {Carnevale}}, \bibinfo {author}
  {\bibfnamefont {G.}~\bibnamefont {Keefe}}, \bibinfo {author} {\bibfnamefont
  {D.}~\bibnamefont {Klaus}}, \bibinfo {author} {\bibfnamefont
  {O.}~\bibnamefont {Dial}},\ and\ \bibinfo {author} {\bibfnamefont
  {D.}~\bibnamefont {McKay}},\ }\bibfield  {title} {\bibinfo {title}
  {Demonstration of a {High}-{Fidelity} cnot {Gate} for {Fixed}-{Frequency}
  {Transmons} with {Engineered} \${ZZ}\$ {Suppression}},\ }\href
  {https://doi.org/10.1103/PhysRevLett.127.130501} {\bibfield  {journal}
  {\bibinfo  {journal} {Physical Review Letters}\ }\textbf {\bibinfo {volume}
  {127}},\ \bibinfo {pages} {130501} (\bibinfo {year} {2021})}\BibitemShut
  {NoStop}%
\bibitem [{\citenamefont {Chew}\ \emph {et~al.}(2022)\citenamefont {Chew},
  \citenamefont {Tomita}, \citenamefont {Mahesh}, \citenamefont {Sugawa},
  \citenamefont {de~Léséleuc},\ and\ \citenamefont
  {Ohmori}}]{chew_ultrafast_2022}%
  \BibitemOpen
  \bibfield  {author} {\bibinfo {author} {\bibfnamefont {Y.}~\bibnamefont
  {Chew}}, \bibinfo {author} {\bibfnamefont {T.}~\bibnamefont {Tomita}},
  \bibinfo {author} {\bibfnamefont {T.~P.}\ \bibnamefont {Mahesh}}, \bibinfo
  {author} {\bibfnamefont {S.}~\bibnamefont {Sugawa}}, \bibinfo {author}
  {\bibfnamefont {S.}~\bibnamefont {de~Léséleuc}},\ and\ \bibinfo {author}
  {\bibfnamefont {K.}~\bibnamefont {Ohmori}},\ }\bibfield  {title} {\bibinfo
  {title} {Ultrafast energy exchange between two single {Rydberg} atoms on a
  nanosecond timescale},\ }\href {https://doi.org/10.1038/s41566-022-01047-2}
  {\bibfield  {journal} {\bibinfo  {journal} {Nature Photonics}\ }\textbf
  {\bibinfo {volume} {16}},\ \bibinfo {pages} {724} (\bibinfo {year}
  {2022})}\BibitemShut {NoStop}%
\bibitem [{\citenamefont {Wang}\ \emph {et~al.}(2022)\citenamefont {Wang},
  \citenamefont {Yu}, \citenamefont {Wang}, \citenamefont {Luan}, \citenamefont
  {Zhang},\ and\ \citenamefont {Kim}}]{wang_fast_2022}%
  \BibitemOpen
  \bibfield  {author} {\bibinfo {author} {\bibfnamefont {K.}~\bibnamefont
  {Wang}}, \bibinfo {author} {\bibfnamefont {J.-F.}\ \bibnamefont {Yu}},
  \bibinfo {author} {\bibfnamefont {P.}~\bibnamefont {Wang}}, \bibinfo {author}
  {\bibfnamefont {C.}~\bibnamefont {Luan}}, \bibinfo {author} {\bibfnamefont
  {J.-N.}\ \bibnamefont {Zhang}},\ and\ \bibinfo {author} {\bibfnamefont
  {K.}~\bibnamefont {Kim}},\ }\bibfield  {title} {\bibinfo {title} {Fast
  multi-qubit global-entangling gates without individual addressing of trapped
  ions},\ }\href {https://doi.org/10.1088/2058-9565/ac84a3} {\bibfield
  {journal} {\bibinfo  {journal} {Quantum Science and Technology}\ }\textbf
  {\bibinfo {volume} {7}},\ \bibinfo {pages} {044005} (\bibinfo {year}
  {2022})}\BibitemShut {NoStop}%
\bibitem [{\citenamefont {Saner}\ \emph {et~al.}(2023)\citenamefont {Saner},
  \citenamefont {Băzăvan}, \citenamefont {Minder}, \citenamefont {Drmota},
  \citenamefont {Webb}, \citenamefont {Araneda}, \citenamefont {Srinivas},
  \citenamefont {Lucas},\ and\ \citenamefont {Ballance}}]{saner_breaking_2023}%
  \BibitemOpen
  \bibfield  {author} {\bibinfo {author} {\bibfnamefont {S.}~\bibnamefont
  {Saner}}, \bibinfo {author} {\bibfnamefont {O.}~\bibnamefont {Băzăvan}},
  \bibinfo {author} {\bibfnamefont {M.}~\bibnamefont {Minder}}, \bibinfo
  {author} {\bibfnamefont {P.}~\bibnamefont {Drmota}}, \bibinfo {author}
  {\bibfnamefont {D.}~\bibnamefont {Webb}}, \bibinfo {author} {\bibfnamefont
  {G.}~\bibnamefont {Araneda}}, \bibinfo {author} {\bibfnamefont
  {R.}~\bibnamefont {Srinivas}}, \bibinfo {author} {\bibfnamefont
  {D.}~\bibnamefont {Lucas}},\ and\ \bibinfo {author} {\bibfnamefont
  {C.}~\bibnamefont {Ballance}},\ }\bibfield  {title} {\bibinfo {title}
  {Breaking the {Entangling} {Gate} {Speed} {Limit} for {Trapped}-{Ion}
  {Qubits} {Using} a {Phase}-{Stable} {Standing} {Wave}},\ }\href
  {https://doi.org/10.1103/PhysRevLett.131.220601} {\bibfield  {journal}
  {\bibinfo  {journal} {Physical Review Letters}\ }\textbf {\bibinfo {volume}
  {131}},\ \bibinfo {pages} {220601} (\bibinfo {year} {2023})}\BibitemShut
  {NoStop}%
\bibitem [{\citenamefont {Deffner}\ and\ \citenamefont
  {Campbell}(2017)}]{deffner_quantum_2017}%
  \BibitemOpen
  \bibfield  {author} {\bibinfo {author} {\bibfnamefont {S.}~\bibnamefont
  {Deffner}}\ and\ \bibinfo {author} {\bibfnamefont {S.}~\bibnamefont
  {Campbell}},\ }\bibfield  {title} {\bibinfo {title} {Quantum speed limits:
  from {Heisenberg}’s uncertainty principle to optimal quantum control},\
  }\href {https://doi.org/10.1088/1751-8121/aa86c6} {\bibfield  {journal}
  {\bibinfo  {journal} {Journal of Physics A: Mathematical and Theoretical}\
  }\textbf {\bibinfo {volume} {50}},\ \bibinfo {pages} {453001} (\bibinfo
  {year} {2017})}\BibitemShut {NoStop}%
\bibitem [{\citenamefont {Lekitsch}\ \emph {et~al.}(2017)\citenamefont
  {Lekitsch}, \citenamefont {Weidt}, \citenamefont {Fowler}, \citenamefont
  {Mølmer}, \citenamefont {Devitt}, \citenamefont {Wunderlich},\ and\
  \citenamefont {Hensinger}}]{lekitsch_blueprint_2017}%
  \BibitemOpen
  \bibfield  {author} {\bibinfo {author} {\bibfnamefont {B.}~\bibnamefont
  {Lekitsch}}, \bibinfo {author} {\bibfnamefont {S.}~\bibnamefont {Weidt}},
  \bibinfo {author} {\bibfnamefont {A.~G.}\ \bibnamefont {Fowler}}, \bibinfo
  {author} {\bibfnamefont {K.}~\bibnamefont {Mølmer}}, \bibinfo {author}
  {\bibfnamefont {S.~J.}\ \bibnamefont {Devitt}}, \bibinfo {author}
  {\bibfnamefont {C.}~\bibnamefont {Wunderlich}},\ and\ \bibinfo {author}
  {\bibfnamefont {W.~K.}\ \bibnamefont {Hensinger}},\ }\bibfield  {title}
  {\bibinfo {title} {Blueprint for a microwave trapped ion quantum computer},\
  }\href {https://doi.org/10.1126/sciadv.1601540} {\bibfield  {journal}
  {\bibinfo  {journal} {Science Advances}\ }\textbf {\bibinfo {volume} {3}},\
  \bibinfo {pages} {e1601540} (\bibinfo {year} {2017})}\BibitemShut {NoStop}%
\bibitem [{\citenamefont {Pino}\ \emph {et~al.}(2021)\citenamefont {Pino},
  \citenamefont {Dreiling}, \citenamefont {Figgatt}, \citenamefont {Gaebler},
  \citenamefont {Moses}, \citenamefont {Allman}, \citenamefont {Baldwin},
  \citenamefont {Foss-Feig}, \citenamefont {Hayes}, \citenamefont {Mayer},
  \citenamefont {Ryan-Anderson},\ and\ \citenamefont
  {Neyenhuis}}]{pino_demonstration_2021}%
  \BibitemOpen
  \bibfield  {author} {\bibinfo {author} {\bibfnamefont {J.~M.}\ \bibnamefont
  {Pino}}, \bibinfo {author} {\bibfnamefont {J.~M.}\ \bibnamefont {Dreiling}},
  \bibinfo {author} {\bibfnamefont {C.}~\bibnamefont {Figgatt}}, \bibinfo
  {author} {\bibfnamefont {J.~P.}\ \bibnamefont {Gaebler}}, \bibinfo {author}
  {\bibfnamefont {S.~A.}\ \bibnamefont {Moses}}, \bibinfo {author}
  {\bibfnamefont {M.~S.}\ \bibnamefont {Allman}}, \bibinfo {author}
  {\bibfnamefont {C.~H.}\ \bibnamefont {Baldwin}}, \bibinfo {author}
  {\bibfnamefont {M.}~\bibnamefont {Foss-Feig}}, \bibinfo {author}
  {\bibfnamefont {D.}~\bibnamefont {Hayes}}, \bibinfo {author} {\bibfnamefont
  {K.}~\bibnamefont {Mayer}}, \bibinfo {author} {\bibfnamefont
  {C.}~\bibnamefont {Ryan-Anderson}},\ and\ \bibinfo {author} {\bibfnamefont
  {B.}~\bibnamefont {Neyenhuis}},\ }\bibfield  {title} {\bibinfo {title}
  {Demonstration of the trapped-ion quantum {CCD} computer architecture},\
  }\href {https://doi.org/10.1038/s41586-021-03318-4} {\bibfield  {journal}
  {\bibinfo  {journal} {Nature}\ }\textbf {\bibinfo {volume} {592}},\ \bibinfo
  {pages} {209} (\bibinfo {year} {2021})}\BibitemShut {NoStop}%
\bibitem [{\citenamefont {Zhang}\ \emph {et~al.}(2020)\citenamefont {Zhang},
  \citenamefont {Pokorny}, \citenamefont {Li}, \citenamefont {Higgins},
  \citenamefont {Pöschl}, \citenamefont {Lesanovsky},\ and\ \citenamefont
  {Hennrich}}]{zhang_submicrosecond_2020}%
  \BibitemOpen
  \bibfield  {author} {\bibinfo {author} {\bibfnamefont {C.}~\bibnamefont
  {Zhang}}, \bibinfo {author} {\bibfnamefont {F.}~\bibnamefont {Pokorny}},
  \bibinfo {author} {\bibfnamefont {W.}~\bibnamefont {Li}}, \bibinfo {author}
  {\bibfnamefont {G.}~\bibnamefont {Higgins}}, \bibinfo {author} {\bibfnamefont
  {A.}~\bibnamefont {Pöschl}}, \bibinfo {author} {\bibfnamefont
  {I.}~\bibnamefont {Lesanovsky}},\ and\ \bibinfo {author} {\bibfnamefont
  {M.}~\bibnamefont {Hennrich}},\ }\bibfield  {title} {\bibinfo {title}
  {Submicrosecond entangling gate between trapped ions via {Rydberg}
  interaction},\ }\href {https://doi.org/10.1038/s41586-020-2152-9} {\bibfield
  {journal} {\bibinfo  {journal} {Nature}\ }\textbf {\bibinfo {volume} {580}},\
  \bibinfo {pages} {345} (\bibinfo {year} {2020})}\BibitemShut {NoStop}%
\bibitem [{\citenamefont {Flannery}\ and\ \citenamefont
  {Vrinceanu}(2003)}]{flannery_quantal_2003}%
  \BibitemOpen
  \bibfield  {author} {\bibinfo {author} {\bibfnamefont {M.~R.}\ \bibnamefont
  {Flannery}}\ and\ \bibinfo {author} {\bibfnamefont {D.}~\bibnamefont
  {Vrinceanu}},\ }\bibfield  {title} {\bibinfo {title} {Quantal and classical
  radiative cascade in {Rydberg} plasmas},\ }\href
  {https://doi.org/10.1103/PhysRevA.68.030502} {\bibfield  {journal} {\bibinfo
  {journal} {Physical Review A}\ }\textbf {\bibinfo {volume} {68}},\ \bibinfo
  {pages} {030502} (\bibinfo {year} {2003})}\BibitemShut {NoStop}%
\bibitem [{\citenamefont {Cohen}\ and\ \citenamefont
  {Thompson}(2021)}]{cohen_quantum_2021}%
  \BibitemOpen
  \bibfield  {author} {\bibinfo {author} {\bibfnamefont {S.~R.}\ \bibnamefont
  {Cohen}}\ and\ \bibinfo {author} {\bibfnamefont {J.~D.}\ \bibnamefont
  {Thompson}},\ }\bibfield  {title} {\bibinfo {title} {Quantum {Computing} with
  {Circular} {Rydberg} {Atoms}},\ }\href
  {https://doi.org/10.1103/PRXQuantum.2.030322} {\bibfield  {journal} {\bibinfo
   {journal} {PRX Quantum}\ }\textbf {\bibinfo {volume} {2}},\ \bibinfo {pages}
  {030322} (\bibinfo {year} {2021})}\BibitemShut {NoStop}%
\bibitem [{\citenamefont {Wineland}\ \emph {et~al.}(1983)\citenamefont
  {Wineland}, \citenamefont {Bollinger},\ and\ \citenamefont
  {Itano}}]{wineland_laser-fluorescence_1983}%
  \BibitemOpen
  \bibfield  {author} {\bibinfo {author} {\bibfnamefont {D.~J.}\ \bibnamefont
  {Wineland}}, \bibinfo {author} {\bibfnamefont {J.~J.}\ \bibnamefont
  {Bollinger}},\ and\ \bibinfo {author} {\bibfnamefont {W.~M.}\ \bibnamefont
  {Itano}},\ }\bibfield  {title} {\bibinfo {title} {Laser-{Fluorescence} {Mass}
  {Spectroscopy}},\ }\href {https://doi.org/10.1103/PhysRevLett.50.628}
  {\bibfield  {journal} {\bibinfo  {journal} {Physical Review Letters}\
  }\textbf {\bibinfo {volume} {50}},\ \bibinfo {pages} {628} (\bibinfo {year}
  {1983})}\BibitemShut {NoStop}%
\bibitem [{\citenamefont {Matthiesen}\ \emph {et~al.}(2021)\citenamefont
  {Matthiesen}, \citenamefont {Yu}, \citenamefont {Guo}, \citenamefont
  {Alonso},\ and\ \citenamefont {Häffner}}]{matthiesen_trapping_2021}%
  \BibitemOpen
  \bibfield  {author} {\bibinfo {author} {\bibfnamefont {C.}~\bibnamefont
  {Matthiesen}}, \bibinfo {author} {\bibfnamefont {Q.}~\bibnamefont {Yu}},
  \bibinfo {author} {\bibfnamefont {J.}~\bibnamefont {Guo}}, \bibinfo {author}
  {\bibfnamefont {A.~M.}\ \bibnamefont {Alonso}},\ and\ \bibinfo {author}
  {\bibfnamefont {H.}~\bibnamefont {Häffner}},\ }\bibfield  {title} {\bibinfo
  {title} {Trapping {Electrons} in a {Room}-{Temperature} {Microwave} {Paul}
  {Trap}},\ }\href {https://doi.org/10.1103/PhysRevX.11.011019} {\bibfield
  {journal} {\bibinfo  {journal} {Physical Review X}\ }\textbf {\bibinfo
  {volume} {11}},\ \bibinfo {pages} {011019} (\bibinfo {year}
  {2021})}\BibitemShut {NoStop}%
\bibitem [{\citenamefont {Yu}\ \emph {et~al.}(2022)\citenamefont {Yu},
  \citenamefont {Alonso}, \citenamefont {Caminiti}, \citenamefont {Beck},
  \citenamefont {Sutherland}, \citenamefont {Leibfried}, \citenamefont
  {Rodriguez}, \citenamefont {Dhital}, \citenamefont {Hemmerling},\ and\
  \citenamefont {Häffner}}]{yu_feasibility_2022}%
  \BibitemOpen
  \bibfield  {author} {\bibinfo {author} {\bibfnamefont {Q.}~\bibnamefont
  {Yu}}, \bibinfo {author} {\bibfnamefont {A.~M.}\ \bibnamefont {Alonso}},
  \bibinfo {author} {\bibfnamefont {J.}~\bibnamefont {Caminiti}}, \bibinfo
  {author} {\bibfnamefont {K.~M.}\ \bibnamefont {Beck}}, \bibinfo {author}
  {\bibfnamefont {R.~T.}\ \bibnamefont {Sutherland}}, \bibinfo {author}
  {\bibfnamefont {D.}~\bibnamefont {Leibfried}}, \bibinfo {author}
  {\bibfnamefont {K.~J.}\ \bibnamefont {Rodriguez}}, \bibinfo {author}
  {\bibfnamefont {M.}~\bibnamefont {Dhital}}, \bibinfo {author} {\bibfnamefont
  {B.}~\bibnamefont {Hemmerling}},\ and\ \bibinfo {author} {\bibfnamefont
  {H.}~\bibnamefont {Häffner}},\ }\bibfield  {title} {\bibinfo {title}
  {Feasibility study of quantum computing using trapped electrons},\ }\href
  {https://doi.org/10.1103/PhysRevA.105.022420} {\bibfield  {journal} {\bibinfo
   {journal} {Physical Review A}\ }\textbf {\bibinfo {volume} {105}},\ \bibinfo
  {pages} {022420} (\bibinfo {year} {2022})}\BibitemShut {NoStop}%
\bibitem [{\citenamefont {Kotler}\ \emph {et~al.}(2017)\citenamefont {Kotler},
  \citenamefont {Simmonds}, \citenamefont {Leibfried},\ and\ \citenamefont
  {Wineland}}]{kotler_hybrid_2017}%
  \BibitemOpen
  \bibfield  {author} {\bibinfo {author} {\bibfnamefont {S.}~\bibnamefont
  {Kotler}}, \bibinfo {author} {\bibfnamefont {R.~W.}\ \bibnamefont
  {Simmonds}}, \bibinfo {author} {\bibfnamefont {D.}~\bibnamefont
  {Leibfried}},\ and\ \bibinfo {author} {\bibfnamefont {D.~J.}\ \bibnamefont
  {Wineland}},\ }\bibfield  {title} {\bibinfo {title} {Hybrid quantum systems
  with trapped charged particles},\ }\href
  {https://doi.org/10.1103/PhysRevA.95.022327} {\bibfield  {journal} {\bibinfo
  {journal} {Physical Review A}\ }\textbf {\bibinfo {volume} {95}},\ \bibinfo
  {pages} {022327} (\bibinfo {year} {2017})}\BibitemShut {NoStop}%
\bibitem [{\citenamefont {Osada}\ \emph {et~al.}(2022)\citenamefont {Osada},
  \citenamefont {Taniguchi}, \citenamefont {Shigefuji},\ and\ \citenamefont
  {Noguchi}}]{osada_feasibility_2022}%
  \BibitemOpen
  \bibfield  {author} {\bibinfo {author} {\bibfnamefont {A.}~\bibnamefont
  {Osada}}, \bibinfo {author} {\bibfnamefont {K.}~\bibnamefont {Taniguchi}},
  \bibinfo {author} {\bibfnamefont {M.}~\bibnamefont {Shigefuji}},\ and\
  \bibinfo {author} {\bibfnamefont {A.}~\bibnamefont {Noguchi}},\ }\bibfield
  {title} {\bibinfo {title} {Feasibility study on ground-state cooling and
  single-phonon readout of trapped electrons using hybrid quantum systems},\
  }\href {https://doi.org/10.1103/PhysRevResearch.4.033245} {\bibfield
  {journal} {\bibinfo  {journal} {Physical Review Research}\ }\textbf {\bibinfo
  {volume} {4}},\ \bibinfo {pages} {033245} (\bibinfo {year}
  {2022})}\BibitemShut {NoStop}%
\bibitem [{\citenamefont {Hammond}(2012)}]{hammond_spin_2012}%
  \BibitemOpen
  \bibfield  {author} {\bibinfo {author} {\bibfnamefont {R.~T.}\ \bibnamefont
  {Hammond}},\ }\bibfield  {title} {\bibinfo {title} {Spin flip probability of
  electron in a uniform magnetic field},\ }\href
  {https://doi.org/10.1063/1.3691937} {\bibfield  {journal} {\bibinfo
  {journal} {Applied Physics Letters}\ }\textbf {\bibinfo {volume} {100}},\
  \bibinfo {pages} {121112} (\bibinfo {year} {2012})}\BibitemShut {NoStop}%
\bibitem [{\citenamefont {Siegele-Brown}\ \emph {et~al.}(2022)\citenamefont
  {Siegele-Brown}, \citenamefont {Hong}, \citenamefont {Lebrun-Gallagher},
  \citenamefont {Hile}, \citenamefont {Weidt},\ and\ \citenamefont
  {Hensinger}}]{siegele-brown_fabrication_2022}%
  \BibitemOpen
  \bibfield  {author} {\bibinfo {author} {\bibfnamefont {M.}~\bibnamefont
  {Siegele-Brown}}, \bibinfo {author} {\bibfnamefont {S.}~\bibnamefont {Hong}},
  \bibinfo {author} {\bibfnamefont {F.~R.}\ \bibnamefont {Lebrun-Gallagher}},
  \bibinfo {author} {\bibfnamefont {S.~J.}\ \bibnamefont {Hile}}, \bibinfo
  {author} {\bibfnamefont {S.}~\bibnamefont {Weidt}},\ and\ \bibinfo {author}
  {\bibfnamefont {W.~K.}\ \bibnamefont {Hensinger}},\ }\bibfield  {title}
  {\bibinfo {title} {Fabrication of surface ion traps with integrated current
  carrying wires enabling high magnetic field gradients},\ }\href
  {https://doi.org/10.1088/2058-9565/ac66fc} {\bibfield  {journal} {\bibinfo
  {journal} {Quantum Science and Technology}\ }\textbf {\bibinfo {volume}
  {7}},\ \bibinfo {pages} {034003} (\bibinfo {year} {2022})}\BibitemShut
  {NoStop}%
\bibitem [{\citenamefont {Peng}\ \emph {et~al.}(2017)\citenamefont {Peng},
  \citenamefont {Matthiesen},\ and\ \citenamefont {Häffner}}]{peng_spin_2017}%
  \BibitemOpen
  \bibfield  {author} {\bibinfo {author} {\bibfnamefont {P.}~\bibnamefont
  {Peng}}, \bibinfo {author} {\bibfnamefont {C.}~\bibnamefont {Matthiesen}},\
  and\ \bibinfo {author} {\bibfnamefont {H.}~\bibnamefont {Häffner}},\
  }\bibfield  {title} {\bibinfo {title} {Spin readout of trapped electron
  qubits},\ }\href {https://doi.org/10.1103/PhysRevA.95.012312} {\bibfield
  {journal} {\bibinfo  {journal} {Physical Review A}\ }\textbf {\bibinfo
  {volume} {95}},\ \bibinfo {pages} {012312} (\bibinfo {year}
  {2017})}\BibitemShut {NoStop}%
\bibitem [{\citenamefont {Huang}\ \emph {et~al.}(2025)\citenamefont {Huang},
  \citenamefont {Hausten}, \citenamefont {Yu}, \citenamefont {Taniguchi},
  \citenamefont {Yadav}, \citenamefont {Sacksteder}, \citenamefont {Noguchi},
  \citenamefont {Schneider},\ and\ \citenamefont
  {Haeffner}}]{huangNumericalInvestigationsElectron2025}%
  \BibitemOpen
  \bibfield  {author} {\bibinfo {author} {\bibfnamefont {A.}~\bibnamefont
  {Huang}}, \bibinfo {author} {\bibfnamefont {E.}~\bibnamefont {Hausten}},
  \bibinfo {author} {\bibfnamefont {Q.}~\bibnamefont {Yu}}, \bibinfo {author}
  {\bibfnamefont {K.}~\bibnamefont {Taniguchi}}, \bibinfo {author}
  {\bibfnamefont {N.}~\bibnamefont {Yadav}}, \bibinfo {author} {\bibfnamefont
  {I.}~\bibnamefont {Sacksteder}}, \bibinfo {author} {\bibfnamefont
  {A.}~\bibnamefont {Noguchi}}, \bibinfo {author} {\bibfnamefont
  {R.}~\bibnamefont {Schneider}},\ and\ \bibinfo {author} {\bibfnamefont
  {H.}~\bibnamefont {Haeffner}},\ }\href
  {https://doi.org/10.48550/arXiv.2503.12379} {\bibinfo {title} {Numerical
  {{Investigations}} of {{Electron Dynamics}} in a {{Linear Paul Trap}}}}
  (\bibinfo {year} {2025}),\ \Eprint {https://arxiv.org/abs/2503.12379}
  {arXiv:2503.12379 [quant-ph]} \BibitemShut {NoStop}%
\bibitem [{\citenamefont {Itano}\ \emph {et~al.}(1995)\citenamefont {Itano},
  \citenamefont {Bergquist}, \citenamefont {Bollinger},\ and\ \citenamefont
  {Wineland}}]{itano_cooling_1995}%
  \BibitemOpen
  \bibfield  {author} {\bibinfo {author} {\bibfnamefont {W.~M.}\ \bibnamefont
  {Itano}}, \bibinfo {author} {\bibfnamefont {J.~C.}\ \bibnamefont
  {Bergquist}}, \bibinfo {author} {\bibfnamefont {J.~J.}\ \bibnamefont
  {Bollinger}},\ and\ \bibinfo {author} {\bibfnamefont {D.~J.}\ \bibnamefont
  {Wineland}},\ }\bibfield  {title} {\bibinfo {title} {Cooling methods in ion
  traps},\ }\href {https://doi.org/10.1088/0031-8949/1995/T59/013} {\bibfield
  {journal} {\bibinfo  {journal} {Physica Scripta}\ }\textbf {\bibinfo {volume}
  {T59}},\ \bibinfo {pages} {106} (\bibinfo {year} {1995})}\BibitemShut
  {NoStop}%
\bibitem [{\citenamefont {Mehta}\ \emph {et~al.}(2020)\citenamefont {Mehta},
  \citenamefont {Zhang}, \citenamefont {Malinowski}, \citenamefont {Nguyen},
  \citenamefont {Stadler},\ and\ \citenamefont {Home}}]{mehta_integrated_2020}%
  \BibitemOpen
  \bibfield  {author} {\bibinfo {author} {\bibfnamefont {K.~K.}\ \bibnamefont
  {Mehta}}, \bibinfo {author} {\bibfnamefont {C.}~\bibnamefont {Zhang}},
  \bibinfo {author} {\bibfnamefont {M.}~\bibnamefont {Malinowski}}, \bibinfo
  {author} {\bibfnamefont {T.-L.}\ \bibnamefont {Nguyen}}, \bibinfo {author}
  {\bibfnamefont {M.}~\bibnamefont {Stadler}},\ and\ \bibinfo {author}
  {\bibfnamefont {J.~P.}\ \bibnamefont {Home}},\ }\bibfield  {title} {\bibinfo
  {title} {Integrated optical multi-ion quantum logic},\ }\href
  {https://doi.org/10.1038/s41586-020-2823-6} {\bibfield  {journal} {\bibinfo
  {journal} {Nature}\ }\textbf {\bibinfo {volume} {586}},\ \bibinfo {pages}
  {533} (\bibinfo {year} {2020})}\BibitemShut {NoStop}%
\bibitem [{\citenamefont {Gulati}\ \emph {et~al.}(2017)\citenamefont {Gulati},
  \citenamefont {Takahashi}, \citenamefont {Podoliak}, \citenamefont {Horak},\
  and\ \citenamefont {Keller}}]{gulati_fiber_2017}%
  \BibitemOpen
  \bibfield  {author} {\bibinfo {author} {\bibfnamefont {G.~K.}\ \bibnamefont
  {Gulati}}, \bibinfo {author} {\bibfnamefont {H.}~\bibnamefont {Takahashi}},
  \bibinfo {author} {\bibfnamefont {N.}~\bibnamefont {Podoliak}}, \bibinfo
  {author} {\bibfnamefont {P.}~\bibnamefont {Horak}},\ and\ \bibinfo {author}
  {\bibfnamefont {M.}~\bibnamefont {Keller}},\ }\bibfield  {title} {\bibinfo
  {title} {Fiber cavities with integrated mode matching optics},\ }\href
  {https://doi.org/10.1038/s41598-017-05729-8} {\bibfield  {journal} {\bibinfo
  {journal} {Scientific Reports}\ }\textbf {\bibinfo {volume} {7}},\ \bibinfo
  {pages} {5556} (\bibinfo {year} {2017})}\BibitemShut {NoStop}%
\bibitem [{\citenamefont {Doherty}\ \emph {et~al.}(2023)\citenamefont
  {Doherty}, \citenamefont {Kuhn},\ and\ \citenamefont
  {Kassa}}]{doherty_multi-resonant_2023}%
  \BibitemOpen
  \bibfield  {author} {\bibinfo {author} {\bibfnamefont {T.~H.}\ \bibnamefont
  {Doherty}}, \bibinfo {author} {\bibfnamefont {A.}~\bibnamefont {Kuhn}},\ and\
  \bibinfo {author} {\bibfnamefont {E.}~\bibnamefont {Kassa}},\ }\bibfield
  {title} {\bibinfo {title} {Multi-resonant open-access microcavity arrays for
  light matter interaction},\ }\href {https://doi.org/10.1364/OE.475921}
  {\bibfield  {journal} {\bibinfo  {journal} {Optics Express}\ }\textbf
  {\bibinfo {volume} {31}},\ \bibinfo {pages} {6342} (\bibinfo {year}
  {2023})}\BibitemShut {NoStop}%
\bibitem [{\citenamefont {Pogorelov}\ \emph {et~al.}(2021)\citenamefont
  {Pogorelov}, \citenamefont {Feldker}, \citenamefont {Marciniak},
  \citenamefont {Postler}, \citenamefont {Jacob}, \citenamefont
  {Krieglsteiner}, \citenamefont {Podlesnic}, \citenamefont {Meth},
  \citenamefont {Negnevitsky}, \citenamefont {Stadler} \emph
  {et~al.}}]{pogorelov2021compact}%
  \BibitemOpen
  \bibfield  {author} {\bibinfo {author} {\bibfnamefont {I.}~\bibnamefont
  {Pogorelov}}, \bibinfo {author} {\bibfnamefont {T.}~\bibnamefont {Feldker}},
  \bibinfo {author} {\bibfnamefont {C.~D.}\ \bibnamefont {Marciniak}}, \bibinfo
  {author} {\bibfnamefont {L.}~\bibnamefont {Postler}}, \bibinfo {author}
  {\bibfnamefont {G.}~\bibnamefont {Jacob}}, \bibinfo {author} {\bibfnamefont
  {O.}~\bibnamefont {Krieglsteiner}}, \bibinfo {author} {\bibfnamefont
  {V.}~\bibnamefont {Podlesnic}}, \bibinfo {author} {\bibfnamefont
  {M.}~\bibnamefont {Meth}}, \bibinfo {author} {\bibfnamefont {V.}~\bibnamefont
  {Negnevitsky}}, \bibinfo {author} {\bibfnamefont {M.}~\bibnamefont
  {Stadler}}, \emph {et~al.},\ }\bibfield  {title} {\bibinfo {title} {Compact
  ion-trap quantum computing demonstrator},\ }\href
  {https://doi.org/10.1103/PRXQuantum.2.020343} {\bibfield  {journal} {\bibinfo
   {journal} {PRX Quantum}\ }\textbf {\bibinfo {volume} {2}},\ \bibinfo {pages}
  {020343} (\bibinfo {year} {2021})}\BibitemShut {NoStop}%
\bibitem [{\citenamefont {Yu}\ \emph {et~al.}(2024)\citenamefont {Yu},
  \citenamefont {Betzholz},\ and\ \citenamefont {Cai}}]{yu_strong_2024}%
  \BibitemOpen
  \bibfield  {author} {\bibinfo {author} {\bibfnamefont {B.}~\bibnamefont
  {Yu}}, \bibinfo {author} {\bibfnamefont {R.}~\bibnamefont {Betzholz}},\ and\
  \bibinfo {author} {\bibfnamefont {J.}~\bibnamefont {Cai}},\ }\bibfield
  {title} {\bibinfo {title} {Strong coherent ion-electron coupling using a wire
  data bus},\ }\href {https://doi.org/10.1103/PhysRevApplied.22.024032}
  {\bibfield  {journal} {\bibinfo  {journal} {Physical Review Applied}\
  }\textbf {\bibinfo {volume} {22}},\ \bibinfo {pages} {024032} (\bibinfo
  {year} {2024})}\BibitemShut {NoStop}%
\bibitem [{\citenamefont {An}\ \emph {et~al.}(2022)\citenamefont {An},
  \citenamefont {Alonso}, \citenamefont {Matthiesen},\ and\ \citenamefont
  {Häffner}}]{an_coupling_2022}%
  \BibitemOpen
  \bibfield  {author} {\bibinfo {author} {\bibfnamefont {D.}~\bibnamefont
  {An}}, \bibinfo {author} {\bibfnamefont {A.~M.}\ \bibnamefont {Alonso}},
  \bibinfo {author} {\bibfnamefont {C.}~\bibnamefont {Matthiesen}},\ and\
  \bibinfo {author} {\bibfnamefont {H.}~\bibnamefont {Häffner}},\ }\bibfield
  {title} {\bibinfo {title} {Coupling {Two} {Laser}-{Cooled} {Ions} via a
  {Room}-{Temperature} {Conductor}},\ }\href
  {https://doi.org/10.1103/PhysRevLett.128.063201} {\bibfield  {journal}
  {\bibinfo  {journal} {Physical Review Letters}\ }\textbf {\bibinfo {volume}
  {128}},\ \bibinfo {pages} {063201} (\bibinfo {year} {2022})}\BibitemShut
  {NoStop}%
\bibitem [{\citenamefont {Fan}\ \emph {et~al.}(2022)\citenamefont {Fan},
  \citenamefont {Gabrielse}, \citenamefont {Graham}, \citenamefont {Harnik},
  \citenamefont {Myers}, \citenamefont {Ramani}, \citenamefont {Sukra},
  \citenamefont {Wong},\ and\ \citenamefont {Xiao}}]{fan_one-electron_2022}%
  \BibitemOpen
  \bibfield  {author} {\bibinfo {author} {\bibfnamefont {X.}~\bibnamefont
  {Fan}}, \bibinfo {author} {\bibfnamefont {G.}~\bibnamefont {Gabrielse}},
  \bibinfo {author} {\bibfnamefont {P.~W.}\ \bibnamefont {Graham}}, \bibinfo
  {author} {\bibfnamefont {R.}~\bibnamefont {Harnik}}, \bibinfo {author}
  {\bibfnamefont {T.~G.}\ \bibnamefont {Myers}}, \bibinfo {author}
  {\bibfnamefont {H.}~\bibnamefont {Ramani}}, \bibinfo {author} {\bibfnamefont
  {B.~A.}\ \bibnamefont {Sukra}}, \bibinfo {author} {\bibfnamefont {S.~S.}\
  \bibnamefont {Wong}},\ and\ \bibinfo {author} {\bibfnamefont
  {Y.}~\bibnamefont {Xiao}},\ }\bibfield  {title} {\bibinfo {title}
  {One-{Electron} {Quantum} {Cyclotron} as a {Milli}-{eV} {Dark}-{Photon}
  {Detector}},\ }\href {https://doi.org/10.1103/PhysRevLett.129.261801}
  {\bibfield  {journal} {\bibinfo  {journal} {Physical Review Letters}\
  }\textbf {\bibinfo {volume} {129}},\ \bibinfo {pages} {261801} (\bibinfo
  {year} {2022})}\BibitemShut {NoStop}%
\bibitem [{\citenamefont {Kawakami}\ \emph {et~al.}(2023)\citenamefont
  {Kawakami}, \citenamefont {Chen}, \citenamefont {Benito},\ and\ \citenamefont
  {Konstantinov}}]{kawakami_blueprint_2023}%
  \BibitemOpen
  \bibfield  {author} {\bibinfo {author} {\bibfnamefont {E.}~\bibnamefont
  {Kawakami}}, \bibinfo {author} {\bibfnamefont {J.}~\bibnamefont {Chen}},
  \bibinfo {author} {\bibfnamefont {M.}~\bibnamefont {Benito}},\ and\ \bibinfo
  {author} {\bibfnamefont {D.}~\bibnamefont {Konstantinov}},\ }\bibfield
  {title} {\bibinfo {title} {Blueprint for quantum computing using electrons on
  helium},\ }\href {https://doi.org/10.1103/PhysRevApplied.20.054022}
  {\bibfield  {journal} {\bibinfo  {journal} {Physical Review Applied}\
  }\textbf {\bibinfo {volume} {20}},\ \bibinfo {pages} {054022} (\bibinfo
  {year} {2023})}\BibitemShut {NoStop}%
\bibitem [{\citenamefont {Zhou}\ \emph {et~al.}(2022)\citenamefont {Zhou},
  \citenamefont {Koolstra}, \citenamefont {Zhang}, \citenamefont {Yang},
  \citenamefont {Han}, \citenamefont {Dizdar}, \citenamefont {Li},
  \citenamefont {Divan}, \citenamefont {Guo}, \citenamefont {Murch},
  \citenamefont {Schuster},\ and\ \citenamefont {Jin}}]{zhou_single_2022}%
  \BibitemOpen
  \bibfield  {author} {\bibinfo {author} {\bibfnamefont {X.}~\bibnamefont
  {Zhou}}, \bibinfo {author} {\bibfnamefont {G.}~\bibnamefont {Koolstra}},
  \bibinfo {author} {\bibfnamefont {X.}~\bibnamefont {Zhang}}, \bibinfo
  {author} {\bibfnamefont {G.}~\bibnamefont {Yang}}, \bibinfo {author}
  {\bibfnamefont {X.}~\bibnamefont {Han}}, \bibinfo {author} {\bibfnamefont
  {B.}~\bibnamefont {Dizdar}}, \bibinfo {author} {\bibfnamefont
  {X.}~\bibnamefont {Li}}, \bibinfo {author} {\bibfnamefont {R.}~\bibnamefont
  {Divan}}, \bibinfo {author} {\bibfnamefont {W.}~\bibnamefont {Guo}}, \bibinfo
  {author} {\bibfnamefont {K.~W.}\ \bibnamefont {Murch}}, \bibinfo {author}
  {\bibfnamefont {D.~I.}\ \bibnamefont {Schuster}},\ and\ \bibinfo {author}
  {\bibfnamefont {D.}~\bibnamefont {Jin}},\ }\bibfield  {title} {\bibinfo
  {title} {Single electrons on solid neon as a solid-state qubit platform},\
  }\href {https://doi.org/10.1038/s41586-022-04539-x} {\bibfield  {journal}
  {\bibinfo  {journal} {Nature}\ }\textbf {\bibinfo {volume} {605}},\ \bibinfo
  {pages} {46} (\bibinfo {year} {2022})}\BibitemShut {NoStop}%
\bibitem [{\citenamefont {Walz}\ \emph {et~al.}(1995)\citenamefont {Walz},
  \citenamefont {Ross}, \citenamefont {Zimmermann}, \citenamefont {Ricci},
  \citenamefont {Prevedelli},\ and\ \citenamefont
  {Hänsch}}]{walz_combined_1995}%
  \BibitemOpen
  \bibfield  {author} {\bibinfo {author} {\bibfnamefont {J.}~\bibnamefont
  {Walz}}, \bibinfo {author} {\bibfnamefont {S.~B.}\ \bibnamefont {Ross}},
  \bibinfo {author} {\bibfnamefont {C.}~\bibnamefont {Zimmermann}}, \bibinfo
  {author} {\bibfnamefont {L.}~\bibnamefont {Ricci}}, \bibinfo {author}
  {\bibfnamefont {M.}~\bibnamefont {Prevedelli}},\ and\ \bibinfo {author}
  {\bibfnamefont {T.~W.}\ \bibnamefont {Hänsch}},\ }\bibfield  {title}
  {\bibinfo {title} {Combined {Trap} with the {Potential} for {Antihydrogen}
  {Production}},\ }\href {https://doi.org/10.1103/PhysRevLett.75.3257}
  {\bibfield  {journal} {\bibinfo  {journal} {Physical Review Letters}\
  }\textbf {\bibinfo {volume} {75}},\ \bibinfo {pages} {3257} (\bibinfo {year}
  {1995})}\BibitemShut {NoStop}%
\bibitem [{\citenamefont {Foot}\ \emph {et~al.}(2018)\citenamefont {Foot},
  \citenamefont {Trypogeorgos}, \citenamefont {Bentine}, \citenamefont
  {Gardner},\ and\ \citenamefont {Keller}}]{foot_two-frequency_2018}%
  \BibitemOpen
  \bibfield  {author} {\bibinfo {author} {\bibfnamefont {C.~J.}\ \bibnamefont
  {Foot}}, \bibinfo {author} {\bibfnamefont {D.}~\bibnamefont {Trypogeorgos}},
  \bibinfo {author} {\bibfnamefont {E.}~\bibnamefont {Bentine}}, \bibinfo
  {author} {\bibfnamefont {A.}~\bibnamefont {Gardner}},\ and\ \bibinfo {author}
  {\bibfnamefont {M.}~\bibnamefont {Keller}},\ }\bibfield  {title} {\bibinfo
  {title} {Two-frequency operation of a {Paul} trap to optimise confinement of
  two species of ions},\ }\href {https://doi.org/10.1016/j.ijms.2018.05.007}
  {\bibfield  {journal} {\bibinfo  {journal} {International Journal of Mass
  Spectrometry}\ }\textbf {\bibinfo {volume} {430}},\ \bibinfo {pages} {117}
  (\bibinfo {year} {2018})}\BibitemShut {NoStop}%
\bibitem [{\citenamefont {Leefer}\ \emph {et~al.}(2016)\citenamefont {Leefer},
  \citenamefont {Krimmel}, \citenamefont {Bertsche}, \citenamefont {Budker},
  \citenamefont {Fajans}, \citenamefont {Folman}, \citenamefont {Häffner},\
  and\ \citenamefont {Schmidt-Kaler}}]{leefer_investigation_2016}%
  \BibitemOpen
  \bibfield  {author} {\bibinfo {author} {\bibfnamefont {N.}~\bibnamefont
  {Leefer}}, \bibinfo {author} {\bibfnamefont {K.}~\bibnamefont {Krimmel}},
  \bibinfo {author} {\bibfnamefont {W.}~\bibnamefont {Bertsche}}, \bibinfo
  {author} {\bibfnamefont {D.}~\bibnamefont {Budker}}, \bibinfo {author}
  {\bibfnamefont {J.}~\bibnamefont {Fajans}}, \bibinfo {author} {\bibfnamefont
  {R.}~\bibnamefont {Folman}}, \bibinfo {author} {\bibfnamefont
  {H.}~\bibnamefont {Häffner}},\ and\ \bibinfo {author} {\bibfnamefont
  {F.}~\bibnamefont {Schmidt-Kaler}},\ }\bibfield  {title} {\bibinfo {title}
  {Investigation of two-frequency {Paul} traps for antihydrogen production},\
  }\href {https://doi.org/10.1007/s10751-016-1388-0} {\bibfield  {journal}
  {\bibinfo  {journal} {Hyperfine Interactions}\ }\textbf {\bibinfo {volume}
  {238}},\ \bibinfo {pages} {12} (\bibinfo {year} {2016})}\BibitemShut
  {NoStop}%
\bibitem [{\citenamefont {Bykov}\ \emph {et~al.}(2024)\citenamefont {Bykov},
  \citenamefont {Dania}, \citenamefont {Goschin},\ and\ \citenamefont
  {Northup}}]{bykov_nanoparticle_2024}%
  \BibitemOpen
  \bibfield  {author} {\bibinfo {author} {\bibfnamefont {D.~S.}\ \bibnamefont
  {Bykov}}, \bibinfo {author} {\bibfnamefont {L.}~\bibnamefont {Dania}},
  \bibinfo {author} {\bibfnamefont {F.}~\bibnamefont {Goschin}},\ and\ \bibinfo
  {author} {\bibfnamefont {T.~E.}\ \bibnamefont {Northup}},\ }\href
  {https://doi.org/10.48550/arXiv.2403.02034} {\bibinfo {title} {A nanoparticle
  stored with an atomic ion in a linear {Paul} trap}} (\bibinfo {year}
  {2024})\BibitemShut {NoStop}%
\bibitem [{\citenamefont {Gonzalez-Ballestero}\ \emph
  {et~al.}(2021)\citenamefont {Gonzalez-Ballestero}, \citenamefont
  {Aspelmeyer}, \citenamefont {Novotny}, \citenamefont {Quidant},\ and\
  \citenamefont {Romero-Isart}}]{gonzalez-ballestero_levitodynamics_2021}%
  \BibitemOpen
  \bibfield  {author} {\bibinfo {author} {\bibfnamefont {C.}~\bibnamefont
  {Gonzalez-Ballestero}}, \bibinfo {author} {\bibfnamefont {M.}~\bibnamefont
  {Aspelmeyer}}, \bibinfo {author} {\bibfnamefont {L.}~\bibnamefont {Novotny}},
  \bibinfo {author} {\bibfnamefont {R.}~\bibnamefont {Quidant}},\ and\ \bibinfo
  {author} {\bibfnamefont {O.}~\bibnamefont {Romero-Isart}},\ }\bibfield
  {title} {\bibinfo {title} {Levitodynamics: {Levitation} and control of
  microscopic objects in vacuum},\ }\bibfield  {journal} {\bibinfo  {journal}
  {Science}\ }\textbf {\bibinfo {volume} {374}},\ \href
  {https://doi.org/10.1126/science.abg3027} {10.1126/science.abg3027} (\bibinfo
  {year} {2021})\BibitemShut {NoStop}%
\bibitem [{\citenamefont {O’Neil}(1981)}]{oneil_centrifugal_1981}%
  \BibitemOpen
  \bibfield  {author} {\bibinfo {author} {\bibfnamefont {T.~M.}\ \bibnamefont
  {O’Neil}},\ }\bibfield  {title} {\bibinfo {title} {Centrifugal separation
  of a multispecies pure ion plasma},\ }\href
  {https://doi.org/10.1063/1.863565} {\bibfield  {journal} {\bibinfo  {journal}
  {The Physics of Fluids}\ }\textbf {\bibinfo {volume} {24}},\ \bibinfo {pages}
  {1447} (\bibinfo {year} {1981})}\BibitemShut {NoStop}%
\bibitem [{\citenamefont {Kumar}\ \emph {et~al.}(2025)\citenamefont {Kumar},
  \citenamefont {Lausti}, \citenamefont {Hajnyš}, \citenamefont {Hudák},
  \citenamefont {Motyčka}, \citenamefont {Jelínek},\ and\ \citenamefont
  {Hejduk}}]{kumar_3d-printed_2025}%
  \BibitemOpen
  \bibfield  {author} {\bibinfo {author} {\bibfnamefont {V.}~\bibnamefont
  {Kumar}}, \bibinfo {author} {\bibfnamefont {N.~V.}\ \bibnamefont {Lausti}},
  \bibinfo {author} {\bibfnamefont {J.}~\bibnamefont {Hajnyš}}, \bibinfo
  {author} {\bibfnamefont {I.}~\bibnamefont {Hudák}}, \bibinfo {author}
  {\bibfnamefont {D.}~\bibnamefont {Motyčka}}, \bibinfo {author}
  {\bibfnamefont {A.}~\bibnamefont {Jelínek}},\ and\ \bibinfo {author}
  {\bibfnamefont {M.}~\bibnamefont {Hejduk}},\ }\href
  {https://doi.org/10.48550/arXiv.2509.06537} {\bibinfo {title} {{3D}-printed
  components for electron-ion trapping: {Tests} of functionality and ultra-high
  vacuum compatibility}} (\bibinfo {year} {2025}),\ \bibinfo {note}
  {arXiv:2509.06537 [physics]}\BibitemShut {NoStop}%
\bibitem [{\citenamefont {Jefferts}\ \emph {et~al.}(1995)\citenamefont
  {Jefferts}, \citenamefont {Monroe}, \citenamefont {Bell},\ and\ \citenamefont
  {Wineland}}]{jefferts_coaxial-resonator-driven_1995}%
  \BibitemOpen
  \bibfield  {author} {\bibinfo {author} {\bibfnamefont {S.~R.}\ \bibnamefont
  {Jefferts}}, \bibinfo {author} {\bibfnamefont {C.}~\bibnamefont {Monroe}},
  \bibinfo {author} {\bibfnamefont {E.~W.}\ \bibnamefont {Bell}},\ and\
  \bibinfo {author} {\bibfnamefont {D.~J.}\ \bibnamefont {Wineland}},\
  }\bibfield  {title} {\bibinfo {title} {Coaxial-resonator-driven rf ({Paul})
  trap for strong confinement},\ }\href
  {https://doi.org/10.1103/PhysRevA.51.3112} {\bibfield  {journal} {\bibinfo
  {journal} {Physical Review A}\ }\textbf {\bibinfo {volume} {51}},\ \bibinfo
  {pages} {3112} (\bibinfo {year} {1995})}\BibitemShut {NoStop}%
\bibitem [{\citenamefont {Mikhailovskii}\ \emph {et~al.}(2025)\citenamefont
  {Mikhailovskii}, \citenamefont {Sheth}, \citenamefont {Qu}, \citenamefont
  {Hejduk}, \citenamefont {Lausti}, \citenamefont {Satyajith}, \citenamefont
  {Smorra}, \citenamefont {Werth}, \citenamefont {Yadav}, \citenamefont {Yu},
  \citenamefont {Matthiesen}, \citenamefont {Häffner}, \citenamefont
  {Schmidt-Kaler}, \citenamefont {Bekker},\ and\ \citenamefont
  {Budker}}]{mikhailovskii_trapping_2025}%
  \BibitemOpen
  \bibfield  {author} {\bibinfo {author} {\bibfnamefont {V.}~\bibnamefont
  {Mikhailovskii}}, \bibinfo {author} {\bibfnamefont {N.}~\bibnamefont
  {Sheth}}, \bibinfo {author} {\bibfnamefont {G.}~\bibnamefont {Qu}}, \bibinfo
  {author} {\bibfnamefont {M.}~\bibnamefont {Hejduk}}, \bibinfo {author}
  {\bibfnamefont {N.~V.}\ \bibnamefont {Lausti}}, \bibinfo {author}
  {\bibfnamefont {K.~T.}\ \bibnamefont {Satyajith}}, \bibinfo {author}
  {\bibfnamefont {C.}~\bibnamefont {Smorra}}, \bibinfo {author} {\bibfnamefont
  {G.}~\bibnamefont {Werth}}, \bibinfo {author} {\bibfnamefont
  {N.}~\bibnamefont {Yadav}}, \bibinfo {author} {\bibfnamefont
  {Q.}~\bibnamefont {Yu}}, \bibinfo {author} {\bibfnamefont {C.}~\bibnamefont
  {Matthiesen}}, \bibinfo {author} {\bibfnamefont {H.}~\bibnamefont
  {Häffner}}, \bibinfo {author} {\bibfnamefont {F.}~\bibnamefont
  {Schmidt-Kaler}}, \bibinfo {author} {\bibfnamefont {H.}~\bibnamefont
  {Bekker}},\ and\ \bibinfo {author} {\bibfnamefont {D.}~\bibnamefont
  {Budker}},\ }\href {https://doi.org/10.48550/arXiv.2508.16407} {\bibinfo
  {title} {Trapping of electrons and
  \${\textasciicircum}\{40\}{\textbackslash}textrm\{{Ca}\}{\textasciicircum}+\$
  ions in a dual-frequency {Paul} trap}} (\bibinfo {year} {2025}),\ \bibinfo
  {note} {arXiv:2508.16407 [physics]}\BibitemShut {NoStop}%
\bibitem [{\citenamefont {Ragg}\ \emph {et~al.}(2019)\citenamefont {Ragg},
  \citenamefont {Decaroli}, \citenamefont {Lutz},\ and\ \citenamefont
  {Home}}]{ragg_segmented_2019}%
  \BibitemOpen
  \bibfield  {author} {\bibinfo {author} {\bibfnamefont {S.}~\bibnamefont
  {Ragg}}, \bibinfo {author} {\bibfnamefont {C.}~\bibnamefont {Decaroli}},
  \bibinfo {author} {\bibfnamefont {T.}~\bibnamefont {Lutz}},\ and\ \bibinfo
  {author} {\bibfnamefont {J.~P.}\ \bibnamefont {Home}},\ }\bibfield  {title}
  {\bibinfo {title} {Segmented ion-trap fabrication using high precision
  stacked wafers},\ }\href {https://doi.org/10.1063/1.5119785} {\bibfield
  {journal} {\bibinfo  {journal} {Review of Scientific Instruments}\ }\textbf
  {\bibinfo {volume} {90}},\ \bibinfo {pages} {103203} (\bibinfo {year}
  {2019})}\BibitemShut {NoStop}%
\bibitem [{\citenamefont {Tolpygo}\ \emph {et~al.}(2023)\citenamefont
  {Tolpygo}, \citenamefont {Mallek}, \citenamefont {Bolkhovsky}, \citenamefont
  {Rastogi}, \citenamefont {Golden}, \citenamefont {Weir}, \citenamefont
  {Johnson},\ and\ \citenamefont {Gouker}}]{tolpygo_progress_2023}%
  \BibitemOpen
  \bibfield  {author} {\bibinfo {author} {\bibfnamefont {S.~K.}\ \bibnamefont
  {Tolpygo}}, \bibinfo {author} {\bibfnamefont {J.~L.}\ \bibnamefont {Mallek}},
  \bibinfo {author} {\bibfnamefont {V.}~\bibnamefont {Bolkhovsky}}, \bibinfo
  {author} {\bibfnamefont {R.}~\bibnamefont {Rastogi}}, \bibinfo {author}
  {\bibfnamefont {E.~B.}\ \bibnamefont {Golden}}, \bibinfo {author}
  {\bibfnamefont {T.~J.}\ \bibnamefont {Weir}}, \bibinfo {author}
  {\bibfnamefont {L.~M.}\ \bibnamefont {Johnson}},\ and\ \bibinfo {author}
  {\bibfnamefont {M.~A.}\ \bibnamefont {Gouker}},\ }\bibfield  {title}
  {\bibinfo {title} {Progress {Toward} {Superconductor} {Electronics}
  {Fabrication} {Process} {With} {Planarized} {NbN} and {NbN}/{Nb} {Layers}},\
  }\href {https://doi.org/10.1109/TASC.2023.3246430} {\bibfield  {journal}
  {\bibinfo  {journal} {IEEE Transactions on Applied Superconductivity}\
  }\textbf {\bibinfo {volume} {33}},\ \bibinfo {pages} {1} (\bibinfo {year}
  {2023})}\BibitemShut {NoStop}%
\bibitem [{\citenamefont {Leibfried}\ \emph {et~al.}(2003)\citenamefont
  {Leibfried}, \citenamefont {Blatt}, \citenamefont {Monroe},\ and\
  \citenamefont {Wineland}}]{leibfried_quantum_2003}%
  \BibitemOpen
  \bibfield  {author} {\bibinfo {author} {\bibfnamefont {D.}~\bibnamefont
  {Leibfried}}, \bibinfo {author} {\bibfnamefont {R.}~\bibnamefont {Blatt}},
  \bibinfo {author} {\bibfnamefont {C.}~\bibnamefont {Monroe}},\ and\ \bibinfo
  {author} {\bibfnamefont {D.}~\bibnamefont {Wineland}},\ }\bibfield  {title}
  {\bibinfo {title} {Quantum dynamics of single trapped ions},\ }\href
  {https://doi.org/10.1103/RevModPhys.75.281} {\bibfield  {journal} {\bibinfo
  {journal} {Reviews of Modern Physics}\ }\textbf {\bibinfo {volume} {75}},\
  \bibinfo {pages} {281} (\bibinfo {year} {2003})}\BibitemShut {NoStop}%
\bibitem [{\citenamefont {Andrijauskas}\ \emph {et~al.}(2021)\citenamefont
  {Andrijauskas}, \citenamefont {Vogel}, \citenamefont {Mokhberi},\ and\
  \citenamefont {Schmidt-Kaler}}]{andrijauskas_rydberg_2021}%
  \BibitemOpen
  \bibfield  {author} {\bibinfo {author} {\bibfnamefont {J.}~\bibnamefont
  {Andrijauskas}}, \bibinfo {author} {\bibfnamefont {J.}~\bibnamefont {Vogel}},
  \bibinfo {author} {\bibfnamefont {A.}~\bibnamefont {Mokhberi}},\ and\
  \bibinfo {author} {\bibfnamefont {F.}~\bibnamefont {Schmidt-Kaler}},\
  }\bibfield  {title} {\bibinfo {title} {Rydberg {Series} {Excitation} of a
  {Single} {Trapped} $^{\textrm{40}}${Ca}$^{\textrm{+}}$ {Ion} for {Precision}
  {Measurements} and {Principal} {Quantum} {Number} {Scalings}},\ }\href
  {https://doi.org/10.1103/PhysRevLett.127.203001} {\bibfield  {journal}
  {\bibinfo  {journal} {Physical Review Letters}\ }\textbf {\bibinfo {volume}
  {127}},\ \bibinfo {pages} {203001} (\bibinfo {year} {2021})}\BibitemShut
  {NoStop}%
\bibitem [{\citenamefont {Hayasaka}\ \emph {et~al.}(2000)\citenamefont
  {Hayasaka}, \citenamefont {Urabe},\ and\ \citenamefont
  {Watanabe}}]{hayasaka_laser_2000}%
  \BibitemOpen
  \bibfield  {author} {\bibinfo {author} {\bibfnamefont {K.~H.~K.}\
  \bibnamefont {Hayasaka}}, \bibinfo {author} {\bibfnamefont {S.~U.~S.}\
  \bibnamefont {Urabe}},\ and\ \bibinfo {author} {\bibfnamefont {M.~W.~M.}\
  \bibnamefont {Watanabe}},\ }\bibfield  {title} {\bibinfo {title} {Laser
  {Cooling} of {Ca}+ with an {External}-{Cavity} {Ultraviolet} {Diode}
  {Laser}},\ }\href {https://doi.org/10.1143/JJAP.39.L687} {\bibfield
  {journal} {\bibinfo  {journal} {Japanese Journal of Applied Physics}\
  }\textbf {\bibinfo {volume} {39}},\ \bibinfo {pages} {L687} (\bibinfo {year}
  {2000})}\BibitemShut {NoStop}%
\bibitem [{\citenamefont {Tangtrongbenchasil}\ \emph
  {et~al.}(2006)\citenamefont {Tangtrongbenchasil}, \citenamefont {Ohara},
  \citenamefont {Itagaki}, \citenamefont {Vesarach},\ and\ \citenamefont
  {Nonaka}}]{tangtrongbenchasil_219-nm_2006}%
  \BibitemOpen
  \bibfield  {author} {\bibinfo {author} {\bibfnamefont {C.}~\bibnamefont
  {Tangtrongbenchasil}}, \bibinfo {author} {\bibfnamefont {K.}~\bibnamefont
  {Ohara}}, \bibinfo {author} {\bibfnamefont {T.}~\bibnamefont {Itagaki}},
  \bibinfo {author} {\bibfnamefont {P.}~\bibnamefont {Vesarach}},\ and\
  \bibinfo {author} {\bibfnamefont {K.}~\bibnamefont {Nonaka}},\ }\bibfield
  {title} {\bibinfo {title} {219-nm {Ultra} {Violet} {Generation} {Using}
  {Blue} {Laser} {Diode} and {External} {Cavity}},\ }\href
  {https://doi.org/10.1143/JJAP.45.6315} {\bibfield  {journal} {\bibinfo
  {journal} {Japanese Journal of Applied Physics}\ }\textbf {\bibinfo {volume}
  {45}},\ \bibinfo {pages} {6315} (\bibinfo {year} {2006})}\BibitemShut
  {NoStop}%
\bibitem [{\citenamefont {Wolz}\ \emph {et~al.}(2020)\citenamefont {Wolz},
  \citenamefont {Malbrunot}, \citenamefont {Vieille-Grosjean}, \citenamefont
  {Comparat},\ and\ \citenamefont {{D. Comparat}}}]{wolz_stimulated_2020}%
  \BibitemOpen
  \bibfield  {author} {\bibinfo {author} {\bibfnamefont {T.}~\bibnamefont
  {Wolz}}, \bibinfo {author} {\bibfnamefont {C.}~\bibnamefont {Malbrunot}},
  \bibinfo {author} {\bibfnamefont {M.}~\bibnamefont {Vieille-Grosjean}},
  \bibinfo {author} {\bibfnamefont {D.}~\bibnamefont {Comparat}},\ and\
  \bibinfo {author} {\bibnamefont {{D. Comparat}}},\ }\bibfield  {title}
  {\bibinfo {title} {Stimulated decay and formation of antihydrogen atoms},\
  }\bibfield  {journal} {\bibinfo  {journal} {Physical Review A}\ }\href
  {https://doi.org/10.1103/physreva.101.043412} {10.1103/physreva.101.043412}
  (\bibinfo {year} {2020})\BibitemShut {NoStop}%
\bibitem [{\citenamefont {Mahmoudi}\ \emph {et~al.}(2025)\citenamefont
  {Mahmoudi}, \citenamefont {Tarana},\ and\ \citenamefont
  {Hejduk}}]{mahmoudi_tarana_hejduk}%
  \BibitemOpen
  \bibfield  {author} {\bibinfo {author} {\bibfnamefont {P.}~\bibnamefont
  {Mahmoudi}}, \bibinfo {author} {\bibfnamefont {M.}~\bibnamefont {Tarana}},\
  and\ \bibinfo {author} {\bibfnamefont {M.}~\bibnamefont {Hejduk}},\
  }\bibfield  {title} {\bibinfo {title} {Theoretical study of classical
  dynamics of electron in two-frequency paul trap}} (\bibinfo {year} {2025}),\
  \bibinfo {note} {to be published}\BibitemShut {NoStop}%
\bibitem [{\citenamefont {Foot}(2005)}]{foot_atomic_2005}%
  \BibitemOpen
  \bibfield  {author} {\bibinfo {author} {\bibfnamefont {C.~J.}\ \bibnamefont
  {Foot}},\ }\href@noop {} {\emph {\bibinfo {title} {Atomic physics}}},\
  \bibinfo {series} {Oxford master series in physics}\ No.\ \bibinfo {number}
  {7. Atomic, Optical, and laser physics}\ (\bibinfo  {publisher} {Oxford
  University Press},\ \bibinfo {address} {Oxford ; New York},\ \bibinfo {year}
  {2005})\BibitemShut {NoStop}%
\bibitem [{\citenamefont {Cook}\ \emph {et~al.}(1985)\citenamefont {Cook},
  \citenamefont {Shankland},\ and\ \citenamefont {Wells}}]{Cook_1985}%
  \BibitemOpen
  \bibfield  {author} {\bibinfo {author} {\bibfnamefont {R.~J.}\ \bibnamefont
  {Cook}}, \bibinfo {author} {\bibfnamefont {D.~G.}\ \bibnamefont
  {Shankland}},\ and\ \bibinfo {author} {\bibfnamefont {A.~L.}\ \bibnamefont
  {Wells}},\ }\bibfield  {title} {\bibinfo {title} {Quantum theory of particle
  motion in a rapidly oscillating field},\ }\href
  {https://doi.org/10.1103/physreva.31.564} {\bibfield  {journal} {\bibinfo
  {journal} {Phys. Rev. A}\ }\textbf {\bibinfo {volume} {31}},\ \bibinfo
  {pages} {564} (\bibinfo {year} {1985})}\BibitemShut {NoStop}%
\bibitem [{\citenamefont
  {Qi}(2009)}]{qiElectromagneticallyInducedTransparency2009}%
  \BibitemOpen
  \bibfield  {author} {\bibinfo {author} {\bibfnamefont {J.}~\bibnamefont
  {Qi}},\ }\bibfield  {title} {\bibinfo {title} {Electromagnetically induced
  transparency in an inverted {{Y-type}} four-level system},\ }\href
  {https://doi.org/10.1088/0031-8949/81/01/015402} {\bibfield  {journal}
  {\bibinfo  {journal} {Physica Scripta}\ }\textbf {\bibinfo {volume} {81}},\
  \bibinfo {pages} {015402} (\bibinfo {year} {2009})}\BibitemShut {NoStop}%
\bibitem [{\citenamefont {Finkelstein}\ \emph {et~al.}(2023)\citenamefont
  {Finkelstein}, \citenamefont {Bali}, \citenamefont {Firstenberg},\ and\
  \citenamefont {Novikova}}]{finkelstein_practical_2023}%
  \BibitemOpen
  \bibfield  {author} {\bibinfo {author} {\bibfnamefont {R.}~\bibnamefont
  {Finkelstein}}, \bibinfo {author} {\bibfnamefont {S.}~\bibnamefont {Bali}},
  \bibinfo {author} {\bibfnamefont {O.}~\bibnamefont {Firstenberg}},\ and\
  \bibinfo {author} {\bibfnamefont {I.}~\bibnamefont {Novikova}},\ }\bibfield
  {title} {\bibinfo {title} {A practical guide to electromagnetically induced
  transparency in atomic vapor},\ }\href
  {https://doi.org/10.1088/1367-2630/acbc40} {\bibfield  {journal} {\bibinfo
  {journal} {New Journal of Physics}\ }\textbf {\bibinfo {volume} {25}},\
  \bibinfo {pages} {035001} (\bibinfo {year} {2023})}\BibitemShut {NoStop}%
\bibitem [{\citenamefont {Lesanovsky}\ \emph {et~al.}(2009)\citenamefont
  {Lesanovsky}, \citenamefont {Müller},\ and\ \citenamefont
  {Zoller}}]{lesanovsky_trap-assisted_2009}%
  \BibitemOpen
  \bibfield  {author} {\bibinfo {author} {\bibfnamefont {I.}~\bibnamefont
  {Lesanovsky}}, \bibinfo {author} {\bibfnamefont {M.}~\bibnamefont
  {Müller}},\ and\ \bibinfo {author} {\bibfnamefont {P.}~\bibnamefont
  {Zoller}},\ }\bibfield  {title} {\bibinfo {title} {Trap-assisted creation of
  giant molecules and {Rydberg}-mediated coherent charge transfer in a
  {Penning} trap},\ }\href {https://doi.org/10.1103/PhysRevA.79.010701}
  {\bibfield  {journal} {\bibinfo  {journal} {Physical Review A}\ }\textbf
  {\bibinfo {volume} {79}},\ \bibinfo {pages} {010701} (\bibinfo {year}
  {2009})}\BibitemShut {NoStop}%
\bibitem [{\citenamefont {Vogt}\ \emph {et~al.}(2019)\citenamefont {Vogt},
  \citenamefont {Gross}, \citenamefont {Han}, \citenamefont {Pal},
  \citenamefont {Lam}, \citenamefont {Kiffner},\ and\ \citenamefont
  {Li}}]{vogt_efficient_2019}%
  \BibitemOpen
  \bibfield  {author} {\bibinfo {author} {\bibfnamefont {T.}~\bibnamefont
  {Vogt}}, \bibinfo {author} {\bibfnamefont {C.}~\bibnamefont {Gross}},
  \bibinfo {author} {\bibfnamefont {J.}~\bibnamefont {Han}}, \bibinfo {author}
  {\bibfnamefont {S.~B.}\ \bibnamefont {Pal}}, \bibinfo {author} {\bibfnamefont
  {M.}~\bibnamefont {Lam}}, \bibinfo {author} {\bibfnamefont {M.}~\bibnamefont
  {Kiffner}},\ and\ \bibinfo {author} {\bibfnamefont {W.}~\bibnamefont {Li}},\
  }\bibfield  {title} {\bibinfo {title} {Efficient microwave-to-optical
  conversion using {Rydberg} atoms},\ }\href
  {https://doi.org/10.1103/PhysRevA.99.023832} {\bibfield  {journal} {\bibinfo
  {journal} {Physical Review A}\ }\textbf {\bibinfo {volume} {99}},\ \bibinfo
  {pages} {023832} (\bibinfo {year} {2019})}\BibitemShut {NoStop}%
\bibitem [{\citenamefont {Borówka}\ \emph {et~al.}(2024)\citenamefont
  {Borówka}, \citenamefont {Pylypenko}, \citenamefont {Mazelanik},\ and\
  \citenamefont {Parniak}}]{borowka_continuous_2024}%
  \BibitemOpen
  \bibfield  {author} {\bibinfo {author} {\bibfnamefont {S.}~\bibnamefont
  {Borówka}}, \bibinfo {author} {\bibfnamefont {U.}~\bibnamefont {Pylypenko}},
  \bibinfo {author} {\bibfnamefont {M.}~\bibnamefont {Mazelanik}},\ and\
  \bibinfo {author} {\bibfnamefont {M.}~\bibnamefont {Parniak}},\ }\bibfield
  {title} {\bibinfo {title} {Continuous wideband microwave-to-optical converter
  based on room-temperature {Rydberg} atoms},\ }\href
  {https://doi.org/10.1038/s41566-023-01295-w} {\bibfield  {journal} {\bibinfo
  {journal} {Nature Photonics}\ }\textbf {\bibinfo {volume} {18}},\ \bibinfo
  {pages} {32} (\bibinfo {year} {2024})}\BibitemShut {NoStop}%
\bibitem [{\citenamefont {{International Telecommunication
  Union}}(2015)}]{ITU2015}%
  \BibitemOpen
  \bibfield  {author} {\bibinfo {author} {\bibnamefont {{International
  Telecommunication Union}}},\ }\href {https://www.itu.int/rec/R-REC-V.431/en}
  {\emph {\bibinfo {title} {Recommendation ITU-R V.431-8: Nomenclature of the
  frequency and wavelength bands used in telecommunications}}},\ \bibinfo
  {type} {Recommendation}\ \bibinfo {number} {ITU-R V.431-8}\ (\bibinfo
  {institution} {International Telecommunication Union},\ \bibinfo {address}
  {Geneva},\ \bibinfo {year} {2015})\BibitemShut {NoStop}%
\bibitem [{\citenamefont {Kim}\ \emph {et~al.}(2010)\citenamefont {Kim},
  \citenamefont {Herskind}, \citenamefont {Kim}, \citenamefont {Kim},\ and\
  \citenamefont {Chuang}}]{kim_surface-electrode_2010}%
  \BibitemOpen
  \bibfield  {author} {\bibinfo {author} {\bibfnamefont {T.~H.}\ \bibnamefont
  {Kim}}, \bibinfo {author} {\bibfnamefont {P.~F.}\ \bibnamefont {Herskind}},
  \bibinfo {author} {\bibfnamefont {T.}~\bibnamefont {Kim}}, \bibinfo {author}
  {\bibfnamefont {J.}~\bibnamefont {Kim}},\ and\ \bibinfo {author}
  {\bibfnamefont {I.~L.}\ \bibnamefont {Chuang}},\ }\bibfield  {title}
  {\bibinfo {title} {Surface-electrode point {Paul} trap},\ }\href
  {https://doi.org/10.1103/PhysRevA.82.043412} {\bibfield  {journal} {\bibinfo
  {journal} {Physical Review A}\ }\textbf {\bibinfo {volume} {82}},\ \bibinfo
  {pages} {043412} (\bibinfo {year} {2010})}\BibitemShut {NoStop}%
\bibitem [{\citenamefont {Clark}(2013)}]{clark_ideal_2013}%
  \BibitemOpen
  \bibfield  {author} {\bibinfo {author} {\bibfnamefont {R.~J.}\ \bibnamefont
  {Clark}},\ }\bibfield  {title} {\bibinfo {title} {Ideal multipole ion traps
  from planar ring electrodes},\ }\href
  {https://doi.org/10.1007/s00340-013-5451-0} {\bibfield  {journal} {\bibinfo
  {journal} {Applied Physics B}\ }\textbf {\bibinfo {volume} {113}},\ \bibinfo
  {pages} {171} (\bibinfo {year} {2013})}\BibitemShut {NoStop}%
\bibitem [{\citenamefont {Wang}\ \emph {et~al.}(2015)\citenamefont {Wang},
  \citenamefont {Li}, \citenamefont {Noel}, \citenamefont {Chuang},
  \citenamefont {Zhang},\ and\ \citenamefont {Häffner}}]{wang_surface_2015}%
  \BibitemOpen
  \bibfield  {author} {\bibinfo {author} {\bibfnamefont {P.-J.}\ \bibnamefont
  {Wang}}, \bibinfo {author} {\bibfnamefont {T.}~\bibnamefont {Li}}, \bibinfo
  {author} {\bibfnamefont {C.}~\bibnamefont {Noel}}, \bibinfo {author}
  {\bibfnamefont {A.}~\bibnamefont {Chuang}}, \bibinfo {author} {\bibfnamefont
  {X.}~\bibnamefont {Zhang}},\ and\ \bibinfo {author} {\bibfnamefont
  {H.}~\bibnamefont {Häffner}},\ }\bibfield  {title} {\bibinfo {title}
  {Surface traps for freely rotating ion ring crystals},\ }\href
  {https://doi.org/10.1088/0953-4075/48/20/205002} {\bibfield  {journal}
  {\bibinfo  {journal} {Journal of Physics B: Atomic, Molecular and Optical
  Physics}\ }\textbf {\bibinfo {volume} {48}},\ \bibinfo {pages} {205002}
  (\bibinfo {year} {2015})}\BibitemShut {NoStop}%
\bibitem [{\citenamefont {Schmied}\ \emph {et~al.}(2009)\citenamefont
  {Schmied}, \citenamefont {Wesenberg},\ and\ \citenamefont
  {Leibfried}}]{schmied_optimal_2009}%
  \BibitemOpen
  \bibfield  {author} {\bibinfo {author} {\bibfnamefont {R.}~\bibnamefont
  {Schmied}}, \bibinfo {author} {\bibfnamefont {J.~H.}\ \bibnamefont
  {Wesenberg}},\ and\ \bibinfo {author} {\bibfnamefont {D.}~\bibnamefont
  {Leibfried}},\ }\bibfield  {title} {\bibinfo {title} {Optimal
  {Surface}-{Electrode} {Trap} {Lattices} for {Quantum} {Simulation} with
  {Trapped} {Ions}},\ }\href {https://doi.org/10.1103/PhysRevLett.102.233002}
  {\bibfield  {journal} {\bibinfo  {journal} {Physical Review Letters}\
  }\textbf {\bibinfo {volume} {102}},\ \bibinfo {pages} {233002} (\bibinfo
  {year} {2009})}\BibitemShut {NoStop}%
\bibitem [{\citenamefont {Schmied}(2018)}]{schmied_surfacepattern_nodate}%
  \BibitemOpen
  \bibfield  {author} {\bibinfo {author} {\bibfnamefont {R.}~\bibnamefont
  {Schmied}},\ }\href
  {https://atom.physik.unibas.ch/en/people/group-members/roman-schmied/surfacepattern/}
  {\bibinfo {title} {{SurfacePattern}}} (\bibinfo {year} {2018}),\ \bibinfo
  {note} {retrieved from
  https://atom.physik.unibas.ch/en/people/group-members/roman-schmied/surfacepattern/}\BibitemShut
  {NoStop}%
\bibitem [{\citenamefont {Zhang}\ \emph {et~al.}(2005)\citenamefont {Zhang},
  \citenamefont {Guo}, \citenamefont {Ong},\ and\ \citenamefont
  {Chia}}]{zhang_new_2005}%
  \BibitemOpen
  \bibfield  {author} {\bibinfo {author} {\bibfnamefont {Z.-Y.}\ \bibnamefont
  {Zhang}}, \bibinfo {author} {\bibfnamefont {Y.-X.}\ \bibnamefont {Guo}},
  \bibinfo {author} {\bibfnamefont {L.}~\bibnamefont {Ong}},\ and\ \bibinfo
  {author} {\bibfnamefont {M.}~\bibnamefont {Chia}},\ }\bibfield  {title}
  {\bibinfo {title} {A new planar marchand balun},\ }in\ \href
  {https://doi.org/10.1109/MWSYM.2005.1516893} {\emph {\bibinfo {booktitle}
  {{IEEE} {MTT}-{S} {International} {Microwave} {Symposium} {Digest}, 2005.}}}\
  (\bibinfo {year} {2005})\ pp.\ \bibinfo {pages} {1207--1210}\BibitemShut
  {NoStop}%
\bibitem [{\citenamefont {Rendek}(2022)}]{rendek_simulations_2022}%
  \BibitemOpen
  \bibfield  {author} {\bibinfo {author} {\bibfnamefont {A.}~\bibnamefont
  {Rendek}},\ }\emph {\bibinfo {title} {Simulations of dynamics of ultra-cold
  quantum plasma}},\ \href {http://hdl.handle.net/20.500.11956/175610}
  {\bibinfo {type} {bachelor thesis}},\ \bibinfo  {school} {Charles
  University}, \bibinfo {address} {Praha, Czech Republic} (\bibinfo {year}
  {2022})\BibitemShut {NoStop}%
\bibitem [{\citenamefont {Salahouelhadj}\ \emph {et~al.}(2014)\citenamefont
  {Salahouelhadj}, \citenamefont {Martiny}, \citenamefont {Mercier},
  \citenamefont {Bodin}, \citenamefont {Manteigas},\ and\ \citenamefont
  {Stephan}}]{salahouelhadj_reliability_2014}%
  \BibitemOpen
  \bibfield  {author} {\bibinfo {author} {\bibfnamefont {A.}~\bibnamefont
  {Salahouelhadj}}, \bibinfo {author} {\bibfnamefont {M.}~\bibnamefont
  {Martiny}}, \bibinfo {author} {\bibfnamefont {S.}~\bibnamefont {Mercier}},
  \bibinfo {author} {\bibfnamefont {L.}~\bibnamefont {Bodin}}, \bibinfo
  {author} {\bibfnamefont {D.}~\bibnamefont {Manteigas}},\ and\ \bibinfo
  {author} {\bibfnamefont {B.}~\bibnamefont {Stephan}},\ }\bibfield  {title}
  {\bibinfo {title} {Reliability of thermally stressed rigid–flex printed
  circuit boards for {High} {Density} {Interconnect} applications},\ }\href
  {https://doi.org/10.1016/j.microrel.2013.08.005} {\bibfield  {journal}
  {\bibinfo  {journal} {Microelectronics Reliability}\ }\textbf {\bibinfo
  {volume} {54}},\ \bibinfo {pages} {204} (\bibinfo {year} {2014})}\BibitemShut
  {NoStop}%
\bibitem [{\citenamefont {Krupka}\ \emph {et~al.}(2006)\citenamefont {Krupka},
  \citenamefont {Breeze}, \citenamefont {Centeno}, \citenamefont {Alford},
  \citenamefont {Claussen},\ and\ \citenamefont
  {Jensen}}]{krupka_measurements_2006}%
  \BibitemOpen
  \bibfield  {author} {\bibinfo {author} {\bibfnamefont {J.}~\bibnamefont
  {Krupka}}, \bibinfo {author} {\bibfnamefont {J.}~\bibnamefont {Breeze}},
  \bibinfo {author} {\bibfnamefont {A.}~\bibnamefont {Centeno}}, \bibinfo
  {author} {\bibfnamefont {N.}~\bibnamefont {Alford}}, \bibinfo {author}
  {\bibfnamefont {T.}~\bibnamefont {Claussen}},\ and\ \bibinfo {author}
  {\bibfnamefont {L.}~\bibnamefont {Jensen}},\ }\bibfield  {title} {\bibinfo
  {title} {Measurements of {Permittivity}, {Dielectric} {Loss} {Tangent}, and
  {Resistivity} of {Float}-{Zone} {Silicon} at {Microwave} {Frequencies}},\
  }\href {https://doi.org/10.1109/TMTT.2006.883655} {\bibfield  {journal}
  {\bibinfo  {journal} {IEEE Transactions on Microwave Theory and Techniques}\
  }\textbf {\bibinfo {volume} {54}},\ \bibinfo {pages} {3995} (\bibinfo {year}
  {2006})}\BibitemShut {NoStop}%
\bibitem [{\citenamefont {Rodriguez-Cano}\ \emph {et~al.}(2023)\citenamefont
  {Rodriguez-Cano}, \citenamefont {Perini}, \citenamefont {Foley},\ and\
  \citenamefont {Lanagan}}]{rodriguez-cano_broadband_2023}%
  \BibitemOpen
  \bibfield  {author} {\bibinfo {author} {\bibfnamefont {R.}~\bibnamefont
  {Rodriguez-Cano}}, \bibinfo {author} {\bibfnamefont {S.~E.}\ \bibnamefont
  {Perini}}, \bibinfo {author} {\bibfnamefont {B.~M.}\ \bibnamefont {Foley}},\
  and\ \bibinfo {author} {\bibfnamefont {M.}~\bibnamefont {Lanagan}},\
  }\bibfield  {title} {\bibinfo {title} {Broadband {Characterization} of
  {Silicate} {Materials} for {Potential} {5G}/{6G} {Applications}},\ }\href
  {https://doi.org/10.1109/TIM.2023.3256463} {\bibfield  {journal} {\bibinfo
  {journal} {IEEE Transactions on Instrumentation and Measurement}\ }\textbf
  {\bibinfo {volume} {72}},\ \bibinfo {pages} {1} (\bibinfo {year}
  {2023})}\BibitemShut {NoStop}%
\bibitem [{\citenamefont {Antony}\ \emph {et~al.}(2023)\citenamefont {Antony},
  \citenamefont {Hejduk}, \citenamefont {Hrbek}, \citenamefont {Kúš},
  \citenamefont {Bičišťová}, \citenamefont {Hauschwitz},\ and\
  \citenamefont {Cvrček}}]{antony_laser-assisted_2023}%
  \BibitemOpen
  \bibfield  {author} {\bibinfo {author} {\bibfnamefont {A.}~\bibnamefont
  {Antony}}, \bibinfo {author} {\bibfnamefont {M.}~\bibnamefont {Hejduk}},
  \bibinfo {author} {\bibfnamefont {T.}~\bibnamefont {Hrbek}}, \bibinfo
  {author} {\bibfnamefont {P.}~\bibnamefont {Kúš}}, \bibinfo {author}
  {\bibfnamefont {R.}~\bibnamefont {Bičišťová}}, \bibinfo {author}
  {\bibfnamefont {P.}~\bibnamefont {Hauschwitz}},\ and\ \bibinfo {author}
  {\bibfnamefont {L.}~\bibnamefont {Cvrček}},\ }\bibfield  {title} {\bibinfo
  {title} {Laser-assisted two-step glass wafer metallization: {An} experimental
  procedure to improve compatibility between glass and metallic films},\ }\href
  {https://doi.org/10.1016/j.apsusc.2023.157276} {\bibfield  {journal}
  {\bibinfo  {journal} {Applied Surface Science}\ }\textbf {\bibinfo {volume}
  {627}},\ \bibinfo {pages} {157276} (\bibinfo {year} {2023})}\BibitemShut
  {NoStop}%
\bibitem [{\citenamefont {Wilson}\ \emph {et~al.}(2022)\citenamefont {Wilson},
  \citenamefont {Tilles}, \citenamefont {Haltli}, \citenamefont {Ou},
  \citenamefont {Blain}, \citenamefont {Clark},\ and\ \citenamefont
  {Revelle}}]{wilson_situ_2022}%
  \BibitemOpen
  \bibfield  {author} {\bibinfo {author} {\bibfnamefont {J.~M.}\ \bibnamefont
  {Wilson}}, \bibinfo {author} {\bibfnamefont {J.~N.}\ \bibnamefont {Tilles}},
  \bibinfo {author} {\bibfnamefont {R.~A.}\ \bibnamefont {Haltli}}, \bibinfo
  {author} {\bibfnamefont {E.}~\bibnamefont {Ou}}, \bibinfo {author}
  {\bibfnamefont {M.~G.}\ \bibnamefont {Blain}}, \bibinfo {author}
  {\bibfnamefont {S.~M.}\ \bibnamefont {Clark}},\ and\ \bibinfo {author}
  {\bibfnamefont {M.~C.}\ \bibnamefont {Revelle}},\ }\bibfield  {title}
  {\bibinfo {title} {In situ detection of {RF} breakdown on microfabricated
  surface ion traps},\ }\href {https://doi.org/10.1063/5.0082740} {\bibfield
  {journal} {\bibinfo  {journal} {Journal of Applied Physics}\ }\textbf
  {\bibinfo {volume} {131}},\ \bibinfo {pages} {134401} (\bibinfo {year}
  {2022})}\BibitemShut {NoStop}%
\bibitem [{\citenamefont {Dahan}\ \emph {et~al.}(2021)\citenamefont {Dahan},
  \citenamefont {Holdengreber}, \citenamefont {Glassner}, \citenamefont
  {Sorkin}, \citenamefont {Schacham},\ and\ \citenamefont
  {Farber}}]{dahan_measurement_2021}%
  \BibitemOpen
  \bibfield  {author} {\bibinfo {author} {\bibfnamefont {Y.}~\bibnamefont
  {Dahan}}, \bibinfo {author} {\bibfnamefont {E.}~\bibnamefont {Holdengreber}},
  \bibinfo {author} {\bibfnamefont {E.}~\bibnamefont {Glassner}}, \bibinfo
  {author} {\bibfnamefont {O.}~\bibnamefont {Sorkin}}, \bibinfo {author}
  {\bibfnamefont {S.~E.}\ \bibnamefont {Schacham}},\ and\ \bibinfo {author}
  {\bibfnamefont {E.}~\bibnamefont {Farber}},\ }\bibfield  {title} {\bibinfo
  {title} {Measurement of {Electrical} {Properties} of {Superconducting} {YBCO}
  {Thin} {Films} in the {VHF} {Range}},\ }\href
  {https://doi.org/10.3390/ma14123360} {\bibfield  {journal} {\bibinfo
  {journal} {Materials}\ }\textbf {\bibinfo {volume} {14}},\ \bibinfo {pages}
  {3360} (\bibinfo {year} {2021})}\BibitemShut {NoStop}%
\bibitem [{\citenamefont {Moriya}\ \emph {et~al.}(2018)\citenamefont {Moriya},
  \citenamefont {Igarashi}, \citenamefont {Watanabe}, \citenamefont {Hasegawa},
  \citenamefont {Sasaki},\ and\ \citenamefont {Yasuda}}]{moriya_growth_2018}%
  \BibitemOpen
  \bibfield  {author} {\bibinfo {author} {\bibfnamefont {K.}~\bibnamefont
  {Moriya}}, \bibinfo {author} {\bibfnamefont {K.}~\bibnamefont {Igarashi}},
  \bibinfo {author} {\bibfnamefont {H.}~\bibnamefont {Watanabe}}, \bibinfo
  {author} {\bibfnamefont {H.}~\bibnamefont {Hasegawa}}, \bibinfo {author}
  {\bibfnamefont {T.}~\bibnamefont {Sasaki}},\ and\ \bibinfo {author}
  {\bibfnamefont {A.}~\bibnamefont {Yasuda}},\ }\bibfield  {title} {\bibinfo
  {title} {Growth of {YBa2Cu3O7} superconductor thin films using
  ethanolamine-based solutions via simple spin coating},\ }\href
  {https://doi.org/10.1016/j.rinp.2018.09.020} {\bibfield  {journal} {\bibinfo
  {journal} {Results in Physics}\ }\textbf {\bibinfo {volume} {11}},\ \bibinfo
  {pages} {364} (\bibinfo {year} {2018})}\BibitemShut {NoStop}%
\bibitem [{\citenamefont {Studebaker}\ \emph {et~al.}(1995)\citenamefont
  {Studebaker}, \citenamefont {Doubinina}, \citenamefont {Zhang}, \citenamefont
  {Wang}, \citenamefont {Dravid},\ and\ \citenamefont
  {Marks}}]{studebaker_liquid_1995}%
  \BibitemOpen
  \bibfield  {author} {\bibinfo {author} {\bibfnamefont {D.~B.}\ \bibnamefont
  {Studebaker}}, \bibinfo {author} {\bibfnamefont {G.}~\bibnamefont
  {Doubinina}}, \bibinfo {author} {\bibfnamefont {J.}~\bibnamefont {Zhang}},
  \bibinfo {author} {\bibfnamefont {Y.~Y.}\ \bibnamefont {Wang}}, \bibinfo
  {author} {\bibfnamefont {V.~P.}\ \bibnamefont {Dravid}},\ and\ \bibinfo
  {author} {\bibfnamefont {T.~J.}\ \bibnamefont {Marks}},\ }\bibfield  {title}
  {\bibinfo {title} {Liquid {Source} {Mocvd} of {High} {Quality} {Yba2Cu3O7}-x
  {Films} on {Polycrystalline} {And} {Amorphous} {Substrates}},\ }\href
  {https://doi.org/10.1557/PROC-415-255} {\bibfield  {journal} {\bibinfo
  {journal} {MRS Online Proceedings Library (OPL)}\ }\textbf {\bibinfo {volume}
  {415}},\ \bibinfo {pages} {255} (\bibinfo {year} {1995})}\BibitemShut
  {NoStop}%
\bibitem [{\citenamefont {Deyu}\ \emph {et~al.}(2025)\citenamefont {Deyu},
  \citenamefont {Wenskat}, \citenamefont {Díaz-Palacio}, \citenamefont
  {Blick}, \citenamefont {Zierold},\ and\ \citenamefont
  {Hillert}}]{deyu_recent_2025}%
  \BibitemOpen
  \bibfield  {author} {\bibinfo {author} {\bibfnamefont {G.~K.}\ \bibnamefont
  {Deyu}}, \bibinfo {author} {\bibfnamefont {M.}~\bibnamefont {Wenskat}},
  \bibinfo {author} {\bibfnamefont {I.~G.}\ \bibnamefont {Díaz-Palacio}},
  \bibinfo {author} {\bibfnamefont {R.~H.}\ \bibnamefont {Blick}}, \bibinfo
  {author} {\bibfnamefont {R.}~\bibnamefont {Zierold}},\ and\ \bibinfo {author}
  {\bibfnamefont {W.}~\bibnamefont {Hillert}},\ }\bibfield  {title} {\bibinfo
  {title} {Recent advances in atomic layer deposition of superconducting thin
  films: a review},\ }\href {https://doi.org/10.1039/D5MH00323G} {\bibfield
  {journal} {\bibinfo  {journal} {Materials Horizons}\ }\textbf {\bibinfo
  {volume} {12}},\ \bibinfo {pages} {5594} (\bibinfo {year} {2025})},\ \bibinfo
  {note} {publisher: The Royal Society of Chemistry}\BibitemShut {NoStop}%
\bibitem [{\citenamefont {Golden}(1966)}]{goldenComparisonLowEnergyTotal1966}%
  \BibitemOpen
  \bibfield  {author} {\bibinfo {author} {\bibfnamefont {D.~E.}\ \bibnamefont
  {Golden}},\ }\bibfield  {title} {\bibinfo {title} {Comparison of {{Low-Energy
  Total}} and {{Momentum-Transfer Scattering Cross Sections}} for {{Electrons}}
  on {{Helium}} and {{Argon}}},\ }\href
  {https://doi.org/10.1103/PhysRev.151.48} {\bibfield  {journal} {\bibinfo
  {journal} {Physical Review}\ }\textbf {\bibinfo {volume} {151}},\ \bibinfo
  {pages} {48} (\bibinfo {year} {1966})}\BibitemShut {NoStop}%
\bibitem [{\citenamefont {Higgins}\ \emph {et~al.}(2017)\citenamefont
  {Higgins}, \citenamefont {Pokorny}, \citenamefont {Zhang}, \citenamefont
  {Bodart},\ and\ \citenamefont {Hennrich}}]{higginsCoherentControlSingle2017}%
  \BibitemOpen
  \bibfield  {author} {\bibinfo {author} {\bibfnamefont {G.}~\bibnamefont
  {Higgins}}, \bibinfo {author} {\bibfnamefont {F.}~\bibnamefont {Pokorny}},
  \bibinfo {author} {\bibfnamefont {C.}~\bibnamefont {Zhang}}, \bibinfo
  {author} {\bibfnamefont {Q.}~\bibnamefont {Bodart}},\ and\ \bibinfo {author}
  {\bibfnamefont {M.}~\bibnamefont {Hennrich}},\ }\bibfield  {title} {\bibinfo
  {title} {Coherent {{Control}} of a {{Single Trapped Rydberg Ion}}},\ }\href
  {https://doi.org/10.1103/PhysRevLett.119.220501} {\bibfield  {journal}
  {\bibinfo  {journal} {Physical Review Letters}\ }\textbf {\bibinfo {volume}
  {119}},\ \bibinfo {pages} {220501} (\bibinfo {year} {2017})}\BibitemShut
  {NoStop}%
\bibitem [{\citenamefont {Savard}\ \emph {et~al.}(1997)\citenamefont {Savard},
  \citenamefont {O’Hara},\ and\ \citenamefont {Thomas}}]{savard_1997}%
  \BibitemOpen
  \bibfield  {author} {\bibinfo {author} {\bibfnamefont {T.~A.}\ \bibnamefont
  {Savard}}, \bibinfo {author} {\bibfnamefont {K.~M.}\ \bibnamefont
  {O’Hara}},\ and\ \bibinfo {author} {\bibfnamefont {J.~E.}\ \bibnamefont
  {Thomas}},\ }\bibfield  {title} {\bibinfo {title} {Laser-noise-induced
  heating in far-off resonance optical traps},\ }\href
  {https://doi.org/10.1103/physreva.56.r1095} {\bibfield  {journal} {\bibinfo
  {journal} {Physical Review A}\ }\textbf {\bibinfo {volume} {56}},\ \bibinfo
  {pages} {R1095–R1098} (\bibinfo {year} {1997})}\BibitemShut {NoStop}%
\bibitem [{\citenamefont {Brownnutt}\ \emph {et~al.}(2015)\citenamefont
  {Brownnutt}, \citenamefont {Kumph}, \citenamefont {Rabl},\ and\ \citenamefont
  {Blatt}}]{brownnutt_ion-trap_2015}%
  \BibitemOpen
  \bibfield  {author} {\bibinfo {author} {\bibfnamefont {M.}~\bibnamefont
  {Brownnutt}}, \bibinfo {author} {\bibfnamefont {M.}~\bibnamefont {Kumph}},
  \bibinfo {author} {\bibfnamefont {P.}~\bibnamefont {Rabl}},\ and\ \bibinfo
  {author} {\bibfnamefont {R.}~\bibnamefont {Blatt}},\ }\bibfield  {title}
  {\bibinfo {title} {Ion-trap measurements of electric-field noise near
  surfaces},\ }\href {https://doi.org/10.1103/RevModPhys.87.1419} {\bibfield
  {journal} {\bibinfo  {journal} {Reviews of Modern Physics}\ }\textbf
  {\bibinfo {volume} {87}},\ \bibinfo {pages} {1419} (\bibinfo {year}
  {2015})}\BibitemShut {NoStop}%
\bibitem [{\citenamefont {Hudák}\ \emph {et~al.}(2025)\citenamefont {Hudák},
  \citenamefont {Kumar}, \citenamefont {Lausti}, \citenamefont {Honzátko},\
  and\ \citenamefont {Hejduk}}]{hudak_microcavity_2025}%
  \BibitemOpen
  \bibfield  {author} {\bibinfo {author} {\bibfnamefont {I.}~\bibnamefont
  {Hudák}}, \bibinfo {author} {\bibfnamefont {V.}~\bibnamefont {Kumar}},
  \bibinfo {author} {\bibfnamefont {N.}~\bibnamefont {Lausti}}, \bibinfo
  {author} {\bibfnamefont {P.}~\bibnamefont {Honzátko}},\ and\ \bibinfo
  {author} {\bibfnamefont {M.}~\bibnamefont {Hejduk}},\ }\bibfield  {title}
  {\bibinfo {title} {Microcavity integration with {2D} {Paul} trap},\ }in\
  \href {https://doi.org/10.1117/12.3056538} {\emph {\bibinfo {booktitle}
  {Quantum {Optics} and {Photon} {Counting} 2025}}},\ Vol.\ \bibinfo {volume}
  {13525}\ (\bibinfo  {publisher} {SPIE},\ \bibinfo {year} {2025})\ pp.\
  \bibinfo {pages} {40--46}\BibitemShut {NoStop}%
\bibitem [{\citenamefont {Tu}\ \emph {et~al.}(2024)\citenamefont {Tu},
  \citenamefont {Liao}, \citenamefont {Wang}, \citenamefont {Zhu},
  \citenamefont {Qiu}, \citenamefont {Jiang}, \citenamefont {Huang},
  \citenamefont {Bian}, \citenamefont {Yan},\ and\ \citenamefont
  {Zhu}}]{tu_approaching_2024}%
  \BibitemOpen
  \bibfield  {author} {\bibinfo {author} {\bibfnamefont {H.-T.}\ \bibnamefont
  {Tu}}, \bibinfo {author} {\bibfnamefont {K.-Y.}\ \bibnamefont {Liao}},
  \bibinfo {author} {\bibfnamefont {H.-L.}\ \bibnamefont {Wang}}, \bibinfo
  {author} {\bibfnamefont {Y.-F.}\ \bibnamefont {Zhu}}, \bibinfo {author}
  {\bibfnamefont {S.-Y.}\ \bibnamefont {Qiu}}, \bibinfo {author} {\bibfnamefont
  {H.}~\bibnamefont {Jiang}}, \bibinfo {author} {\bibfnamefont
  {W.}~\bibnamefont {Huang}}, \bibinfo {author} {\bibfnamefont
  {W.}~\bibnamefont {Bian}}, \bibinfo {author} {\bibfnamefont {H.}~\bibnamefont
  {Yan}},\ and\ \bibinfo {author} {\bibfnamefont {S.-L.}\ \bibnamefont {Zhu}},\
  }\bibfield  {title} {\bibinfo {title} {Approaching the standard quantum limit
  of a {Rydberg}-atom microwave electrometer},\ }\bibfield  {journal} {\bibinfo
   {journal} {Science Advances}\ }\textbf {\bibinfo {volume} {10}},\ \href
  {https://doi.org/10.1126/sciadv.ads0683} {10.1126/sciadv.ads0683} (\bibinfo
  {year} {2024})\BibitemShut {NoStop}%
\bibitem [{\citenamefont {Degen}\ \emph {et~al.}(2017)\citenamefont {Degen},
  \citenamefont {Reinhard},\ and\ \citenamefont
  {Cappellaro}}]{degen_quantum_2017}%
  \BibitemOpen
  \bibfield  {author} {\bibinfo {author} {\bibfnamefont {C.}~\bibnamefont
  {Degen}}, \bibinfo {author} {\bibfnamefont {F.}~\bibnamefont {Reinhard}},\
  and\ \bibinfo {author} {\bibfnamefont {P.}~\bibnamefont {Cappellaro}},\
  }\bibfield  {title} {\bibinfo {title} {Quantum sensing},\ }\href
  {https://doi.org/10.1103/RevModPhys.89.035002} {\bibfield  {journal}
  {\bibinfo  {journal} {Reviews of Modern Physics}\ }\textbf {\bibinfo {volume}
  {89}},\ \bibinfo {pages} {035002} (\bibinfo {year} {2017})},\ \bibinfo {note}
  {publisher: American Physical Society}\BibitemShut {NoStop}%
\bibitem [{\citenamefont {Gao}\ \emph {et~al.}(2023)\citenamefont {Gao},
  \citenamefont {Sun}, \citenamefont {Yu}, \citenamefont {Feng}, \citenamefont
  {Ren}, \citenamefont {Zhang}, \citenamefont {Wu},\ and\ \citenamefont
  {Xiao}}]{gao_investigation_2023}%
  \BibitemOpen
  \bibfield  {author} {\bibinfo {author} {\bibfnamefont {M.}~\bibnamefont
  {Gao}}, \bibinfo {author} {\bibfnamefont {J.}~\bibnamefont {Sun}}, \bibinfo
  {author} {\bibfnamefont {S.}~\bibnamefont {Yu}}, \bibinfo {author}
  {\bibfnamefont {J.}~\bibnamefont {Feng}}, \bibinfo {author} {\bibfnamefont
  {X.}~\bibnamefont {Ren}}, \bibinfo {author} {\bibfnamefont {Y.}~\bibnamefont
  {Zhang}}, \bibinfo {author} {\bibfnamefont {X.}~\bibnamefont {Wu}},\ and\
  \bibinfo {author} {\bibfnamefont {D.}~\bibnamefont {Xiao}},\ }\bibfield
  {title} {\bibinfo {title} {Investigation of the {Charge} {Accumulation}
  {Based} on {Stiffness} {Variation} of the {Micro}-{Shell} {Resonator}
  {Gyroscope}},\ }\href {https://doi.org/10.3390/mi14091755} {\bibfield
  {journal} {\bibinfo  {journal} {Micromachines}\ }\textbf {\bibinfo {volume}
  {14}},\ \bibinfo {pages} {1755} (\bibinfo {year} {2023})}\BibitemShut
  {NoStop}%
\bibitem [{\citenamefont {Koszewski}\ \emph {et~al.}(2013)\citenamefont
  {Koszewski}, \citenamefont {Souchon}, \citenamefont {Dieppedale},
  \citenamefont {Bloch},\ and\ \citenamefont
  {Ouisse}}]{koszewski_physical_2013}%
  \BibitemOpen
  \bibfield  {author} {\bibinfo {author} {\bibfnamefont {A.}~\bibnamefont
  {Koszewski}}, \bibinfo {author} {\bibfnamefont {F.}~\bibnamefont {Souchon}},
  \bibinfo {author} {\bibfnamefont {C.}~\bibnamefont {Dieppedale}}, \bibinfo
  {author} {\bibfnamefont {D.}~\bibnamefont {Bloch}},\ and\ \bibinfo {author}
  {\bibfnamefont {T.}~\bibnamefont {Ouisse}},\ }\bibfield  {title} {\bibinfo
  {title} {Physical model of dielectric charging in {MEMS}},\ }\href
  {https://doi.org/10.1088/0960-1317/23/4/045019} {\bibfield  {journal}
  {\bibinfo  {journal} {Journal of Micromechanics and Microengineering}\
  }\textbf {\bibinfo {volume} {23}},\ \bibinfo {pages} {045019} (\bibinfo
  {year} {2013})}\BibitemShut {NoStop}%
\bibitem [{\citenamefont {Bahl}\ \emph {et~al.}(2010)\citenamefont {Bahl},
  \citenamefont {Salvia}, \citenamefont {Bargatin}, \citenamefont {Yoneoka},
  \citenamefont {Melamud}, \citenamefont {Kim}, \citenamefont {Chandorkar},
  \citenamefont {Hopcroft}, \citenamefont {Bahl}, \citenamefont {Howe},\ and\
  \citenamefont {Kenny}}]{bahl_charge-drift_2010}%
  \BibitemOpen
  \bibfield  {author} {\bibinfo {author} {\bibfnamefont {G.}~\bibnamefont
  {Bahl}}, \bibinfo {author} {\bibfnamefont {J.}~\bibnamefont {Salvia}},
  \bibinfo {author} {\bibfnamefont {I.}~\bibnamefont {Bargatin}}, \bibinfo
  {author} {\bibfnamefont {S.}~\bibnamefont {Yoneoka}}, \bibinfo {author}
  {\bibfnamefont {R.}~\bibnamefont {Melamud}}, \bibinfo {author} {\bibfnamefont
  {B.}~\bibnamefont {Kim}}, \bibinfo {author} {\bibfnamefont {S.}~\bibnamefont
  {Chandorkar}}, \bibinfo {author} {\bibfnamefont {M.~A.}\ \bibnamefont
  {Hopcroft}}, \bibinfo {author} {\bibfnamefont {R.}~\bibnamefont {Bahl}},
  \bibinfo {author} {\bibfnamefont {R.~T.}\ \bibnamefont {Howe}},\ and\
  \bibinfo {author} {\bibfnamefont {T.~W.}\ \bibnamefont {Kenny}},\ }\bibfield
  {title} {\bibinfo {title} {Charge-drift elimination in resonant electrostatic
  {MEMS}},\ }in\ \href {https://doi.org/10.1109/MEMSYS.2010.5442555} {\emph
  {\bibinfo {booktitle} {2010 {IEEE} 23rd {International} {Conference} on
  {Micro} {Electro} {Mechanical} {Systems} ({MEMS})}}}\ (\bibinfo {year}
  {2010})\ pp.\ \bibinfo {pages} {108--111}\BibitemShut {NoStop}%
\bibitem [{\citenamefont {Ziemba}\ \emph {et~al.}(2025)\citenamefont {Ziemba},
  \citenamefont {Phrompao}, \citenamefont {Jung}, \citenamefont {Rabey},\ and\
  \citenamefont {Rempe}}]{ziemba_removal_2025}%
  \BibitemOpen
  \bibfield  {author} {\bibinfo {author} {\bibfnamefont {M.~T.}\ \bibnamefont
  {Ziemba}}, \bibinfo {author} {\bibfnamefont {J.}~\bibnamefont {Phrompao}},
  \bibinfo {author} {\bibfnamefont {F.}~\bibnamefont {Jung}}, \bibinfo {author}
  {\bibfnamefont {I.~M.}\ \bibnamefont {Rabey}},\ and\ \bibinfo {author}
  {\bibfnamefont {G.}~\bibnamefont {Rempe}},\ }\bibfield  {title} {\bibinfo
  {title} {Removal of high-voltage-induced surface charges by ultraviolet
  light},\ }\href {https://doi.org/10.1063/5.0269038} {\bibfield  {journal}
  {\bibinfo  {journal} {Review of Scientific Instruments}\ }\textbf {\bibinfo
  {volume} {96}},\ \bibinfo {pages} {073201} (\bibinfo {year}
  {2025})}\BibitemShut {NoStop}%
\bibitem [{\citenamefont {Quinn}\ \emph {et~al.}(2022)\citenamefont {Quinn},
  \citenamefont {Brown}, \citenamefont {Gardner},\ and\ \citenamefont
  {Allcock}}]{quinn_geometries_2022}%
  \BibitemOpen
  \bibfield  {author} {\bibinfo {author} {\bibfnamefont {A.}~\bibnamefont
  {Quinn}}, \bibinfo {author} {\bibfnamefont {M.}~\bibnamefont {Brown}},
  \bibinfo {author} {\bibfnamefont {T.~J.}\ \bibnamefont {Gardner}},\ and\
  \bibinfo {author} {\bibfnamefont {D.~T.~C.}\ \bibnamefont {Allcock}},\ }\href
  {https://doi.org/10.48550/ARXIV.2205.15892} {\bibinfo {title} {Geometries and
  fabrication methods for {3D} printing ion traps}} (\bibinfo {year}
  {2022})\BibitemShut {NoStop}%
\bibitem [{\citenamefont {Xu}\ \emph {et~al.}(2025)\citenamefont {Xu},
  \citenamefont {Xia}, \citenamefont {Yu}, \citenamefont {Parakh},
  \citenamefont {Khan}, \citenamefont {Megidish}, \citenamefont {You},
  \citenamefont {Hemmerling}, \citenamefont {Jayich}, \citenamefont {Beck},
  \citenamefont {Biener},\ and\ \citenamefont
  {Häffner}}]{xu_3d-printed_2025-nature}%
  \BibitemOpen
  \bibfield  {author} {\bibinfo {author} {\bibfnamefont {S.}~\bibnamefont
  {Xu}}, \bibinfo {author} {\bibfnamefont {X.}~\bibnamefont {Xia}}, \bibinfo
  {author} {\bibfnamefont {Q.}~\bibnamefont {Yu}}, \bibinfo {author}
  {\bibfnamefont {A.}~\bibnamefont {Parakh}}, \bibinfo {author} {\bibfnamefont
  {S.}~\bibnamefont {Khan}}, \bibinfo {author} {\bibfnamefont {E.}~\bibnamefont
  {Megidish}}, \bibinfo {author} {\bibfnamefont {B.}~\bibnamefont {You}},
  \bibinfo {author} {\bibfnamefont {B.}~\bibnamefont {Hemmerling}}, \bibinfo
  {author} {\bibfnamefont {A.}~\bibnamefont {Jayich}}, \bibinfo {author}
  {\bibfnamefont {K.}~\bibnamefont {Beck}}, \bibinfo {author} {\bibfnamefont
  {J.}~\bibnamefont {Biener}},\ and\ \bibinfo {author} {\bibfnamefont
  {H.}~\bibnamefont {Häffner}},\ }\bibfield  {title} {\bibinfo {title}
  {{3D}-printed micro ion trap technology for quantum information
  applications},\ }\href {https://doi.org/10.1038/s41586-025-09474-1}
  {\bibfield  {journal} {\bibinfo  {journal} {Nature}\ }\textbf {\bibinfo
  {volume} {645}},\ \bibinfo {pages} {362} (\bibinfo {year}
  {2025})}\BibitemShut {NoStop}%
\bibitem [{\citenamefont {Hejduk}\ and\ \citenamefont
  {Lausti}(2024)}]{hejduk_supply_2024}%
  \BibitemOpen
  \bibfield  {author} {\bibinfo {author} {\bibfnamefont {M.}~\bibnamefont
  {Hejduk}}\ and\ \bibinfo {author} {\bibfnamefont {N.~V.}\ \bibnamefont
  {Lausti}},\ }\href@noop {} {\bibinfo {title} {A supply circuit system of a
  planar {Paul} trap}} (\bibinfo {year} {2024}),\ \bibinfo {note} {patent
  number CZ310234B6}\BibitemShut {NoStop}%
\bibitem [{\citenamefont {{Niklas Lausti}}\ \emph {et~al.}()\citenamefont
  {{Niklas Lausti}}, \citenamefont {{Michal Hejduk}},\ and\ \citenamefont
  {{Michal Tarana}}}]{niklaslaustiSupportingDataRoadmap}%
  \BibitemOpen
  \bibfield  {author} {\bibinfo {author} {\bibnamefont {{Niklas Lausti}}},
  \bibinfo {author} {\bibnamefont {{Michal Hejduk}}},\ and\ \bibinfo {author}
  {\bibnamefont {{Michal Tarana}}},\ }\href
  {https://doi.org/10.48700/datst.8n8wk-52567} {\bibinfo {title} {Supporting
  {{Data}} for "{{Roadmap}} to planar electron-ion point {{Paul}}
  trap"}}\BibitemShut {NoStop}%
\bibitem [{\citenamefont {Xin}\ \emph {et~al.}(2000)\citenamefont {Xin},
  \citenamefont {Oates}, \citenamefont {Anderson}, \citenamefont {Slattery},
  \citenamefont {Dresselhaus},\ and\ \citenamefont
  {Dresselhaus}}]{xin_comparison_2000}%
  \BibitemOpen
  \bibfield  {author} {\bibinfo {author} {\bibfnamefont {H.}~\bibnamefont
  {Xin}}, \bibinfo {author} {\bibfnamefont {D.}~\bibnamefont {Oates}}, \bibinfo
  {author} {\bibfnamefont {A.}~\bibnamefont {Anderson}}, \bibinfo {author}
  {\bibfnamefont {A.}~\bibnamefont {Slattery}}, \bibinfo {author}
  {\bibfnamefont {G.}~\bibnamefont {Dresselhaus}},\ and\ \bibinfo {author}
  {\bibfnamefont {M.}~\bibnamefont {Dresselhaus}},\ }\bibfield  {title}
  {\bibinfo {title} {Comparison of power dependence of microwave surface
  resistance of unpatterned and patterned {YBCO} thin film},\ }\href
  {https://doi.org/10.1109/22.853465} {\bibfield  {journal} {\bibinfo
  {journal} {IEEE Transactions on Microwave Theory and Techniques}\ }\textbf
  {\bibinfo {volume} {48}},\ \bibinfo {pages} {1221} (\bibinfo {year}
  {2000})}\BibitemShut {NoStop}%
\bibitem [{\citenamefont {Turchette}\ \emph {et~al.}(2000)\citenamefont
  {Turchette}, \citenamefont {Kielpinski}, \citenamefont {King}, \citenamefont
  {Leibfried}, \citenamefont {Meekhof}, \citenamefont {Myatt}, \citenamefont
  {Rowe}, \citenamefont {Sackett}, \citenamefont {Wood}, \citenamefont {Itano},
  \citenamefont {Monroe},\ and\ \citenamefont {Wineland}}]{turchette_2000}%
  \BibitemOpen
  \bibfield  {author} {\bibinfo {author} {\bibfnamefont {Q.~A.}\ \bibnamefont
  {Turchette}}, \bibinfo {author} {\bibnamefont {Kielpinski}}, \bibinfo
  {author} {\bibfnamefont {B.~E.}\ \bibnamefont {King}}, \bibinfo {author}
  {\bibfnamefont {D.}~\bibnamefont {Leibfried}}, \bibinfo {author}
  {\bibfnamefont {D.~M.}\ \bibnamefont {Meekhof}}, \bibinfo {author}
  {\bibfnamefont {C.~J.}\ \bibnamefont {Myatt}}, \bibinfo {author}
  {\bibfnamefont {M.~A.}\ \bibnamefont {Rowe}}, \bibinfo {author}
  {\bibfnamefont {C.~A.}\ \bibnamefont {Sackett}}, \bibinfo {author}
  {\bibfnamefont {C.~S.}\ \bibnamefont {Wood}}, \bibinfo {author}
  {\bibfnamefont {W.~M.}\ \bibnamefont {Itano}}, \bibinfo {author}
  {\bibfnamefont {C.}~\bibnamefont {Monroe}},\ and\ \bibinfo {author}
  {\bibfnamefont {D.~J.}\ \bibnamefont {Wineland}},\ }\bibfield  {title}
  {\bibinfo {title} {Heating of trapped ions from the quantum ground state},\
  }\bibfield  {journal} {\bibinfo  {journal} {Physical Review A}\ }\textbf
  {\bibinfo {volume} {61}},\ \href {https://doi.org/10.1103/physreva.61.063418}
  {10.1103/physreva.61.063418} (\bibinfo {year} {2000})\BibitemShut {NoStop}%
\bibitem [{\citenamefont {Lin}\ \emph {et~al.}(2016)\citenamefont {Lin},
  \citenamefont {Low},\ and\ \citenamefont {Chuang}}]{lin_effects_2016}%
  \BibitemOpen
  \bibfield  {author} {\bibinfo {author} {\bibfnamefont {K.-Y.}\ \bibnamefont
  {Lin}}, \bibinfo {author} {\bibfnamefont {G.~H.}\ \bibnamefont {Low}},\ and\
  \bibinfo {author} {\bibfnamefont {I.~L.}\ \bibnamefont {Chuang}},\ }\bibfield
   {title} {\bibinfo {title} {Effects of electrode surface roughness on
  motional heating of trapped ions},\ }\href
  {https://doi.org/10.1103/PhysRevA.94.013418} {\bibfield  {journal} {\bibinfo
  {journal} {Physical Review A}\ }\textbf {\bibinfo {volume} {94}},\ \bibinfo
  {pages} {013418} (\bibinfo {year} {2016})}\BibitemShut {NoStop}%
\bibitem [{\citenamefont {Wang}\ \emph {et~al.}(2010)\citenamefont {Wang},
  \citenamefont {Ge}, \citenamefont {Labaziewicz}, \citenamefont {Dauler},
  \citenamefont {Berggren},\ and\ \citenamefont
  {Chuang}}]{wang_superconducting_2010}%
  \BibitemOpen
  \bibfield  {author} {\bibinfo {author} {\bibfnamefont {S.~X.}\ \bibnamefont
  {Wang}}, \bibinfo {author} {\bibfnamefont {Y.}~\bibnamefont {Ge}}, \bibinfo
  {author} {\bibfnamefont {J.}~\bibnamefont {Labaziewicz}}, \bibinfo {author}
  {\bibfnamefont {E.}~\bibnamefont {Dauler}}, \bibinfo {author} {\bibfnamefont
  {K.}~\bibnamefont {Berggren}},\ and\ \bibinfo {author} {\bibfnamefont
  {I.~L.}\ \bibnamefont {Chuang}},\ }\bibfield  {title} {\bibinfo {title}
  {Superconducting microfabricated ion traps},\ }\href
  {https://doi.org/10.1063/1.3526733} {\bibfield  {journal} {\bibinfo
  {journal} {Applied Physics Letters}\ }\textbf {\bibinfo {volume} {97}},\
  \bibinfo {pages} {244102} (\bibinfo {year} {2010})}\BibitemShut {NoStop}%
\end{thebibliography}%

\end{document}